\documentclass[10pt,aps,prc,amsmath,amssymb,twocolumn,floatfix,nofootinbib,
               superscriptaddress]{revtex4-1}

\usepackage[utf8]{inputenc}
\usepackage[english]{babel}
\usepackage{amssymb,amsthm,amsmath,amstext,amsbsy,amsopn,amsfonts}
\usepackage{bbm}
\usepackage{url}
\usepackage{nicefrac}
\usepackage{graphicx}
\usepackage{feynmf}
\usepackage{ifpdf}
\usepackage{pstricks,pst-plot}
\usepackage{leftidx}
\usepackage{environ}
\usepackage{suffix}
\usepackage{xspace}
\usepackage{overpic}
\usepackage{tikz}
\usepackage[math]{cellspace}
\usepackage{makecell}

\usepackage{hyperref}
\graphicspath{{./figures/}}

\newcommand{\beq}{\begin{equation}}
\newcommand{\eeq}{\end{equation}}
\newcommand{\beqstar}{\begin{equation*}}
\newcommand{\eeqstar}{\end{equation*}}
\newcommand{\bea}{\begin{eqnarray}}
\newcommand{\eea}{\end{eqnarray}}
\newcommand{\la}{\langle}
\newcommand{\ra}{\rangle}
\newcommand{\bi}{\begin{itemize}}
\newcommand{\ei}{\end{itemize}}

\newcommand{\I}{\item}

\newcommand{\beal}{\begin{align}}
\newcommand{\eeal}{\end{align}}

\newcommand{\kmax}{k_{\rm max}}

\newcommand{\vecq}{\mathbf{q}}

\newcommand{\dd}{\mathrm{d}}

\newcommand{\ket}[1]{|#1\rangle}

\newcommand{\braket}[2]{\langle #1|#2\rangle}

\newcommand{\mbraket}[3]{\langle #1|#2|#3\rangle}

\NewEnviron{subalign}[1][]{%
\begin{subequations}\begin{align}
  \BODY
\end{align}\label{#1}\end{subequations}
}

\NewEnviron{spliteq}{%
\begin{equation}\begin{split}
  \BODY
\end{split}\end{equation}
}

\newcommand{\fmi}{\ensuremath{\mathrm{fm}^{-1}}}

\newcommand{\CG}[6]{\ensuremath{\braket{#1\,#2\,#3\,#4}{#5\,#6}}}
\WithSuffix\newcommand\CG*[6]{\ensuremath{\braket{#1\,#2}{#3\,#4\,#5\,#6}}}
\newcommand{\mbf}{\mathbf}
\newcommand{\pp}{p^\prime}
\newcommand{\Ep}{E^\prime}

\newcommand{\fL}{f_L}

\newcommand{\msf}{m_{s_f}}
\newcommand{\mJd}{m_{J_d}}
\newcommand{\mJ}{m_{J}}
\newcommand{\wt}{\widetilde}
\newcommand{\mst}{\wt{m}_s}

\newcommand{\thetap}{\theta^{\prime}}
\newcommand{\phip}{\varphi^{\prime}}

\begin{document}

\title{Scale dependence of deuteron electrodisintegration}

\author{S.~N.~More}
\email{more@nscl.msu.edu}
\affiliation{National Superconducting Cyclotron Laboratory and Department 
of Physics and Astronomy, Michigan State University, East Lansing, MI 48824, USA}

\author{S.~K.~Bogner}
\email{bogner@nscl.msu.edu}
\affiliation{National Superconducting Cyclotron Laboratory and Department 
of Physics and Astronomy, Michigan State University, East Lansing, MI 48824, USA}

\author{R.~J.~Furnstahl}
\email{furnstahl.1@osu.edu}
\affiliation{Department of Physics, The Ohio State University,
Columbus, Ohio 43210, USA}

\date{\today}

\begin{abstract}

\begin{description}

	\item[Background] 
Isolating nuclear structure properties from knock-out reactions in a process-independent
manner requires a controlled factorization, which is always to some degree scale and 
scheme dependent.  Understanding this dependence is important
for robust extractions from experiment, to correctly use the structure information
in other processes, and to understand the impact of approximations for 
both.

	\item[Purpose]
We seek insight into scale dependence by exploring a model calculation of deuteron electrodisintegration, which provides a simple
and clean theoretical laboratory. 

	\item[Methods] 
By considering various kinematic regions of the longitudinal structure function,
we can examine how the components---the initial deuteron
wave function, the current operator, and the final state interactions (FSI)---combine at different scales.
We use the similarity renormalization group (SRG) to evolve each component.

	\item[Results]
When evolved to different resolutions, the ingredients 
are all modified, but how they combine depends strongly on the kinematic region.
In some regions, for example, the FSI are largely unaffected by evolution, while
elsewhere FSI are greatly reduced.
For certain kinematics, the impulse approximation at a
high RG resolution gives an intuitive picture in terms of a one-body 
current breaking up a short-range correlated neutron-proton pair, although
FSI distort this simple picture.
With evolution to low resolution, however, the cross section is unchanged but
a very different and arguably simpler intuitive picture emerges, with the evolved current
efficiently represented at low momentum through derivative expansions or low-rank singular value decompositions.

	\item[Conclusions]
The underlying physics of deuteron electrodisintegration is scale dependent and not just
kinematics dependent. 
As a result, intuition about physics such as the role of short-range correlations
or $D$-state mixing in particular kinematic regimes can be strongly scale dependent.
Understanding this dependence is crucial in making use of extracted properties.

\end{description}

\end{abstract}

\maketitle

%%%%%%%%%%%%%%%%%%%%%%%%%%%%%%%%%%%%%%%%%%%%%%%%%%%%%%%%
%%%%%%%%%%%%%%%%%%%%%%%%%%%%%%%%%%%%%%%%%%%%%%%%%%%%%%%%

\section{Introduction}  \label{sec:intro}

Structure information about nuclei is often deduced from knock-out processes,
such as $(p,2p)$ reactions or high-momentum electron scattering.  
Isolating the nuclear structure properties from the reaction dynamics in a 
\emph{process independent} manner requires a controlled
factorization of structure and reaction.
This factorization is always to some degree scale and scheme dependent, because
the dividing point between structure and reaction is not unique.  
Understanding this dependence is important
for robust extractions from experiment, to correctly use the structure information
in other processes, and to understand the impact of approximations for 
both~\cite{Furnstahl:2010wd,Furnstahl:2013dsa}.

Scale and scheme dependence in quantum field theoretic treatments of knock-out
reactions, such as high-energy scattering in quantum 
chromodynamics~\cite{RevModPhys.67.157,Collins:2011zzd}, is
manifested in explicit factorization and renormalization prescriptions.  
In a non-relativistic many-body treatment, scale and scheme dependence 
is hidden in the choice of inter-nucleon potentials and associated currents.
These approaches to knock-out processes are bridged by nuclear effective field theories 
(EFTs)~\cite{Bedaque:2002mn,Epelbaum:2008ga,Machleidt:2011zz,Epelbaum:2012vx},
where the scale is associated with the value of a cutoff parameter and the scheme
with the choice of regulator (and other details of renormalization).

We sometimes refer to this scale dependence as a resolution dependence~\cite{Bogner:2009bt,Furnstahl:2013dsa},
because the wavelengths available are restricted by the potential.    
It is important to distinguish this resolution implied by the potential
from the experimental resolution, which is
dictated by the kinematics of the experiment.  The latter is fixed for
a given experiment while the former can be changed continuously, e.g., by
unitary transformations.

While the cross sections for knock-out reactions are independent of the
factorization scale, the individual components of a theoretical calculation 
--- initial state, interaction current, final state interactions (FSI) --- are not.
As we will show, for some kinematics
the FSI contribution can be substantial at high resolution but largely absent at
a lower resolution.
Furthermore, the physics interpretation of the process can
change with scale.
What is dominantly short-range correlation and/or $D$-state structure physics at
one scale can be mostly low-momentum $S$-state physics at another scale. 

The changing interplay of the different components as the scale
changes is often not immediately intuitive; e.g., how do cross sections
remain invariant as one-body currents become two-body currents
and short-range structure disappears? 
The goal of this paper is to present a clean, focused example that illustrates
the interplay without getting lost in approximations, which opens the door to an intuitive 
understanding of scale-dependent features that we hope can one day be transferred to more
complicated nuclear processes.   

Ultimately a pertinent question is: What is the best choice of scale?
Some motivations for that choice include:
  \bi
   \I To make calculations easier or more convergent.  
   In field theory, this can mean
   choosing the QCD running coupling and scale to improve perturbation theory.
   % or choosing a particular gauge, e.g., Coulomb or Lorentz.
   For many-body problems, this could mean
   using a soft potential to improve many-body convergence,
   or to make microscopic connections to the shell model or density functional
   theory~\cite{Bogner:2009bt}.
   %Or it could mean using a higher-resolution (near-) local potential so that certain quantum Monte Carlo 
   %methods are optimized~\cite{Lynn:2017fxg}.

   \I Better interpretation or intuition, which can lead to more predictability.
     For example, short-range correlation SRC phenomenology for high-momentum 
     transfer electron scattering from nuclei has many successes in explaining
     and predicting experiment~\cite{Alvioli:2013qyz,Vanhalst:2014cqa,Atti:2015eda,Hen:2016kwk}.  
     But, as we will see, a low-resolution scale can also lead to an intuitive picture,
     with complementary advantages.

   \I Allowing for the cleanest extraction from experiment. 
   Final-state interactions are usually a hindrance to extracting structure information
   from electron scattering measurements. 
   Can the choice of scale allow one to ``optimize'' the validity of the impulse 
   approximation or the assumed factorization of structure and reaction?
   Ideally one extracts from a given experiment at the optimal scale
   for the kinematics, then relates to other scales to compare to other experiment or
   theoretical predictions.
   In inclusive high-energy QCD scattering, the optimal scale is typically the four-momentum
   transfer squared of the experiment, but this is not universally true~\cite{Collins:2011zzd}.
  \ei
To study scale dependence and relate nuclear processes at different scales,
the renormalization group (RG) has proven to be a powerful method~\cite{Furnstahl:2012fn,Duguet:2014tua}.
But conventional nuclear knock-out experiments have not been analyzed
using variable resolution with RG methods.  
We strongly advocate embedding the usual approaches in an RG framework, 
which will provide interconnections between calculations with different
potentials and with different approaches such as EFTs and use of
the operator product expansion (OPE)~\cite{Braaten:2008uh,Hofmann:2013oia}.
The present work is a contribution toward realizing this framework.

A candidate RG framework for nuclear applications is
the similarity renormalization group (SRG), which is a useful and versatile
tool for such questions~\cite{Bogner:2006pc,Bogner:2009bt,Furnstahl:2012fn,Furnstahl:2013oba,Binder:2013xaa,Roth:2013fqa,Roth:2014vla,Schuster:2014lga,Neff:2015xda,Neff:2016ajx,Parzuchowski:2017wcq}.  
The SRG has been widely applied for nuclear structure applications, both to
soften the inter-nucleon potential in free space and as a many-body
solution method in the form of the in-medium SRG~\cite{Hergert2017}. %\cite{Hergert:2016iju}.
The improved convergence has also enabled ab initio reaction calculations using
the NCSM-RGM approach~\cite{Bacca:2014tla,Navratil:2016ycn}. 

A recent paper~\cite{More:2015tpa} made the first SRG application to the 
simplest, cleanest knock-out reaction: deuteron electrodisintegration.
This process provides an excellent laboratory for exploring issues
of scale dependence.
Reference~\cite{More:2015tpa} considered various kinematic regions of the longitudinal
structure function $f_L$ and showed how the different
ingredients for calculating this observable --- deuteron wave function, zeroth 
component of the electromagnetic current, relative scattering wave function for the final proton 
and neutron --- evolved under a change of scale.  Each ingredient changed via an SRG unitary
transformation in such a way to leave $f_L$ unchanged.  
The qualitative nature of the changes
varied strongly with the kinematics.

Here we revisit this process for kinematic regions that 
exhibit strong scale dependence of the
ingredients and examine the components at different scales to understand better the physics
behind the evolution.  For high momentum transfers moderately close to threshold, 
we find that an intuitive high-resolution picture of the process in impulse approximation
as a one-body current breaking up a short-range correlated neutron-proton pair
is evolved to a different intuitive picture with simpler initial \emph{and} final
state wavefunctions and a simple two-body current.
The latter combine to allow a simple calculation that is sensitive to different structure
aspects than the high-resolution analysis.
For example, sensitivity to the $D$-state component of the deuteron
at high resolution becomes complete insensitivity at low resolution.

The plan of the paper is as follows.  In Section~\ref{sec:formalism} we recap 
details and results of Ref.~\cite{More:2015tpa}, and review some
relevant SRG formalism and results.  
We analyze deuteron electrodisintegration at different resolution scales for particular
kinematics in Section~\ref{sec:results} that show strong effects of evolution,
including greatly reduced FSI, and demonstrate that the induced two-body current can
be expanded efficiently.
We finish in Section~\ref{sec:summary} with a summary and illustration of how intuition such 
as the sensitivity to the $D$-state probability can change with scale.

%%%%%%%%%%%%%%%%%%%%%%%%%%%%%%%%%%%%%%%%%%%%%%%%%%%%%%%%
%%%%%%%%%%%%%%%%%%%%%%%%%%%%%%%%%%%%%%%%%%%%%%%%%%%%%%%%

\section{Background and formalism}  \label{sec:formalism}

\subsection{Deuteron electrodisintegration formalism}

The deuteron electrodisintegration process is an ideal test ground for a
robust analysis 
of knock-out reaction scale dependence because we are able
to calculate all of the components accurately at different resolutions for
a given approximation~\cite{More:2015tpa}.  
The $d(e,e'p)n$ reaction is the simplest knock-out process and is widely used 
for benchmarking nucleon-nucleon (NN) interactions~\cite{Yang:2013rza}.  
The evolution of the current and wave functions is sufficiently rich  
for a first application of SRG methods to reactions, while
the restriction to a two-body system postpones the complications
of three-body forces and currents. 

To further simplify our analysis, 
we focus on the longitudinal structure function $\fL$, which 
up to some kinematic factors is related to an experimental cross section 
and is therefore an RG-invariant observable.  
The longitudinal structure function is given by~\cite{Yang:2013rza,More:2015tpa} 
\beq
\fL (\pp, \thetap, q) = \mathcal{C} \sum_{m_s, \, m_J} \left|
\mbraket{\psi_f(m_s, \pp, \thetap)}{J_0(q)}{\psi_i(m_J)}\right|^2
\;,
\label{eq:fl_def}
\eeq
where $\pp$ is the magnitude of 3-momentum of the 
outgoing proton, $\thetap$ is the angle that the outgoing proton makes 
with the virtual photon (taken to be along $\hat{z}$) axis, and $q$ is 
the 3-momentum transferred by the virtual photon.  All these quantities 
are in the center-of-mass frame of the outgoing nucleons.  
$\psi_f$ and $\psi_i$ are the final scattering state wave function of 
the outgoing nucleons and the initial deuteron state wave function, 
respectively, with  $m_s$ and $m_J$ corresponding quantum numbers.  
The constant $\mathcal{C}$ in Eq.~\eqref{eq:fl_def} is a kinematic 
factor involving $p$, $q$, deuteron mass, and the nucleon mass~\cite{More:2015tpa}.

The relevant one-body current matrix element is given by
\begin{multline}
 \la \mbf{k}_1 \, T_1| \, J_0(\mbf{q}) \, | \,\mbf{k}_2 \, T=0 \ra \\
 = \frac{1}{2} \big(G_E^p + (-1)^{T_1} G_E^n\big) \,
 \delta(\mbf{k}_1 - \mbf{k}_2 - \mbf{q}/2) \\
 \null + \frac{1}{2} \big((-1)^{T_1} G_E^p +  G_E^n\big) \, \delta(\mbf{k}_1
 - \mbf{k}_2 + \mbf{q}/2) \,.
\label{eq:J0_firstdef}
\end{multline}
Here $G_E^p$ and $G_E^n$ are the proton and neutron electric form factors.  
The deuteron state has isospin $T=0$, and therefore the ket in 
Eq.~\eqref{eq:J0_firstdef} is restricted to $T=0$.  
The final state wave function of the outgoing proton-neutron pair 
with relative momentum $\pp$ is found from
\beq
%\label{eq:LSequation}
	|\psi_{f\, \pp}\ra = | \phi_{\pp} \ra + G_0 (E^\prime) \, t(E^\prime) 
	\,| \phi_{\pp} \ra \,,
  \label{eq:psi_f_def}
\eeq
where $\ket{\phi_{\pp}}$ is a relative plane wave, $G_0$ and $t$ are the 
Green's function and the $t$-matrix, respectively, with outgoing boundary conditions, and
$\Ep = {\pp}^2/ M_{np}$ is the energy of outgoing nucleons. 
The impulse approximation (IA) is defined here by neglecting the interaction 
between the outgoing nucleons (given by the second term in 
Eq.~\eqref{eq:psi_f_def}) and taking $\ket{\psi_{f\, \pp}}_{\rm IA} 
\equiv \ket{\phi_{\pp}}$.  We call this the IA even when we have an induced
two-body current.

We note that in our work the kinematic variables we use are 
$\Ep$, the energy of outgoing nucleons; $\mbf{q}_{\rm c.m.}^2$ (also denoted as
$q^2$), the 
3-momentum transferred by the virtual photon; and 
$\thetap$, the angle that the outgoing proton makes with photon.  All these 
quantities are in the center-of-mass frame of the outgoing nucleons.  
Another set of kinematic variables that are used for electron scattering
from nuclei are Bjorken $x$ and the four-momentum $Q^2$.  In 
Appendix~\ref{app:kinematic_variables_transformation} we relate $x$ and $Q^2$
to $\Ep$ and $\mbf{q}_{\rm c.m.}^2$.

Our formalism implies a non-relativistic treatment, but to ensure clear
demonstrations of scale dependence we will 
apply it for some kinematic regions where that might be questionable.
However, we do so consistently at each resolution,
so the comparison at different scales will always be valid, even though
comparison to experiment will not be so useful or informative (also because we
omit initial two-body currents).

\subsection{Local decoupling with SRG}

  \begin{figure*}[tbh!]
    \includegraphics[width=0.98\textwidth]{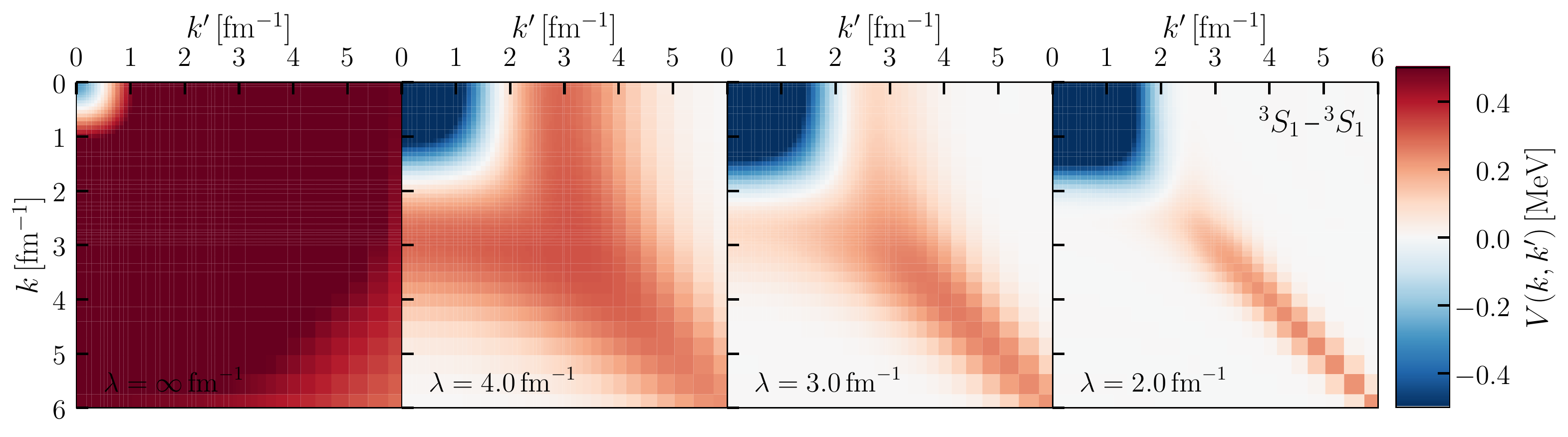}
    \caption{SRG running of the AV18 potential in the $^3S_1-^3\!S_1$ 
    channel.
    }
    \label{fig:potential_3S1_3S1_heat_maps}       
  \end{figure*} 

  \begin{figure*}[tbh!]
    \includegraphics[width=0.98\textwidth]{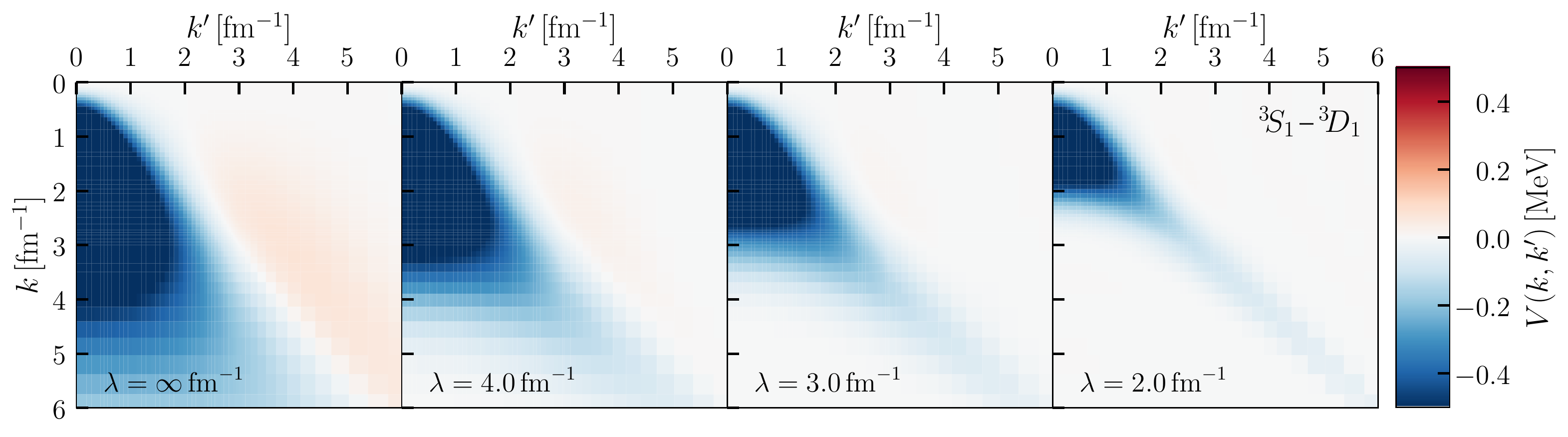}
    \caption{SRG running of the AV18 potential in the $^3S_1-^3\!D_1$ 
    channel.  
    }
    \label{fig:potential_3S1_3D1_heat_maps}       
  \end{figure*}

The SRG for nuclear applications
is well documented in the literature~\cite{Bogner:2006pc,Bogner:2009bt,Furnstahl:2012fn,Furnstahl:2013oba,Dainton:2013axa} but we briefly summarize
the salient points.  The simplest SRG transformations are realized as a flow
equation, which is a differential equation for the Hamiltonian that induces a
continuous series of infinitesimal unitary transformations:
\beq
  \label{eq:SRGflow}
  \frac{dH_s}{ds} = [[G_s, H_s], H_s]  \;,
\eeq
where $s$ is a flow parameter.
Here, and in most nuclear applications to date, the operator $G_s$ is the kinetic
energy operator $T$ and the flow equation becomes (with $V_s \equiv H_s - T$)
\beq    
  \label{eq:SRGflow2}
  \frac{dV_s}{ds} = [[T,V_s],T] + [[T,V_s],V_s] \;.
\eeq
We solve Eq.~\eqref{eq:SRGflow2} in a partial-wave momentum basis, 
where it becomes a set of coupled
differential equations for the matrix elements of the potential.

The SRG equations decouple high-energy from low-energy degrees of
freedom in the Hamiltonian by driving far off-diagonal matrix elements to zero.
(Other choices of $G_s$ also achieve this goal with different decoupling patterns.)
The degree of decoupling is characterized by the scale $\lambda = s^{\frac{1}{4}}$,
which has units of momentum, and we use $\lambda$ to characterize the flow in what
follows and in our notation (so $V_s$ becomes $V_\lambda$).
The first term in Eq.~\eqref{eq:SRGflow2} dominates far off-diagonal matrix elements;
keeping this term only yields the solution for the NN potential (with mass $M=1$)
\beq
  \label{eq:Vlambda_approx}
  V_\lambda(k,k') \approx V_{\lambda=\infty}(k,k') 
      e^{-\bigl(\frac{k^2-k'^2}{\lambda^2}\bigr)^2}
      \;.  
\eeq      
This shows that $\lambda^2$ is roughly the maximum difference between kinetic
energies of nonzero matrix elements.

\begin{figure}[tbh!]
  \includegraphics[width=0.42\textwidth]{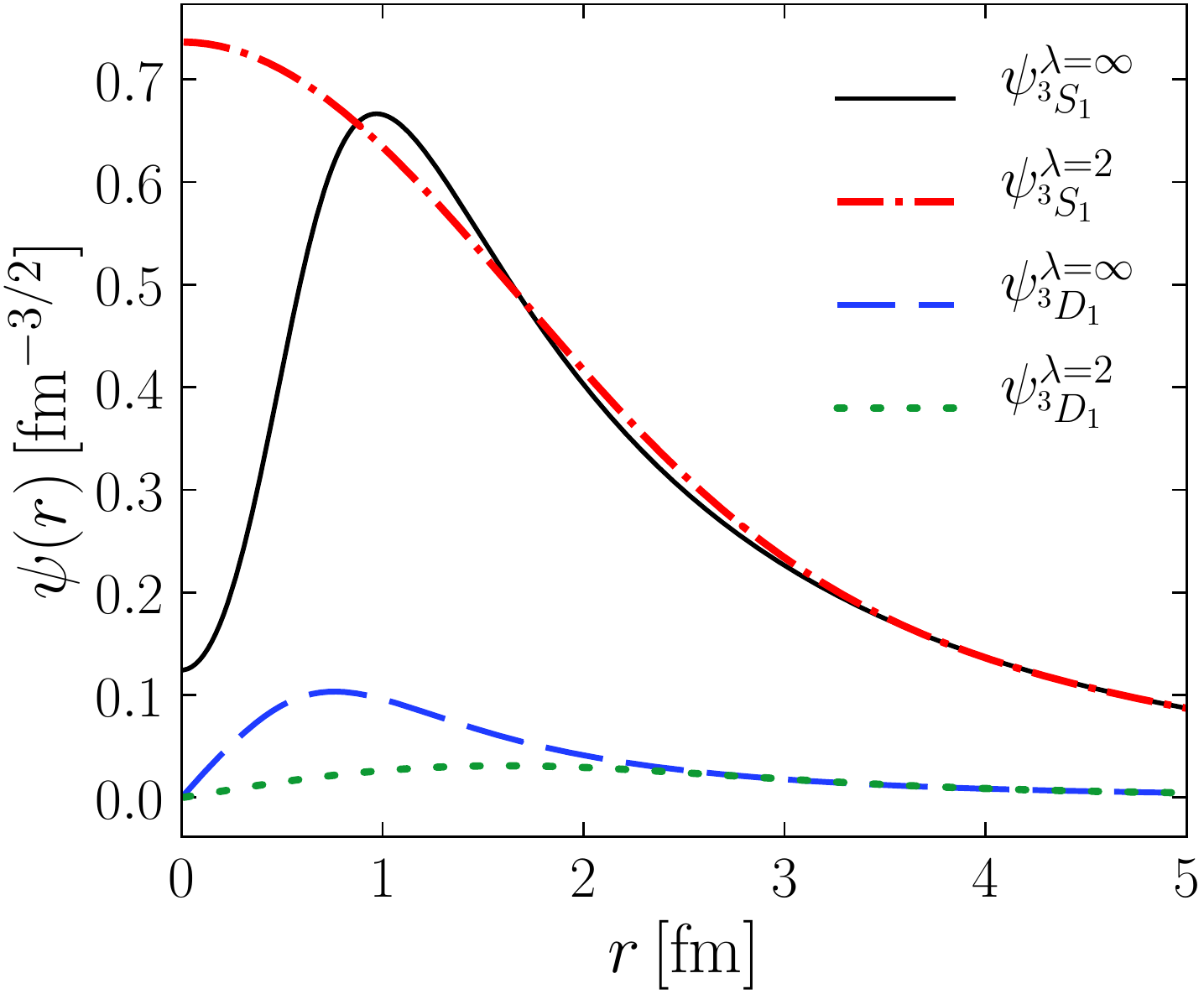}
  \caption{Deuteron wave functions in position space for
  the unevolved ($\lambda = \infty$) and evolved (to $\lambda = 2\,\rm{fm^{-1}}$)
  AV18 potential.}
  \label{fig:deut_wfn_r_space}
\end{figure}

\begin{figure}[tbh!]
  \includegraphics[width=0.42\textwidth]{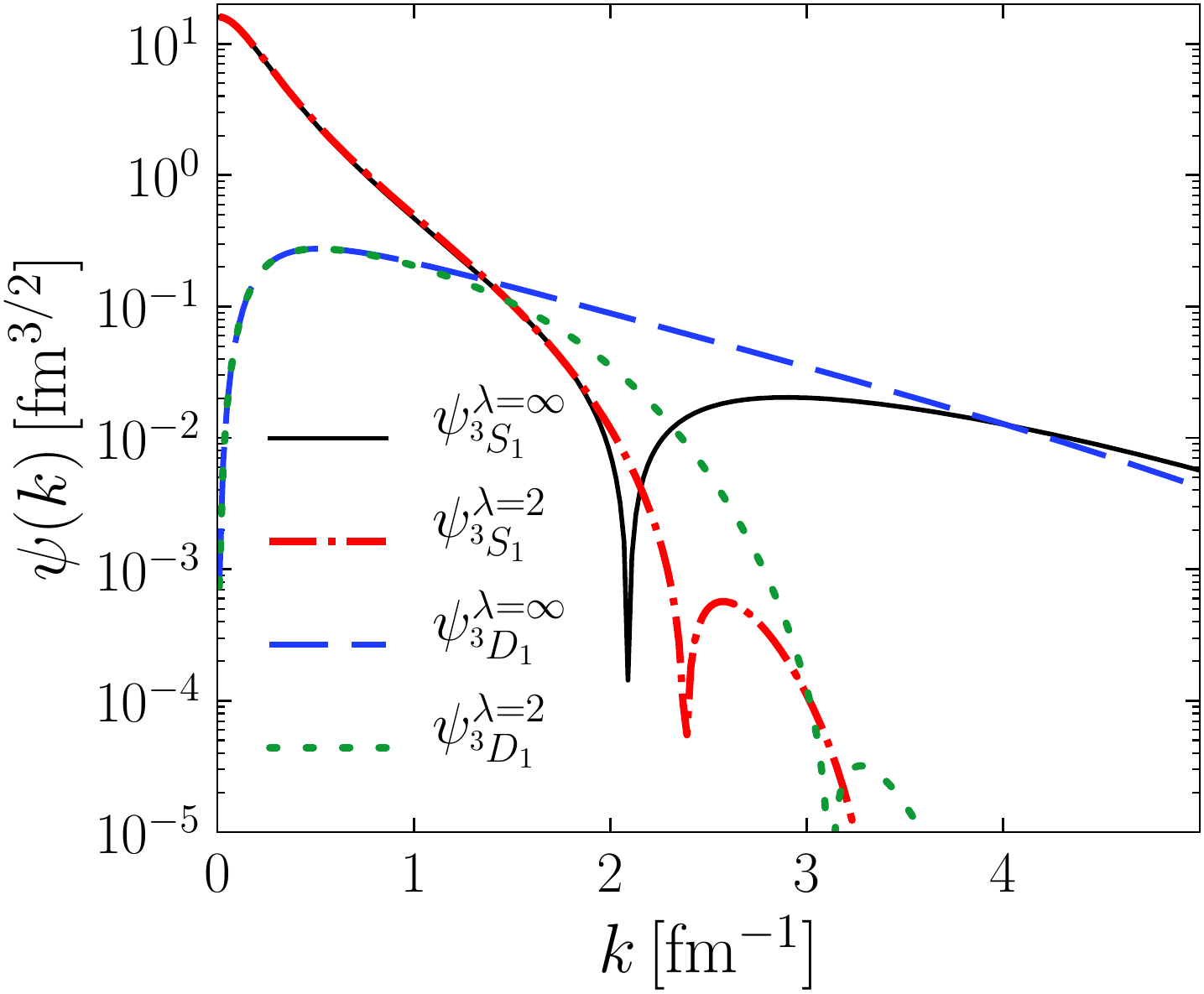}
  \caption{Deuteron wave functions in momentum space for
  the unevolved ($\lambda = \infty$) and evolved (to $\lambda = 2\,\rm{fm^{-1}}$)
  AV18 potential.}
  \label{fig:deut_wfn_k_space}
\end{figure}

Figures~\ref{fig:potential_3S1_3S1_heat_maps} and \ref{fig:potential_3S1_3D1_heat_maps}
provide an intuitive picture of decoupling and the role of $\lambda$ as a decoupling
scale. 
Here $V_{\lambda=\infty}$ is the AV18 potential~\cite{Wiringa:1994wb} and evolution is
shown for two blocks of the $^3S_1$--$^3D_1$ coupled channel.
For convenience in comparing to momentum scales, we plot the potential
using the relative momenta.  This obscures somewhat the uniform decoupling
in $k^2$ that is implied by Eq.~\eqref{eq:Vlambda_approx}; it is manifested for
potentials plotted as functions of $k^2, k'{}^2$ 
(e.g., see Fig.~9 in Ref.~\cite{Furnstahl:2012fn}).

The partial diagonalization of $V_\lambda$ leads to \emph{local} decoupling~\cite{Dainton:2013axa}.
This means that only matrix elements with relative momentum arguments differing by less
than roughly $\lambda$
contribute in the Lippmann-Schwinger equation for the $t$-matrix or wave functions.
This in turn leads to a lower \emph{resolution} as the potential evolves --- 
local decoupling means that only wavelengths in a narrow region are available to 
build wave functions.  We make the association of limited wavelengths and limited 
resolution (as with diffraction), although in past investigations we were
typically restricted to low momentum 
 only.  Here we have the possibility of high-relative-momentum final states.

The impact that local decoupling of the potential has on the deuteron wave function is shown in 
Figs.~\ref{fig:deut_wfn_r_space} and \ref{fig:deut_wfn_k_space}, with complementary effects in position and momentum
space.  For the $S$-wave part, the high-momentum tail from the strong coupling of
low- and high-momentum in the original AV18 potential ($\lambda = \infty$) is evolved
away as $\lambda$ is reduced, with a consequent filling in of the wound 
in the small $r$ part of the wave function.  
For the $D$-wave part, the $D$-state probability is steadily reduced, as implied by the 
reduced $S-D$ tensor coupling in the potential.
This reduction is clearly evident in position space, where the interior part of the D-state
component of the wave function is greatly reduced in evolving from $\lambda = \infty$
to $\lambda = 2\,\fmi$. 
Note that the position-space tails, which are specified by deuteron asymptotic
normalization constants, are RG invariant, as expected because they are exterior
quantities~\cite{Mukhamedzhanov:2010hc}. 

Because $\lambda$ sets a separation scale in the deuteron, we identify the 
subsequent SRG evolution of the wave function
as a change in scale.  The change in the deuteron momentum distribution is
analogous to the RG evolution of the parton distribution function for up or down
quarks in the proton~\cite{Furnstahl:2013dsa}.
In the latter case there is also a clear \emph{scheme} dependence, which refers to the
prescriptions for renormalization and factorization (how the short- and long-distance
parts are divided).  The SRG evolution changes the scale but not the scheme because
these are unitary transformations with a fixed SRG generator.  
The scheme dependence is instead in the choice
of the initial NN potential and the choice of the generator.%
\footnote{Note that the flow to universal potentials can eliminate the scheme dependence
due to the initial potential if all momenta are less than $\lambda$.}  
The difference between scale and scheme dependence
is manifested by different sets of chiral EFT potentials in the literature 
(e.g., Refs.~\cite{Gezerlis:2014zia,Epelbaum:2014sza,Piarulli:2014bda,Entem:2017gor}).  
The sets differ by scheme (e.g., the choice of regulator to be non-local, local, or semi-local)
and within each set differ by scale (determined by the value of the regulator parameter).
The guidance about schemes from the \textit{Handbook of perturbative QCD} applies
equally to low-energy nuclear processes~\cite{RevModPhys.67.157}:
\begin{quote}
  The choice of scheme is a matter of taste and convenience, but it is absolutely 
  crucial to use schemes consistently, and to know in which scheme any given calculation, 
  or comparison to data, is carried out.
\end{quote}

\begin{figure*}[tbh!]
  \includegraphics[width=0.32\textwidth]{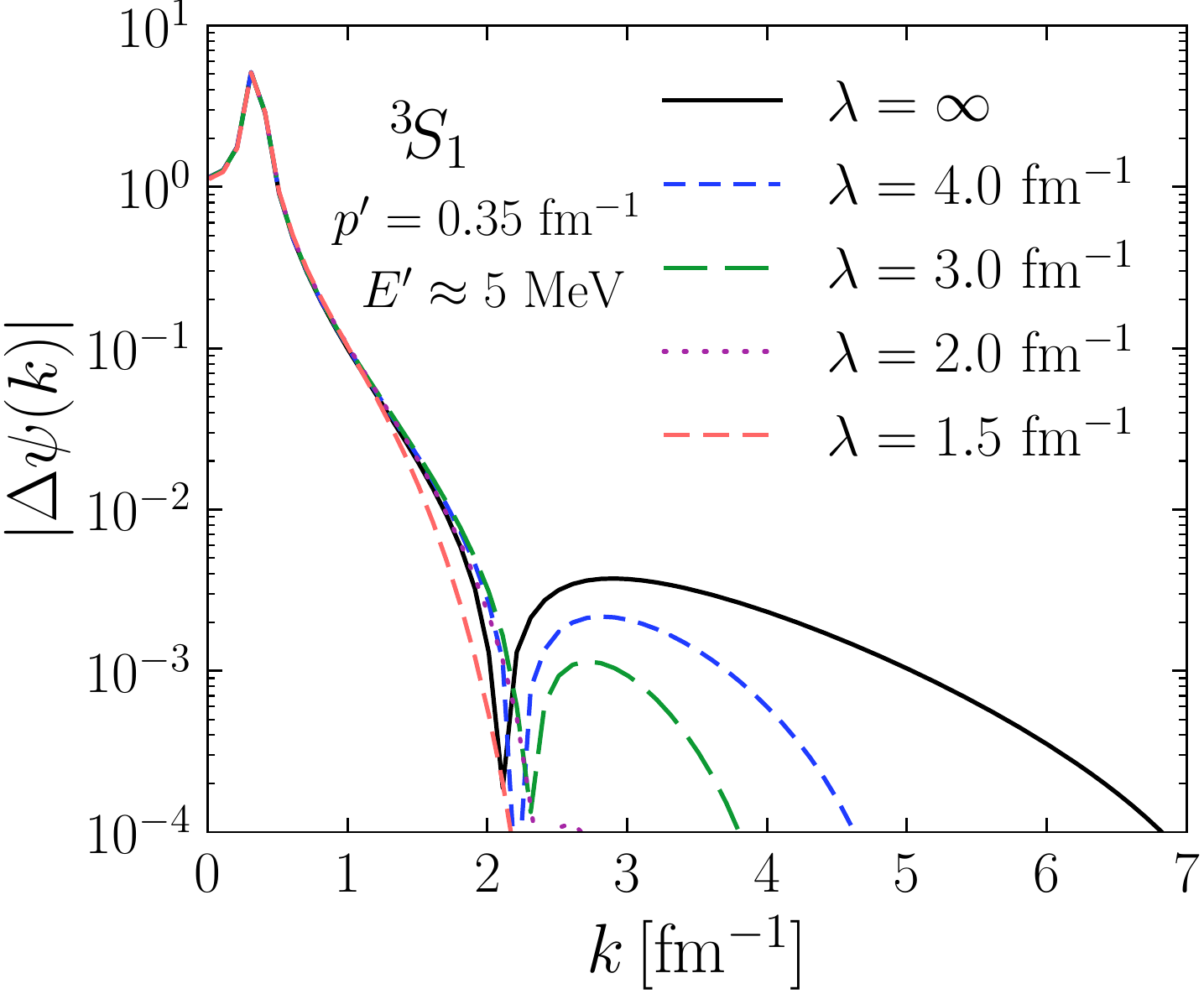}
  \includegraphics[width=0.32\textwidth]{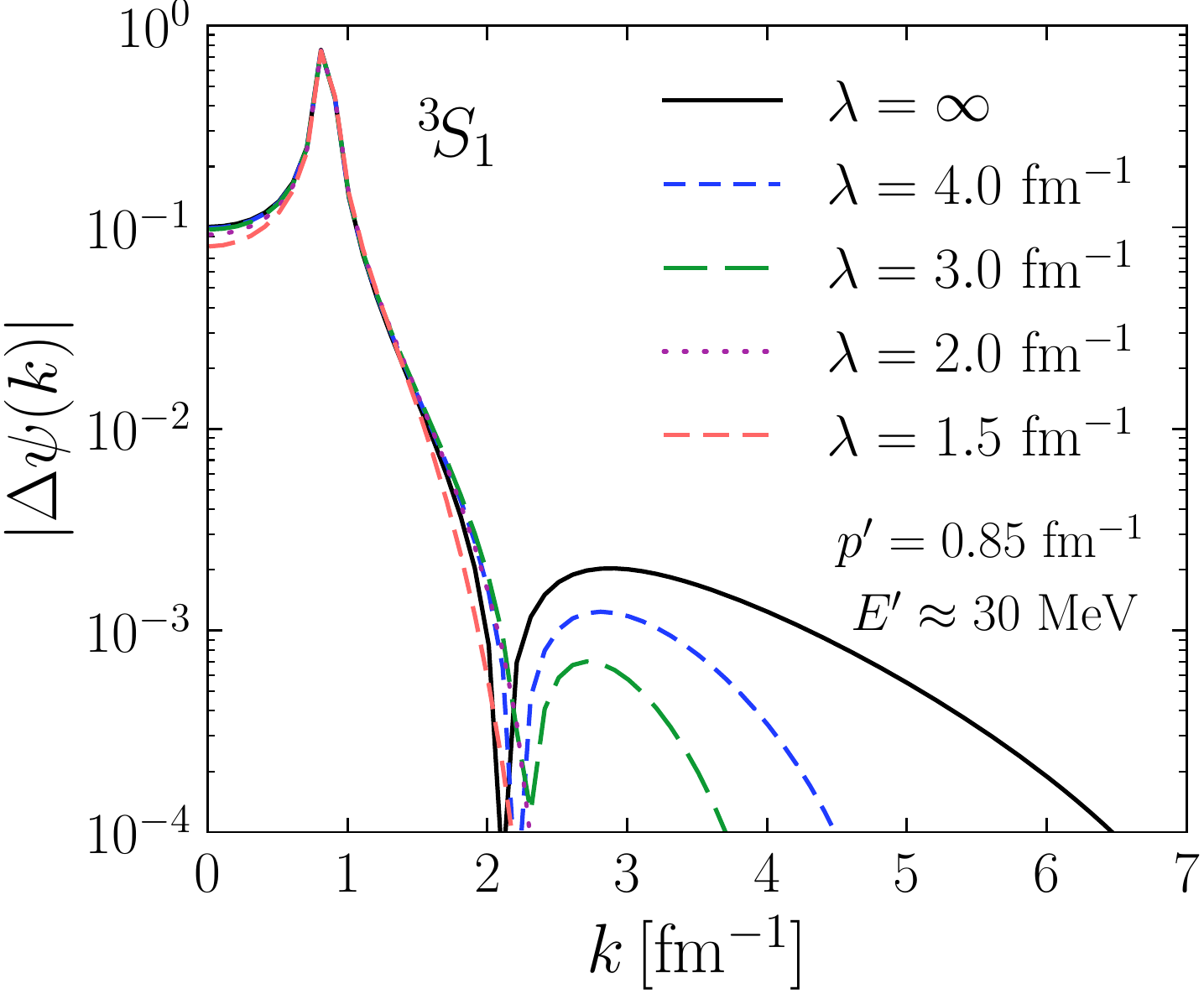}
  \includegraphics[width=0.32\textwidth]{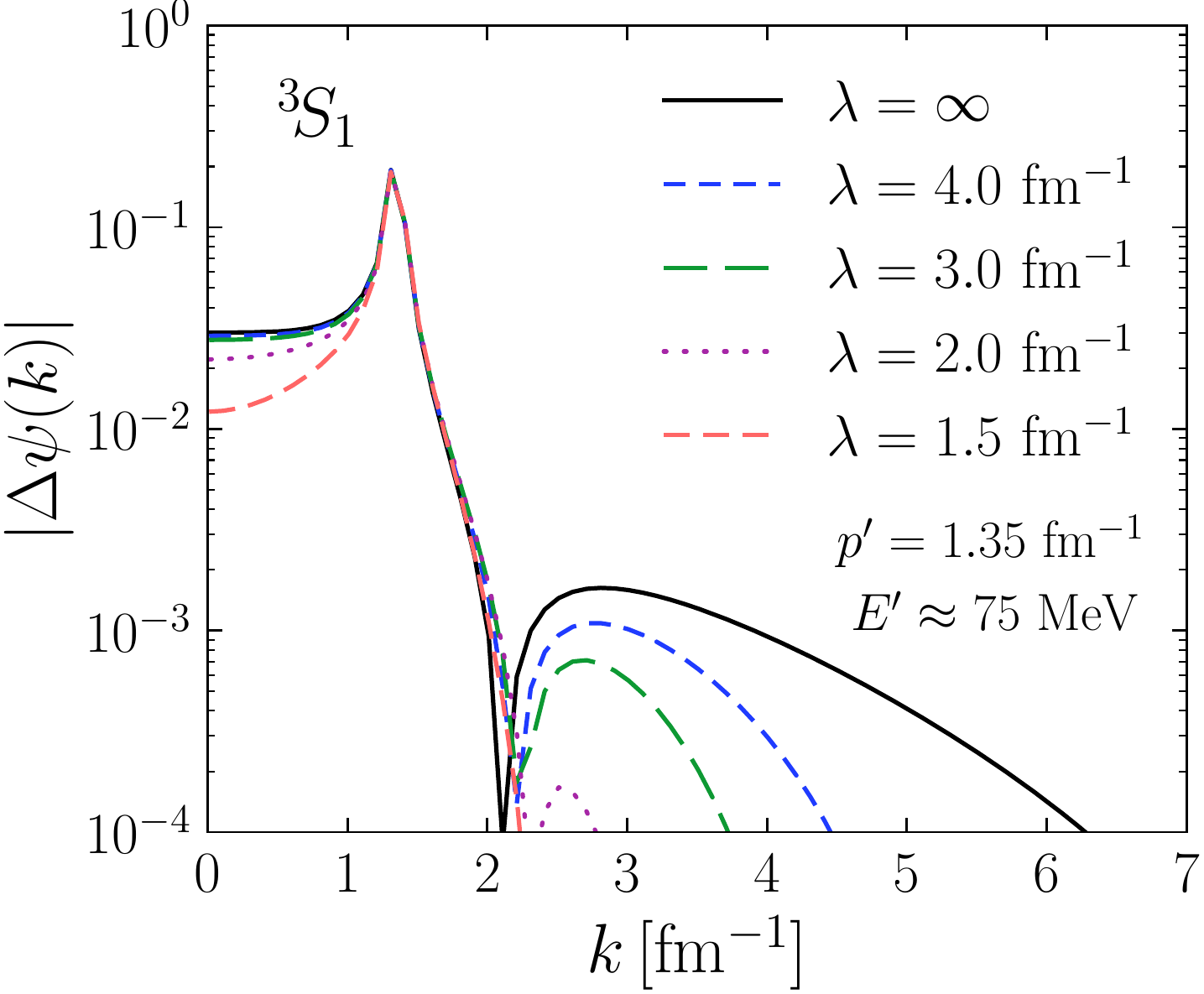}
  \caption{$|\Delta \psi^{\lambda}(\pp; k)|$ in the $^3S_1$ channel 
  for various $\pp$ and SRG $\lambda$ values.  
  $\Delta \psi^{\lambda}(\pp; k)$ in the evolved 
  case is suppressed for large momentum.  For large $\pp$, 
  the values of $\Delta \psi^{\lambda}(\pp; k)$ even at small 
  momenta differ for different SRG $\lambda$s.}
  \label{fig:Delta_psi_mom}       
\end{figure*}
%1

In nuclear reaction applications, one must evolve all components --- initial state, 
current, final state --- of the matrix elements for cross sections to remain 
invariant.
This is conveniently carried out for deuteron electrodisintegration in terms of the 
integrated unitary transformation operator $\widehat U_\lambda$~\cite{More:2015tpa}:
\beq
   |\psi_\lambda\rangle = \widehat U_\lambda | \psi_{\lambda=\infty}\rangle
   \;, \qquad
   \widehat O_\lambda = \widehat U_\lambda \widehat O_{\lambda=\infty} \widehat U_\lambda^\dagger 
   \;,
\eeq
where $|\psi_\lambda\rangle$ is an initial or final state and
$\widehat O_\lambda$ is an operator such as the Hamiltonian or the interaction
current.
If the energy/momentum scale of the external probe is
significantly larger than $\lambda$, then the scale separation with respect to the
ground state wave function (which has only momenta less than roughly $\lambda$) leads to
a factorization of matrix elements of the $\widehat U_\lambda$~\cite{Anderson:2010aq}:
\beq
   U_\lambda(k,q) \longrightarrow K_\lambda(k) Q_\lambda(q)
     \ \mbox{when $k < \lambda$ and $q\gg \lambda$}
     \;.
   \label{eq:U_kq_factorization}  
\eeq
%
% Further, $K_\lambda(k)$ is well represented by a polynomial expansion in $k^2$.
% This leads to a corresponding expansion in contact operators 
% for the \emph{change} in the potential or current with SRG evolution.  
% We will exploit this observation below.
In the next section we will show how the change in the current induced by the unitary
transformation can be efficiently expanded in factorized form with a singular value decomposition.

\begin{figure*}[tbh!]
  \includegraphics[width=0.32\textwidth]{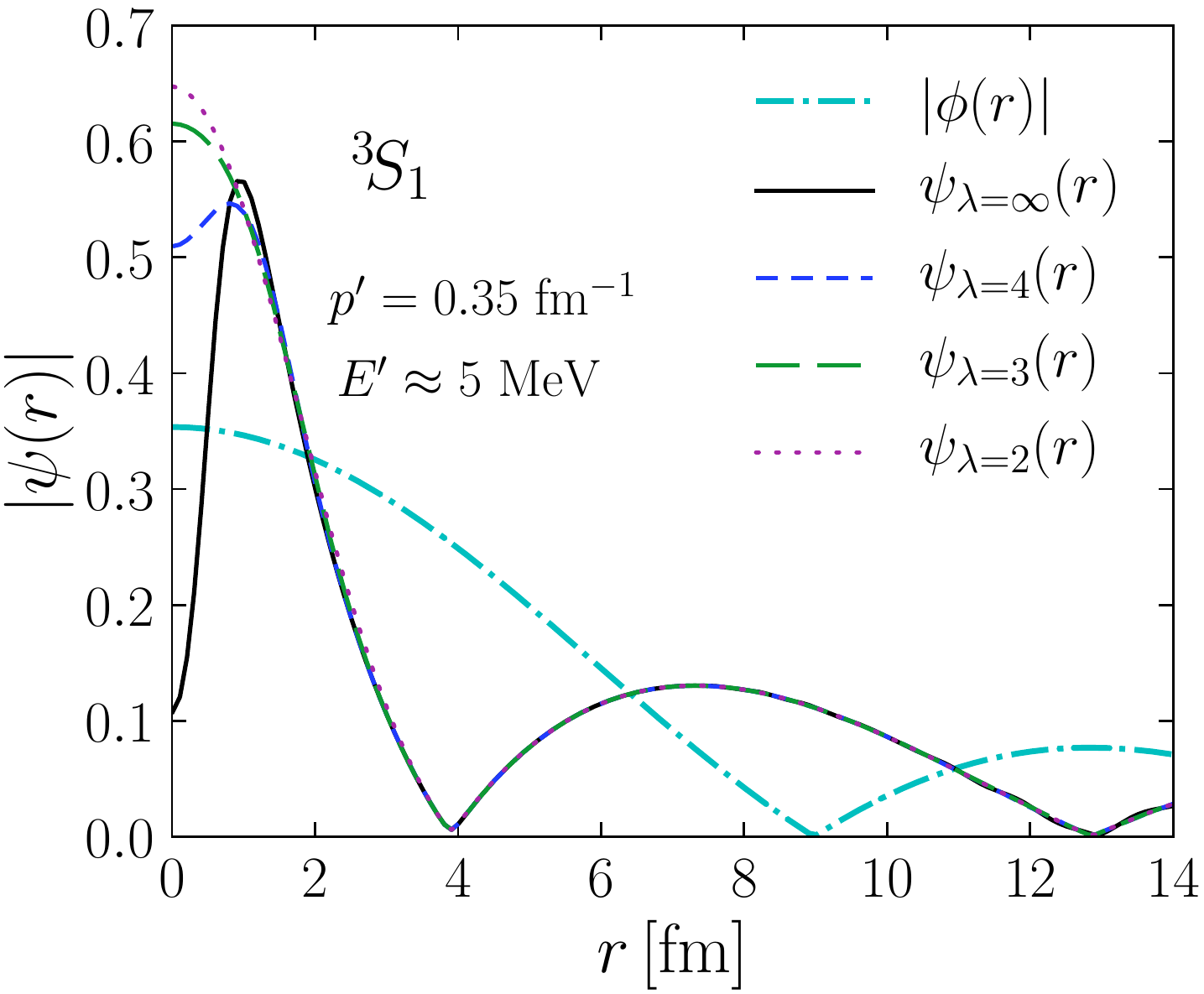}
  \includegraphics[width=0.32\textwidth]{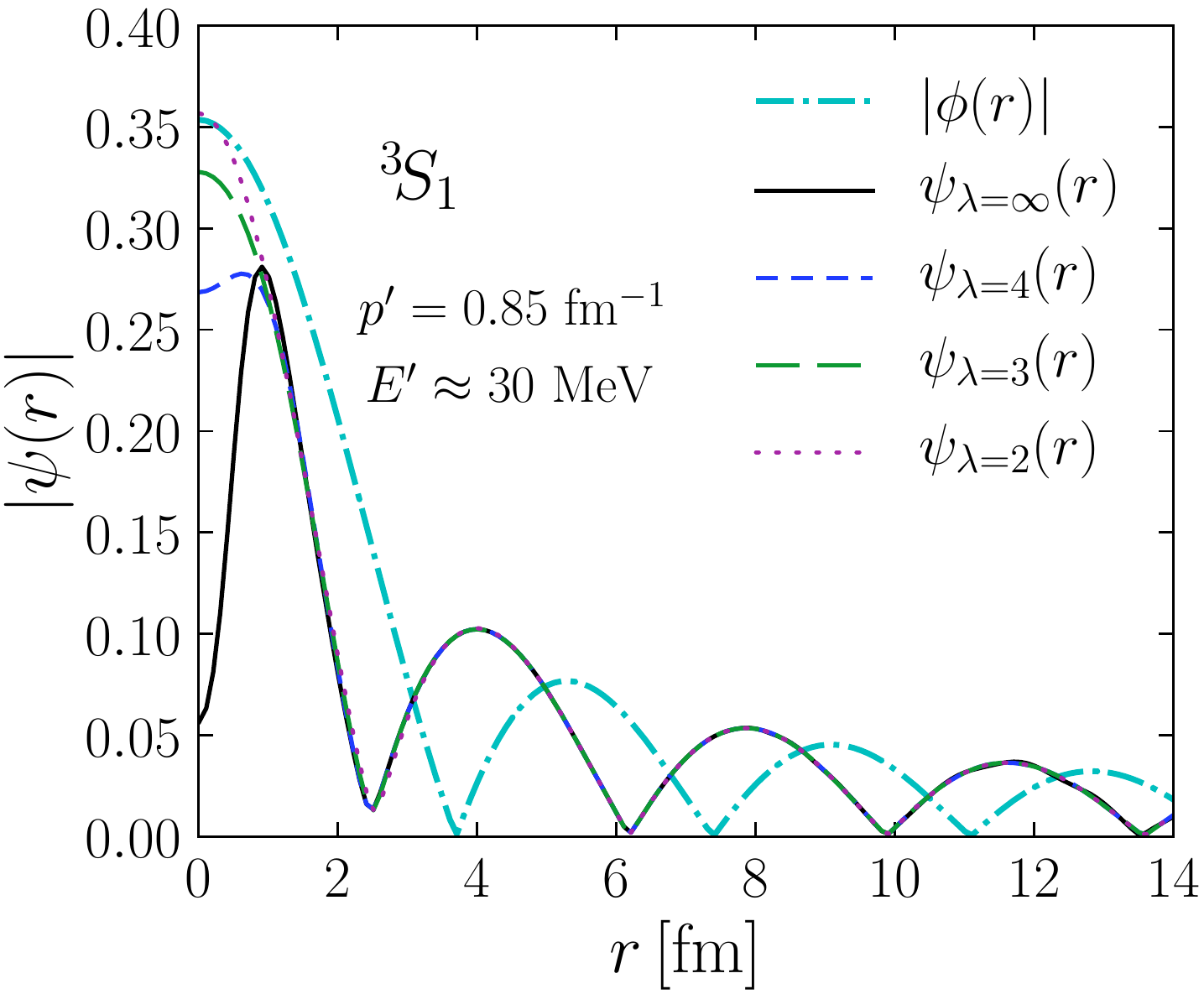}
  \includegraphics[width=0.32\textwidth]{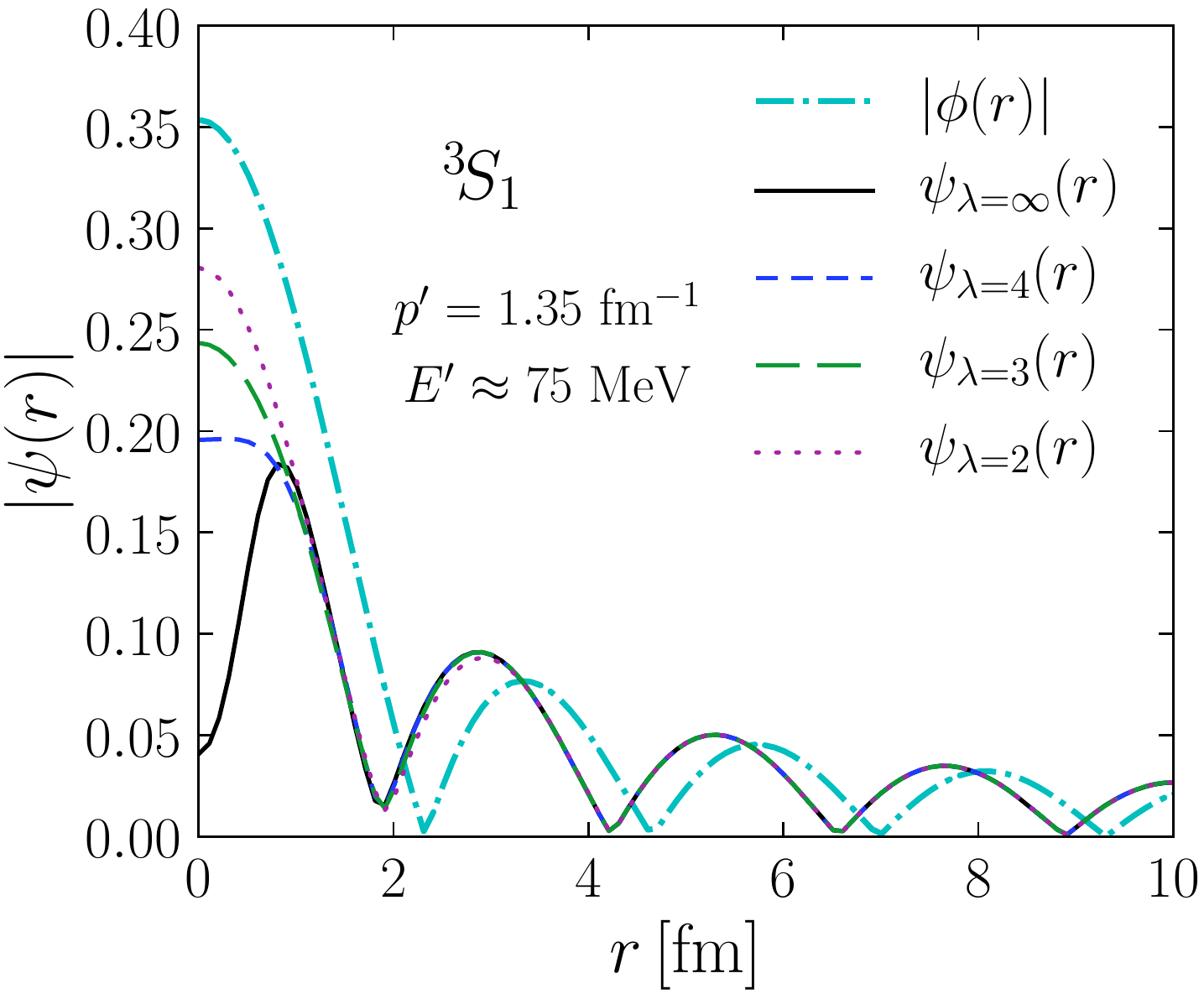}
  \caption{$|\psi^{\lambda}(\pp; r)|$ in the $^3S_1$ channel 
  for various $\pp$ and SRG $\lambda$ values.  $\psi^{\infty}(r)$ 
  has a correlation wound at short distances that is absent for 
  the evolved wave functions.   
  }
  \label{fig:psi_coordinate_space}      
\end{figure*}

%%%%%%%%%%%%%%%%%%%%%%%%%%%%%%%%%%%%%%%%%%%%%%%%%%%%%%%%
%%%%%%%%%%%%%%%%%%%%%%%%%%%%%%%%%%%%%%%%%%%%%%%%%%%%%%%%

\section{Results}  \label{sec:results}

\subsection{Scattering state evolution}

We begin by analyzing how the scattering state 
$\psi^{\lambda}(p';k)\equiv \phi_{p'}(k) + \Delta\psi^{\lambda}(p';k)$ 
of the outgoing proton-neutron pair evolves under the SRG for different values of $p'$ and $\lambda$.  
Multiplying Eq.~\eqref{eq:psi_f_def} from the left by $\langle k|$ and projecting onto partial waves gives
\begin{align}
  & \braket{k_1 \, J_1 \, m_{J_1} \, L_1 \, S_1 \, T_1}
   {\phi_{\pp}} = \frac{1}{2} \sqrt{\frac{2}{\pi}} \frac{\pi}{2} 
   \frac{\delta(\pp - k_1)}{k_1^2}  \notag \\
  & \qquad\ \ \null \times 
   \braket{J_1 \, m_{J_1}}{L_1 \, m_{J_1} - m_{s_f} \, S_1 \,m_{s_f}}  \notag \\
  & \qquad\ \ \null \times 
   \Bigl(1 + (-1)^{T_1}(-1)^{L_1}\Bigr) Y^{\ast}_{L_1 \, m_{J_1} - m_{s_f}}(
   \thetap, \phip) \;,
   \label{eq:phi_pw}
\end{align}
for the free plane-wave, and
\begin{align}
	& \Delta \psi^{\lambda}(\pp; k) =   \notag \\
	& \qquad \mbraket{k \, J_1 \, \mJd \, L_1 \, S_1 \, T_1}
	{G_{0}^{\Ep} t^{\lambda}(\Ep)} {\phi (\pp, J_1, S_1, T_1, \msf)} \notag \\
   &  =
	\frac{1}{2} \sqrt{\frac{2}{\pi}} 
	\frac{1}{{\pp}^2 - k^2 + i \epsilon} \sum_{L_2} t^{\lambda}(
	k, \pp, \Ep, L_1, L_2, J_1, S_1, T_1) \notag \\
  & \qquad\  \null \times 
	\braket{J_1 \, \mJd}{L_2 \, \mJd - \msf \, S_1 \, \msf} \notag \\
  & \qquad\  \null \times 
	\Bigl( 1 + (-1)^{T_1} \, (-1)^{L_2}\Bigr) 
  Y^{\ast}_{L_2 \, \mJd - \msf} (\thetap, \phip)
  \;\,
  \label{eq:Delta_psi_mom_expression}
\end{align}
for the scattered wave that contains the effects of final state interactions between the outgoing nucleons. 
Here and below, $\thetap$ and $\phip$ are the angles of the outgoing proton 
with respect to the virtual photon. Note that $\Delta \psi^{\lambda} (\pp; k)$ 
is singular at the on-shell momentum $k = \pp$, and is in general complex-valued. 

In Fig.~\ref{fig:Delta_psi_mom} we plot the magnitude of 
$\Delta \psi^{\lambda}(\pp; k)$ (omitting the singular point 
$k = \pp$) in the $^3S_1$ channel for various $\pp$ and $\lambda$ 
values.  As expected from SRG decoupling, $\Delta \psi^{\lambda}(\pp; k)$ 
in the evolved case becomes suppressed for large momentum $k\gtrsim\lambda$.  
As $\pp$ increases, the values of $\Delta \psi^{\lambda}(\pp; k)$ at small 
momenta are also suppressed with decreasing SRG $\lambda$s.  
This reflects the local decoupling with SRG evolution for large $\pp$; 
the wave function is suppressed for momenta more than $\lambda$ from $\pp$ in either direction~\cite{Dainton:2013axa}.   

It is instructive to also look at the scattering wave function in  
coordinate space, using 
\beq
  \psi^{\lambda} (\pp; r) = \int \dd k \, k^2 j_l(k\,r) \psi^{\lambda}(\pp; k) \;.
\eeq    	
Figure~\ref{fig:psi_coordinate_space} shows 
$|\psi^{\lambda}(\pp; r)|$ in the $^3S_1$ channel for different 
$\pp$ and SRG $\lambda$s.  As expected, $\psi^{\infty}(r)$ 
has a sizable correlation wound at short distances (up to about 1\,fm) that is
progressively filled in as the wave functions evolve to lower $\lambda$ values.   
Note that beyond the range of the potential, $\psi(r)$ and $\phi(r)$ differ as expected by just a phase 
that is the same for all values of $\lambda$.

\subsection{Operator evolution}

The operator of interest here is the deuteron disintegration current 
operator, which is just the zeroth component of the electromagnetic current.  
The matrix element of the one-body current operator used 
in \cite{More:2015tpa} is given by Eq.~\eqref{eq:J0_firstdef}.  	

We denote the first term in Eq.~\eqref{eq:J0_firstdef} by $J_0^-$.  In the 
following we focus only on results obtained from using $J_0^-$, as it 
was verified in Ref.~\cite{More:2015tpa} 
that $\mbraket{\psi_f}{J_0}{\psi_i} = 2 \mbraket{\psi_f}{J_0^-}{\psi_i}$.  
In the partial-wave basis, the $J_0^-$ matrix element is given by
\begin{align}
  & \mbraket{k_1 \, J_1 \, \mJd \, L_1 \, S\!=\!1 \, T_1}{J_0^-}
  {k_2\, J\!=\!1 \, \mJd \, L_2 \, S\!=\!1 \, T\!=\!0} 
   \notag \\
  & \qquad\quad = \frac{\pi^2}{2}
  \Bigl(G_E^p + (-1)^{T_1} G_E^n\Bigr) \notag \\
  & \qquad\qquad \null \times \sum_{\mst=-1}^{1} \braket{J_1 \, \mJd}
  {L_1 \, \mJd - \wt{m}_s \, S\!=\!1 \, \wt{m}_s} \notag \\
  & \qquad\qquad \null \times P_{L_1}^{\mJd - \mst} \left(\frac{k_1^2 - k_2^2 + q^2/4}{k_1 q}
  \right) \frac{2}{k_1 k_2 q} \notag \\
  & \qquad\qquad \null \times
  P_{L_2}^{\mJd - \mst} \left(\frac{k_1^2 - k_2^2 - q^2/4}{k_2 q}
  \right)  \notag \\
  & \qquad\qquad \null \times
  \CG{L_2}{\mJd - \mst}{S=1}{\mst}{J=1}{\mJd}  
  \label{eq:J0_minus_analytical_delta}
\end{align}
when
\beq
    k_2 \in (|k_1 - q/2|, k_1 + q/2) 
\eeq
and equals zero otherwise.

\begin{figure*}[tbh!]
  \includegraphics[width=0.45\textwidth]
  {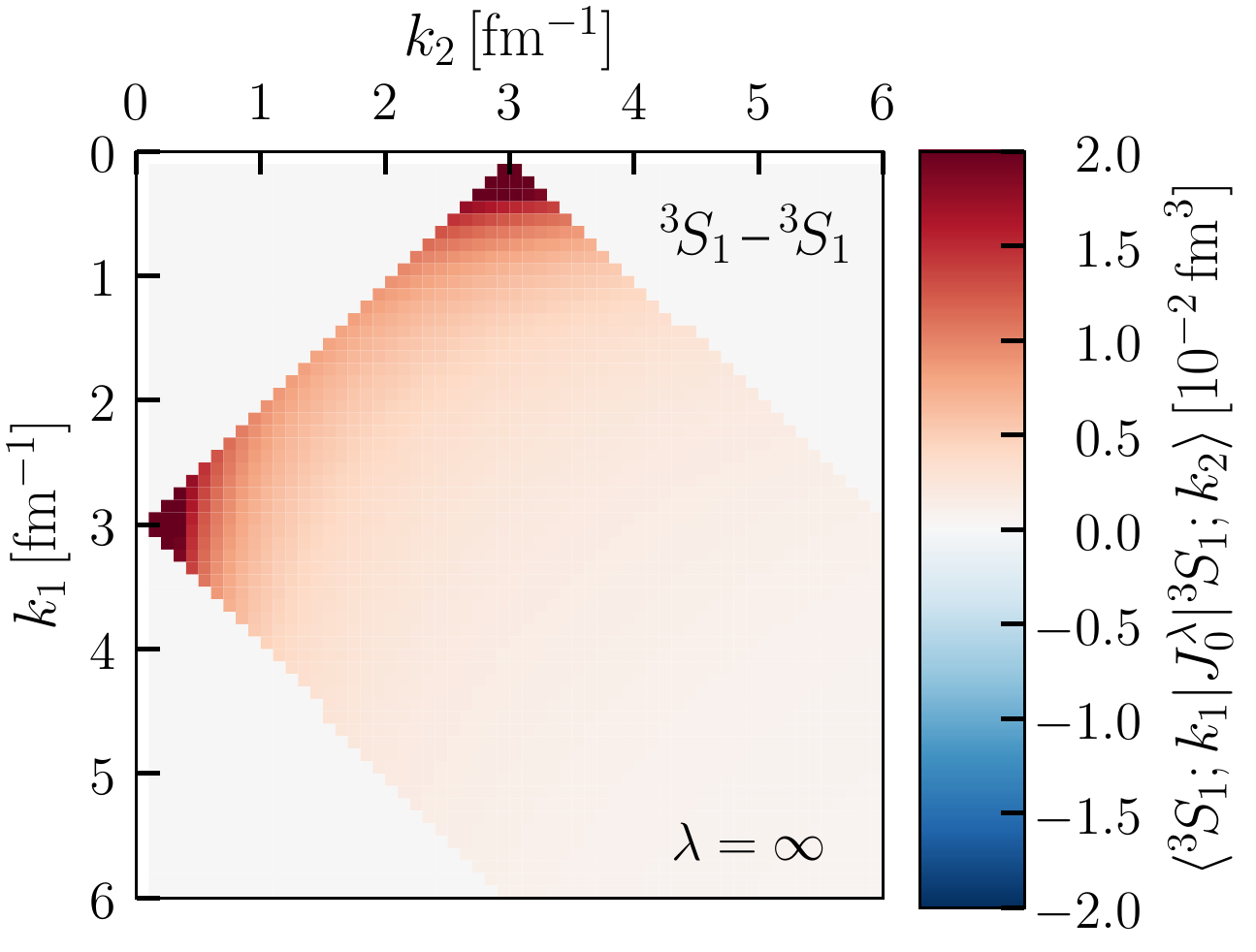}%
  ~~~
  \includegraphics[width=0.45\textwidth]
  {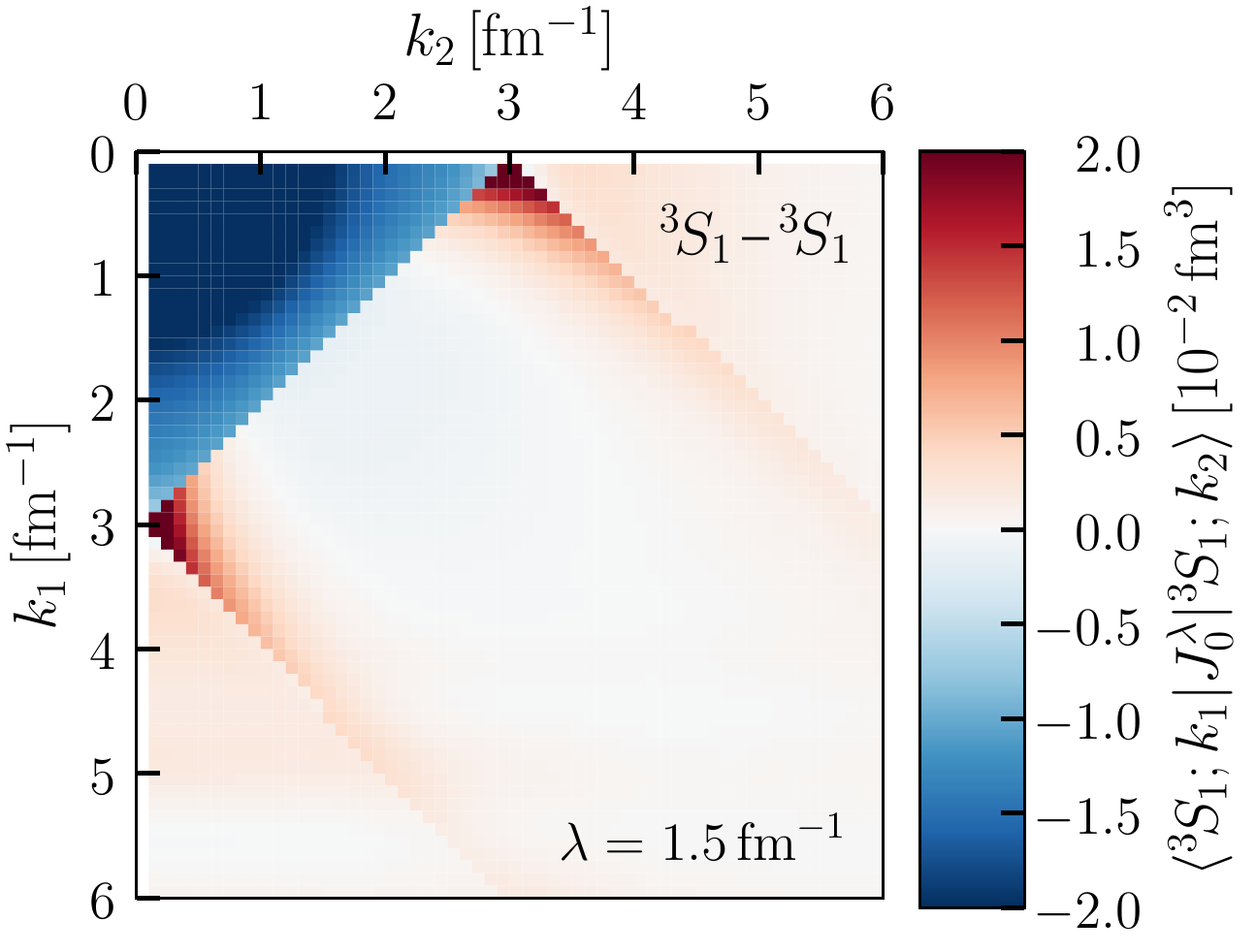}
  \caption{Heatmap plot for the unevolved ($\lambda=\infty$) and the evolved current
  ($\lambda = 1.5\,$\fmi) for $q^2 = 36 {\rm{~fm^{-2}}}$ in the 
  $^3\!S_1-^3\!\!S_1$ block for $\mJd = 0$.  The 
  heatmap plots for $\mJd = 1$ look very similar.  The unevolved current 
  is peaked at $(0, q/2)$ and $(q/2, 0)$.  SRG evolution induces 
  smooth two-body currents in the low-momentum region.    
  }
\label{fig:J_evolution_heatmaps_3S1_3S1_q36}
\end{figure*}

\begin{figure}[tbh!]
  \centering
  \begin{overpic}[width=0.9\columnwidth]
  {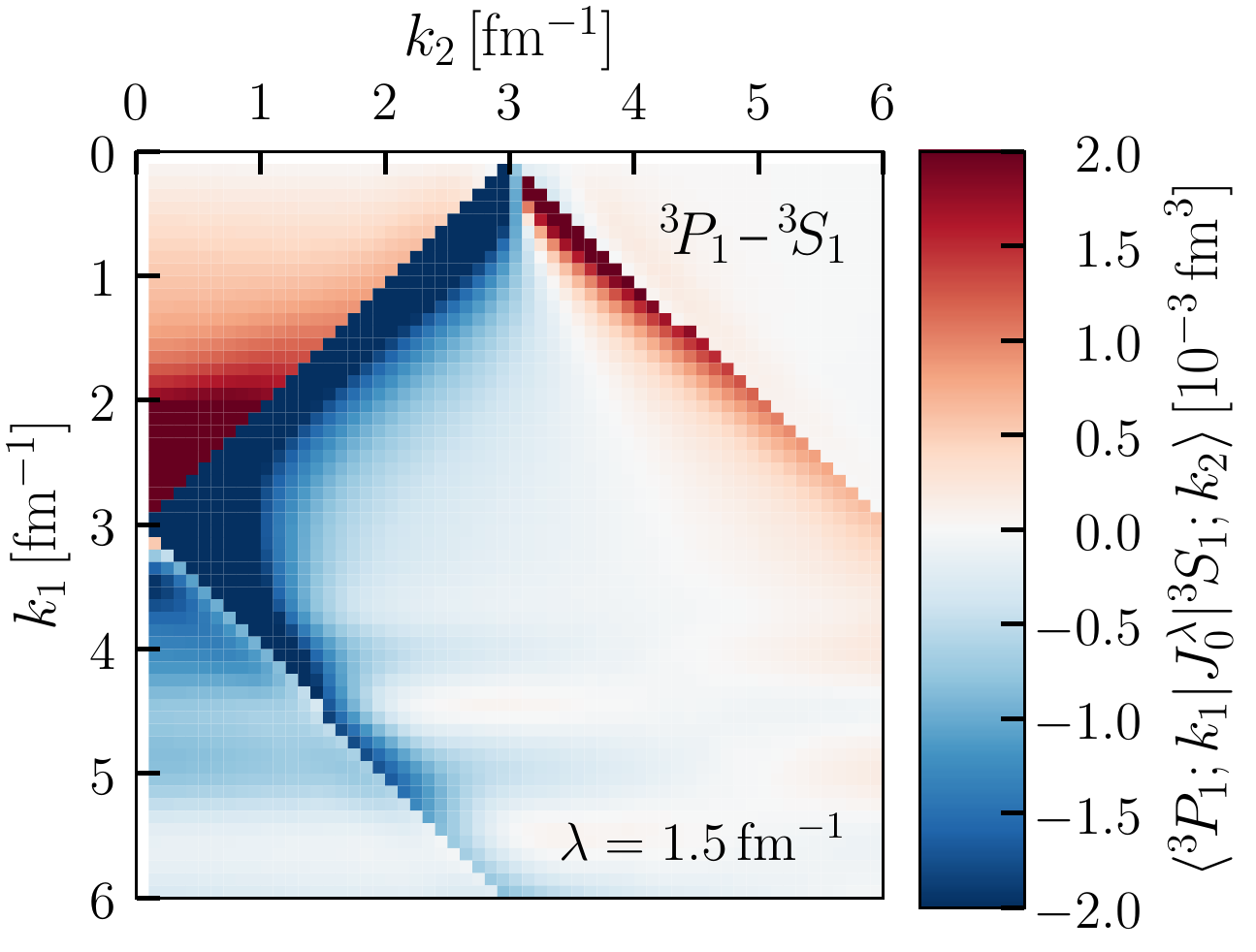}
  \put(11.8,43.4){\tikz\draw[cyan, thick,rounded corners] (0,0) rectangle (1.11,1.51);}
  \end{overpic}
  \caption{Heatmap plot for the evolved current ($\lambda = 1.5\,$\fmi) 
  for $q^2 = 36 {\rm{~fm^{-2}}}$ in the 
  $^3\!P_1-^3\!\!S_1$ channel for $\mJd = 1$.  
  While the changes due to evolution are distributed,  
  only the low-momentum region of the evolved current is 
  relevant for our examples (see the boxed region).  
  In this region, the evolved current is smooth and linear in 
  $k_1$ (cf.~Eq.~\eqref{eq:J0_lam_EFT_P}).    
  }
  \label{fig:J_evolution_heatmap_3P1_3S1}
\end{figure}

In Figs.~\ref{fig:J_evolution_heatmaps_3S1_3S1_q36} and
\ref{fig:J_evolution_heatmap_3P1_3S1}
we show heatmap plots of $J_0^{ \lambda}\equiv U_{\lambda}J_0^{\infty}U_{\lambda}^{\dagger}$ for $q = 6\,\fmi$ 
in the partial wave basis, which are representative of the
characteristics of the current under SRG evolution. 
The unevolved
current is a one-body operator and is peaked at $(0, q/2)$ and $(q/2, 0)$.
Under SRG evolution with $G_s = T$, the one-body part is unchanged but
the current develops two-body components, $\Delta J_0^{\lambda}$.
As seen in Figs.~\ref{fig:J_evolution_heatmaps_3S1_3S1_q36} and 
\ref{fig:J_evolution_heatmap_3P1_3S1}, the changes 
due to evolution are smooth and distributed for momenta less than $q/2$, and  
the evolved current does not become pathologically large
at high momentum.  This is important because for practical calculations
the evolved current will be used in conjunction with the evolved
wave functions.  These wave functions have negligible strength
at high momentum and the absence of pathologies in the evolved current
make calculations with the SRG in a reduced basis possible~\cite{Anderson:2010aq}.

\subsection{Final-state interactions}
\label{subsec:final_state_interactions}

\begin{figure}[tbh!]
  \centering
  \includegraphics[width=0.9\columnwidth]
  {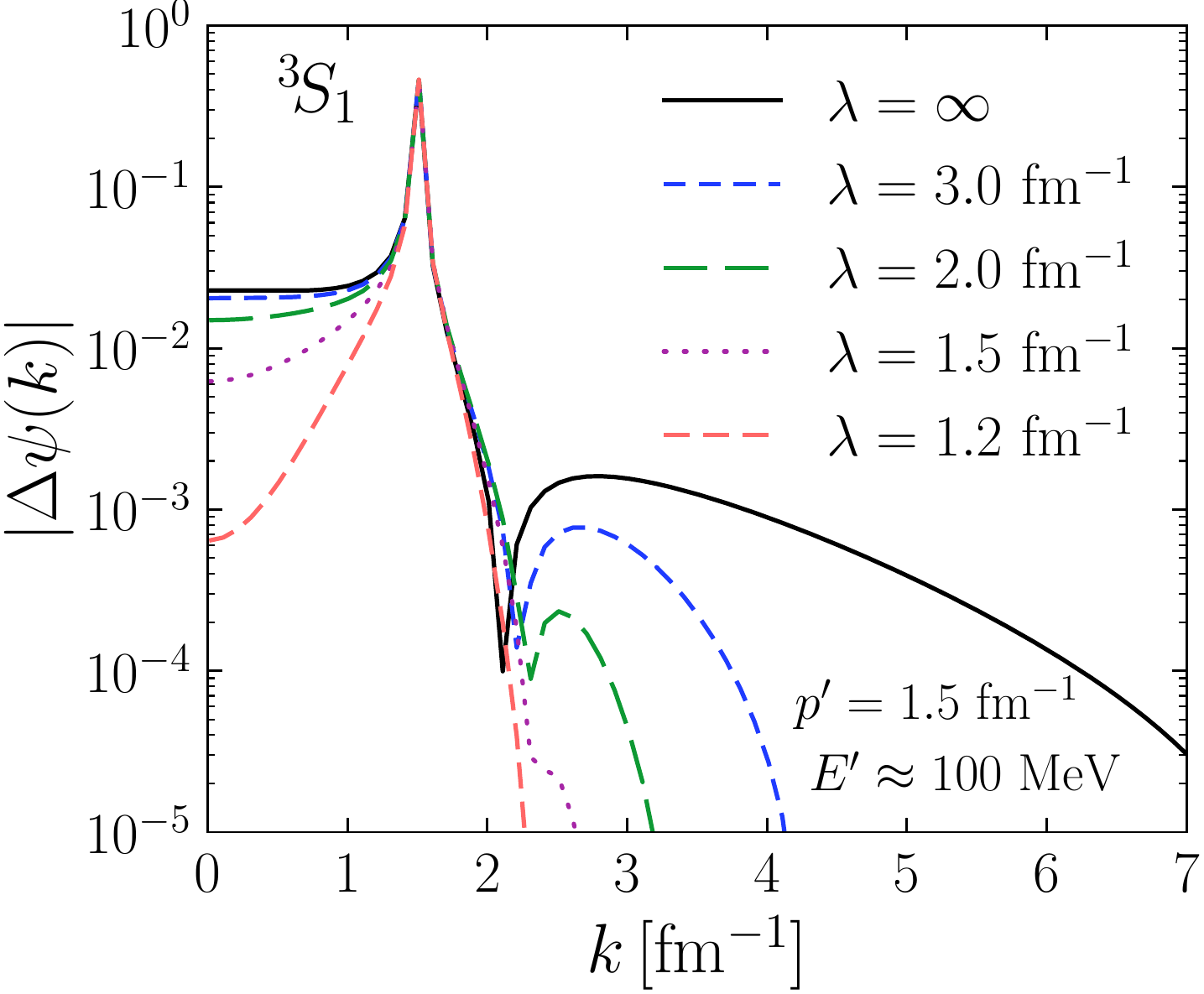}
\caption{Local decoupling in the final state of the $^3$S$_1$ partial wave 
at large momentum $p'=1.5\,$fm.}
\label{fig:local_decoupling_final_state}
\end{figure}

Final-state interactions often complicate the extraction of structure information from electron scattering measurements.
Having shown that the individual ingredients of the observable cross section are by definition scale- and scheme-dependent quantities, it is interesting to ask if one can choose the resolution scale to minimize the importance of FSI in certain kinematics. We use the case of large $\pp$ as an example, taking $\pp = 1.5\,\fmi$ and
considering $\lambda$ values both larger and smaller than $\pp$.  
Recall from Eq.~\eqref{eq:psi_f_def} that the scattering wave function 
is given by the sum of a free plane wave and the 
modification $\Delta \psi(k)$ due to the potential. 
The evolution of $|\Delta \psi(k)|$ is shown in Fig.~\ref{fig:local_decoupling_final_state},
where we see a common peak near the on-shell point $k = \pp$, 
but very different contributions
at large and small $k$ as a function of $\lambda$.
For $\pp \gtrsim \lambda$, $\Delta \psi(k)$ is well localized 
around $k = \pp$ due to the local decoupling properties of the SRG with the present choice of generator.

\begin{figure}[tbh!]
  \centering
  \includegraphics[width=0.9\columnwidth]
  {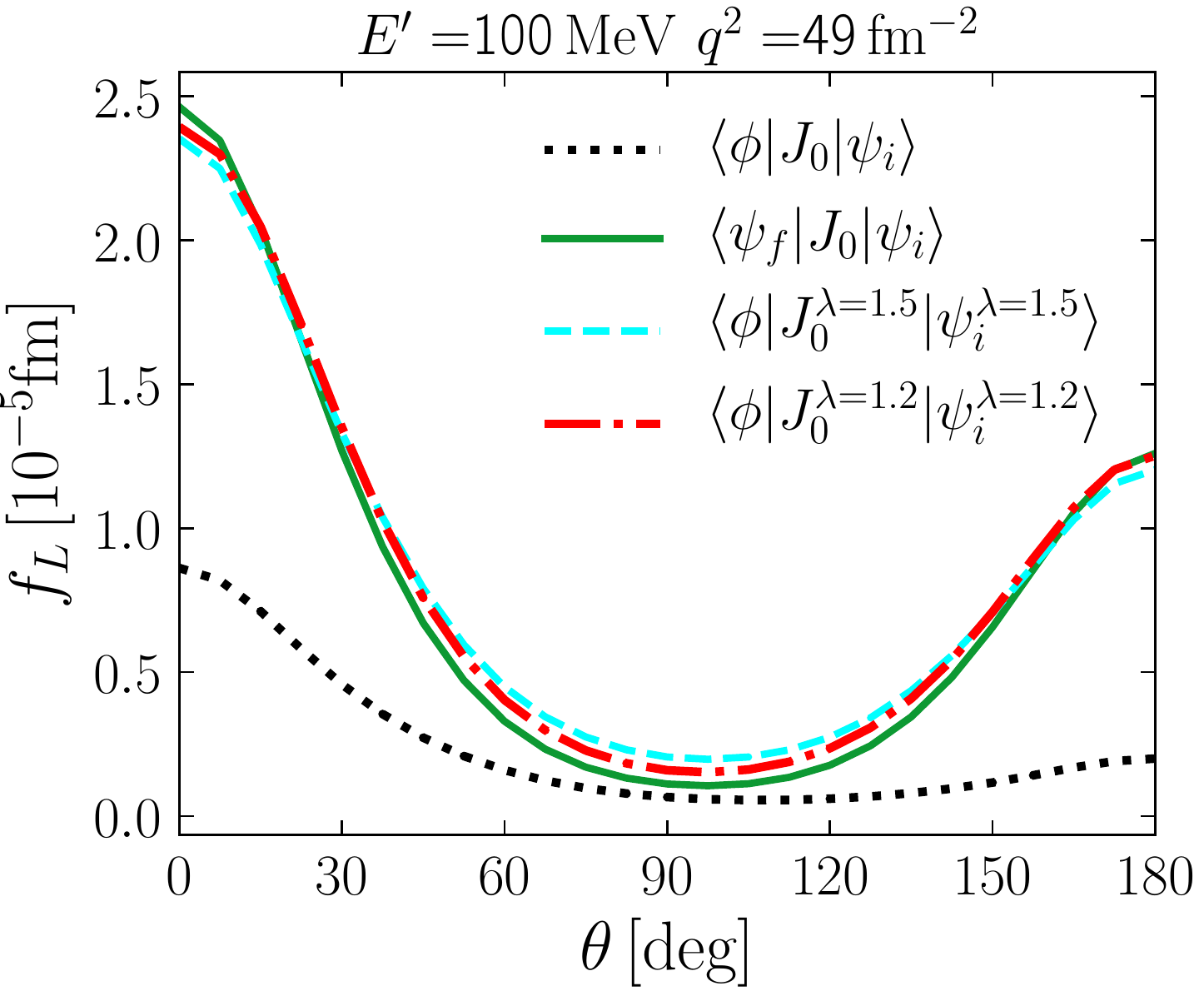}
\caption{Comparing the unevolved IA and the full calculations
of $f_L$ at $p' = 1.5\,\fmi$ (or $E' = 100\,$MeV) and $q^2 = 49\,\mbox{fm}^{-2}$ with the evolved 
IA result at $\lambda = 1.5\,\fmi$ and $\lambda = 1.2\,\fmi$.  
We find that the contribution of FSI is minimal in the evolved picture at this 
kinematics corresponding to
$x_d = 1.64$, $Q^2 = 1.78 ~\rm{GeV^2}$.}
\label{fig:evolved_IA_pic_49p0}
\end{figure}

\begin{figure}[tbh!]
  \centering
  \includegraphics[width=0.9\columnwidth]
  {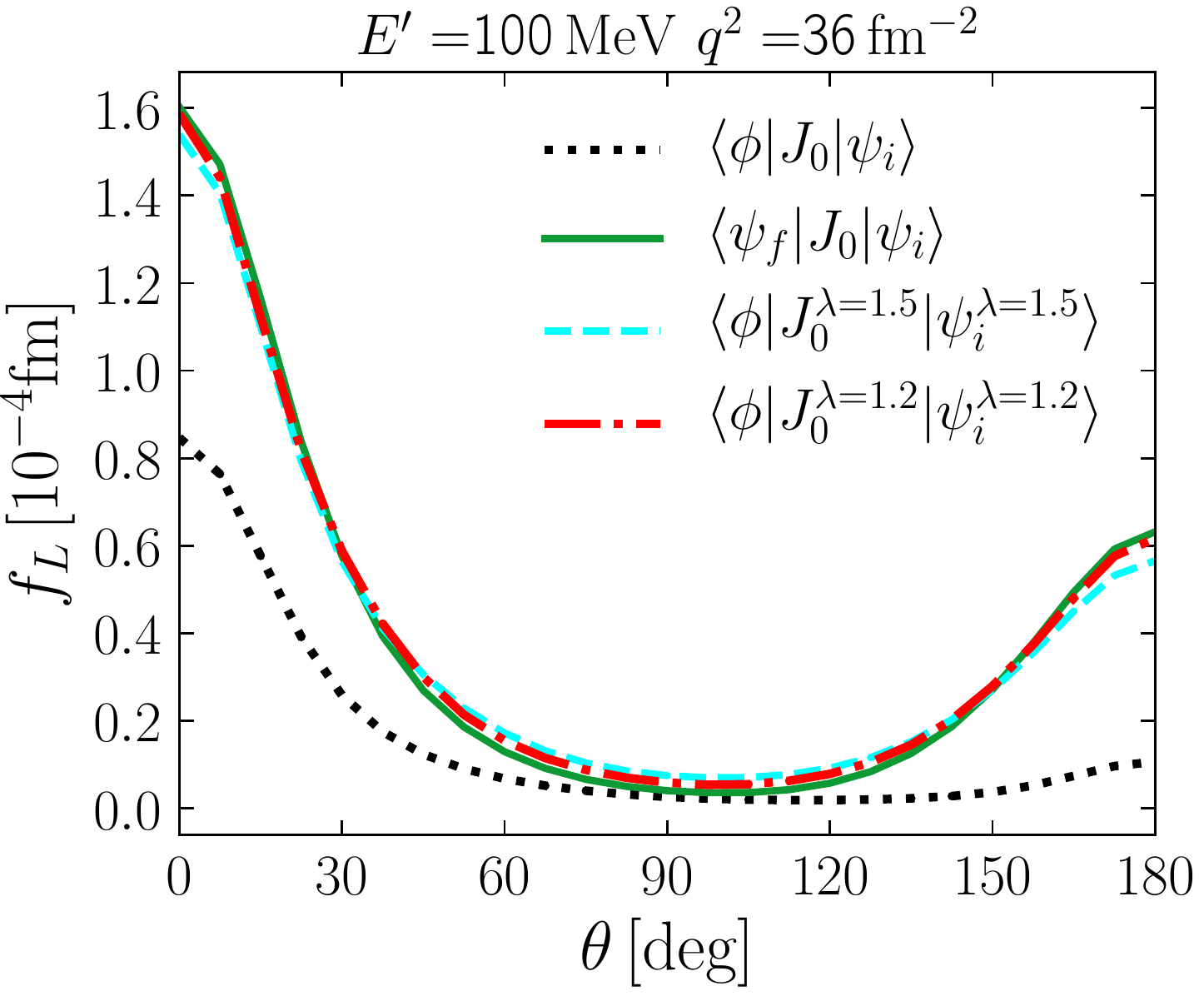}
\caption{Same as Fig.~\ref{fig:evolved_IA_pic_49p0} but at $q^2 = 36\,\mbox{fm}^{-2}$. 
Here $x_d = 1.55$, $Q^2 = 1.34 ~\rm{GeV^2}$.}
\label{fig:evolved_IA_pic_36p0}
\end{figure}

\begin{figure}[tbh!]
  \centering
  \includegraphics[width=0.9\columnwidth]
  {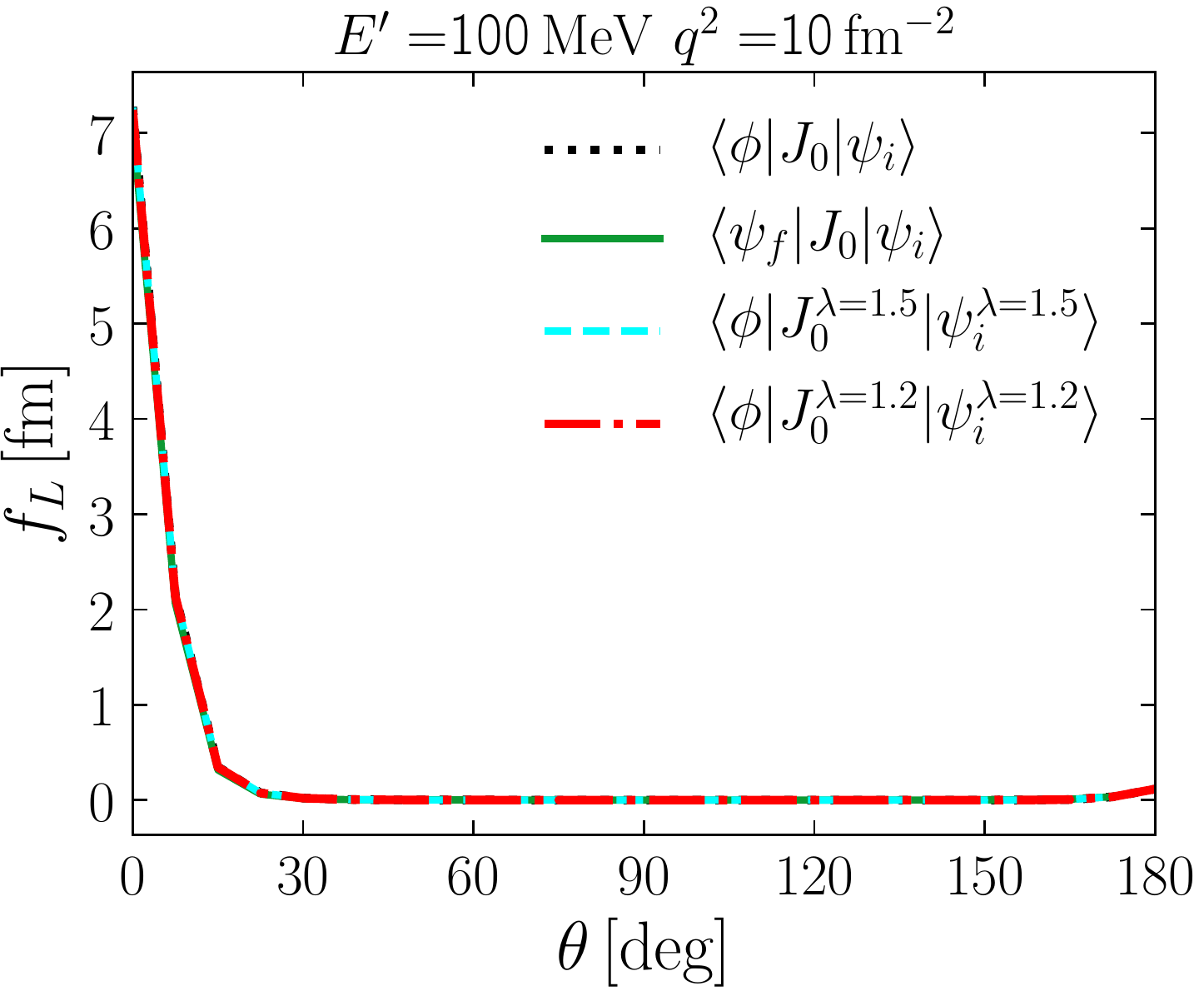}
\caption{Comparing the unevolved IA and the full calculations
of $f_L$ at $E' = 100\,$MeV and $q^2 = 10\,\mbox{fm}^{-2}$ with the evolved 
IA result at $\lambda = 1.5\,\fmi$ and $\lambda = 1.2\,\fmi$.  
This kinematics with $x_d = 0.99$ and $Q^2 = 0.39 ~\rm{GeV^2}$ 
corresponds to the quasi-free ridge.  At the quasi-free ridge, the 
contributions from the FSI are small for all SRG scales.}
\label{fig:evolved_IA_pic_10p0}
\end{figure}

\begin{figure}[tbh!]
  \centering
  \includegraphics[width=0.9\columnwidth]
  {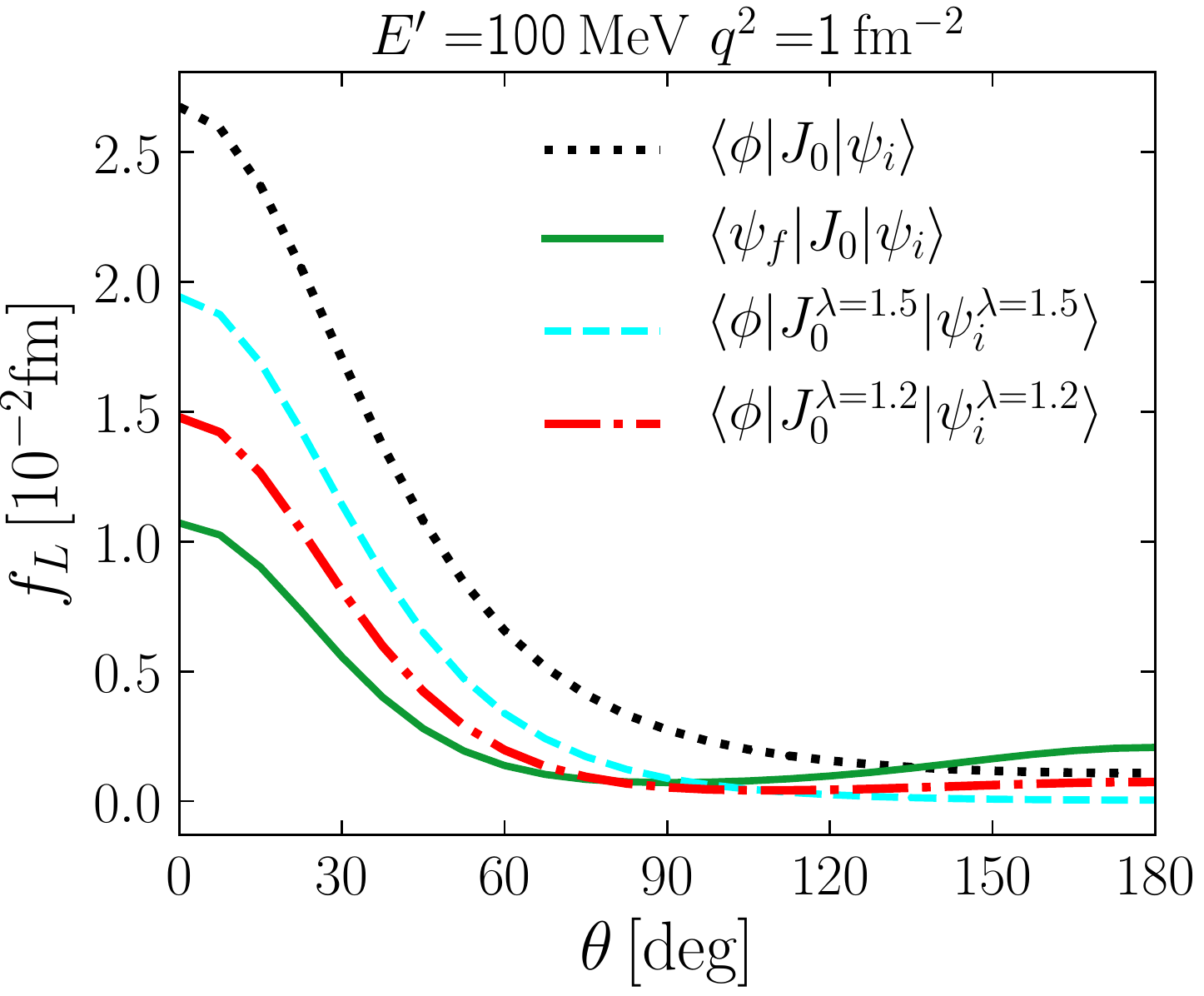}
\caption{Comparing the unevolved IA and the full calculations
of $f_L$ at $p' = 1.5\,\fmi$ (or $E' = 100\,$MeV) and $q^2 = 1\,\mbox{fm}^{-2}$ with the evolved 
IA result at $\lambda = 1.5\,\fmi$ and $\lambda = 1.2\,\fmi$.  
In the region $\pp > q/2$, $\psi_f^{\lambda}$ picks contribution from the 
non-smooth high-momentum region of $J_0^{\lambda}(k, k^\prime)$.  
Here $x_d = 0.14$, $Q^2 = 0.03 ~\rm{GeV^2}$.}
\label{fig:evolved_IA_pic_1p0}
\end{figure}

In Figs.~\ref{fig:evolved_IA_pic_49p0} through \ref{fig:evolved_IA_pic_1p0}, we scan through a
range of $q^2$ and show $f_L$ for $\lambda = \infty$, 1.5, and $1.2\,\fmi$.
The efficacy of the IA for the unevolved ($\lambda=\infty$) calculations
is highly dependent on $q^2$.
At the large $q^2$ values shown in Figs.~\ref{fig:evolved_IA_pic_49p0} and \ref{fig:evolved_IA_pic_36p0},  
which falls in the kinematic range commonly used to probe short-range correlations 
in nuclei~\cite{Atti:2015eda,Hen:2016kwk}, explicit calculations show that in the 
unevolved case, the high-momentum tail 
in the final-state wave function gives a sizable contribution to $f_L$, 
roughly a 100\% correction at $q^2 = 49\,\mbox{fm}^{-2}$.  
As $q^2$ is lowered, the correction decreases and for quasi-free
kinematics (here with $q^2 = 10\,\mbox{fm}^{-2}$), there is only a small 
contribution from the FSI, as expected~\cite{Fabian:1979kx}.
With a further lowering of the momentum transfer to $q^2 = 1\,\mbox{fm}^{-2}$,
the FSI correction is again very large.

In strong contrast,  as we scan through the range of $q^2$, the IA answer in 
the \emph{evolved} picture closely tracks the full FSI answer.  
This can be qualitatively understood from the contour plots of $J_0^{\lambda}$ in 
Figs.~\ref{fig:J_evolution_heatmaps_3S1_3S1_q36} and 
\ref{fig:J_evolution_heatmap_3P1_3S1}, in conjunction with 
Figs.~\ref{fig:deut_wfn_k_space} and \ref{fig:local_decoupling_final_state}.  
Consider the two cases shown in Figs.~\ref{fig:evolved_IA_pic_49p0} and \ref{fig:evolved_IA_pic_36p0} where $p'\sim \lambda < q/2$.  The evolved 
deuteron wave function restricts the contribution from 
$J_0^{\lambda}(k, k^{\prime})$ 
(cf.\ Fig.~\ref{fig:J_evolution_heatmaps_3S1_3S1_q36}) to 
$k^{\prime} \lesssim \lambda$, while 
the evolved final state primarily picks up contributions from 
$J_0^{\lambda}(k, k^{\prime})$ for $k \approx \pp$.  
For $\pp < q/2$, this is the region where $J_0^{\lambda}$ is smooth and
well-approximated by simple derivative and low-rank SVD expansions 
(see Sections~\ref{subsec:Derivative_expansion} and 
\ref{subsec:SVD_analysis}).  Because $\Delta\psi_{p'}(k)$ is localized 
about $k\approx p'$, the final-state interaction is proportional to the 
on-shell $t$-matrix, which is small at large momentum. 

As we decrease $q$, we approach the quasi-free ridge
(cf.~Fig.~\ref{fig:evolved_IA_pic_10p0}) 
characterized by $\pp \approx q/2$ (also note that 
$x_d \approx 1$).  Note from 
Fig.~\ref{fig:J_evolution_heatmaps_3S1_3S1_q36}  
that most of the strength of 
$J_0^{\lambda}$ is concentrated around $\pp \approx q/2$.  
This is the region where the FSI contributions are small 
regardless of the choice of the SRG scale~\cite{More:2015tpa}.  
As we decrease $q$ even further, we are in the region 
where $\pp > q/2$.  Here $\psi_f^{\lambda}$ picks contributions 
from $J_0^{\lambda}(k, k^{\prime})$ in the region 
$k > q/2$.  Unlike the $k < q/2$ region, the form of 
$J_0^{\lambda}$ at large momentum is not smooth in momentum 
(cf.~Fig.~\ref{fig:J_evolution_heatmap_3P1_3S1}) and 
therefore little can be said about the effect of evolution on the 
IA.  Moreover as seen from the $x_d$ and $Q^2$ values, 
unlike the kinematics in 
Figs.~\ref{fig:evolved_IA_pic_49p0} and 
\ref{fig:evolved_IA_pic_36p0}, the kinematics in 
Fig.~\ref{fig:evolved_IA_pic_1p0} is not of much experimental interest.

\subsection{Derivative expansion at low resolution}
\label{subsec:Derivative_expansion}

Conventional wisdom holds that simple low-resolution wave functions inevitably lead to complicated reaction calculations 
(and interpretations) because the relevant transition operators are transformed to more complicated forms.  One might expect these complications to be especially severe for operators that probe the 
high-momentum structure of nuclear wave functions, 
such as the  current operator $J_0({\bf q})$ at large $q^2$, since such components 
are highly suppressed or completely absent from low-resolution wave functions~\cite{Hen:2016kwk}.  
However, in this regime new simplifications emerge due to the separation of scales 
$\lambda \ll q$, with the induced terms taking the form of an EFT derivative expansion, 
where each term consists of a $\lambda$-dependent coupling constant that encodes the 
effects of decoupled high-momentum states, multiplied by a regulated contact interaction 
that is the same for all operators with the same symmetries~\cite{Bogner:2012zm}. 

In the case of deuteron electrodisintegration for $p< \lambda \ll q/2$, 
the initial- and final-state wave functions predominantly probe the low-momentum 
components of the evolved current $J^{\lambda}_0(q)$. 
Because the one-body component of the current doesn't evolve under the SRG and is sharply 
peaked at $(k,k')=(0,q/2)$ and $(q/2,0)$, the transition matrix element is only 
sensitive to the induced two-body current $\Delta J_0^{\lambda}$. 
The low-momentum part of $\Delta J_0^{\lambda}$ can be expanded as
\begin{align}
  \mbraket{^3S_1; k_1}{\Delta J_0^{\lambda}(q)}{^3S_1; k_2}
  & = g_0^{\mJ}(q) + g_{2}^{\mJ}(q)(k_1^2 + k_2^2) \notag
   \\ & \qquad\null + \cdots \label{eq:J0_lam_EFT_S} \;, \\
  \mbraket{^3P_1; k_1}{\Delta J_0^{\lambda}(q)}{^3S_1; k_2} 
  & = g_1^{\mJ}(q) \, k_1 + g_3^{\mJ}(q) \,k_1 \, k_2^3 \notag
   \\ & \qquad\null + \cdots \;,
  \label{eq:J0_lam_EFT_P}
\end{align}
and similarly for higher partial waves. The ``low-energy constants'' (LECs) $g_0^{\mJ}, g_{2}^{\mJ}$, etc., 
are in principle calculable as described in Ref.~\cite{Bogner:2012zm}, although in the present work we extract them 
simply by fitting to the exact  $\Delta J_0^{\lambda}(k',k; q)$ in each channel. 

As a proof-of-principle, we calculate $f_L$ at $\lambda = 1.5$ fm$^{-1}$ 
using the derivative expansion for $\Delta J_0^{\lambda}$ at kinematics 
($E'=20$ MeV, $q^2 = 36$ fm$^{-2}$ or $x_d = 1.88$, $Q^2 = 1.3~\rm{GeV^2}$) that are sensitive to short-range correlations 
at high resolution scales. 
Here, we only include the $S$ state in deuteron in the evolved picture.  
As we will see in Sec.~\ref{sec:summary}, this is a 
very good approximation for small $\pp$ and large $q^2$, i.e., 
$\mbraket{\psi_f^{\lambda}}{J_0^{\lambda}(q)}{\psi_i^{\lambda}} 
\approx \mbraket{\psi_f^{\lambda}}{J_0^{\lambda}(q)}{{\psi_i^{\lambda}}_{^3S_1}}$.

Evaluation of the matrix element 
$\mbraket{\psi_f^{\lambda}}{J_0^{\lambda}(q)}{{\psi_i^{\lambda}}_{^3S_1}}$ 
involves sums over partial wave channels for the final state, i.e., 
\begin{align}
  & \mbraket{\psi_f^{\lambda}}{J_0^{\lambda}}{{\psi_i^{\lambda}}_{^3S_1}} 
   = \braket{\psi_f^{\lambda}}{^3S_1} 
    \underbrace{\mbraket{^3S_1}{\Delta J_0^{\lambda}}{^3S_1}}_{\rm use~der.~exp.}
    \braket{^3S_1}{{\psi_i^{\lambda}}_{^3S_1}} \notag \\
   & \qquad \null + \sum_{J = 0, 1, 2} 
    \braket{\psi_f^{\lambda}}{^3P_J} 
    \underbrace{\mbraket{^3P_J}{\Delta J_0^{\lambda}}{^3S_1}}_{\rm use~der.~exp.}
    \braket{^3S_1}{{\psi_i^{\lambda}}_{^3S_1}} \notag \\ 
   & \qquad \null +  
    \sum_{J = 1, 2, 3} 
    \braket{\psi_f^{\lambda}}{^3D_J} 
    \underbrace{\mbraket{^3D_J}{\Delta J_0^{\lambda}}{^3S_1}}_{\rm use~der.~exp.}
    \braket{^3S_1}{{\psi_i^{\lambda}}_{^3S_1}} \notag \\
   & \qquad \null + \cdots\,,
  \label{eq:final_state_sum_over_pw}       
\end{align}   
where we used that the contribution of the one-body part of the current is exponentially suppressed at these kinematics for small $\lambda$ values. For the matrix elements of the evolved current in the given partial wave channel, 
we use the derivative expansion for that channel as in 
Eqs.~\eqref{eq:J0_lam_EFT_S} and \eqref{eq:J0_lam_EFT_P}.  

\begin{figure}[htbp]
\centering
\includegraphics[width=0.9\columnwidth]
{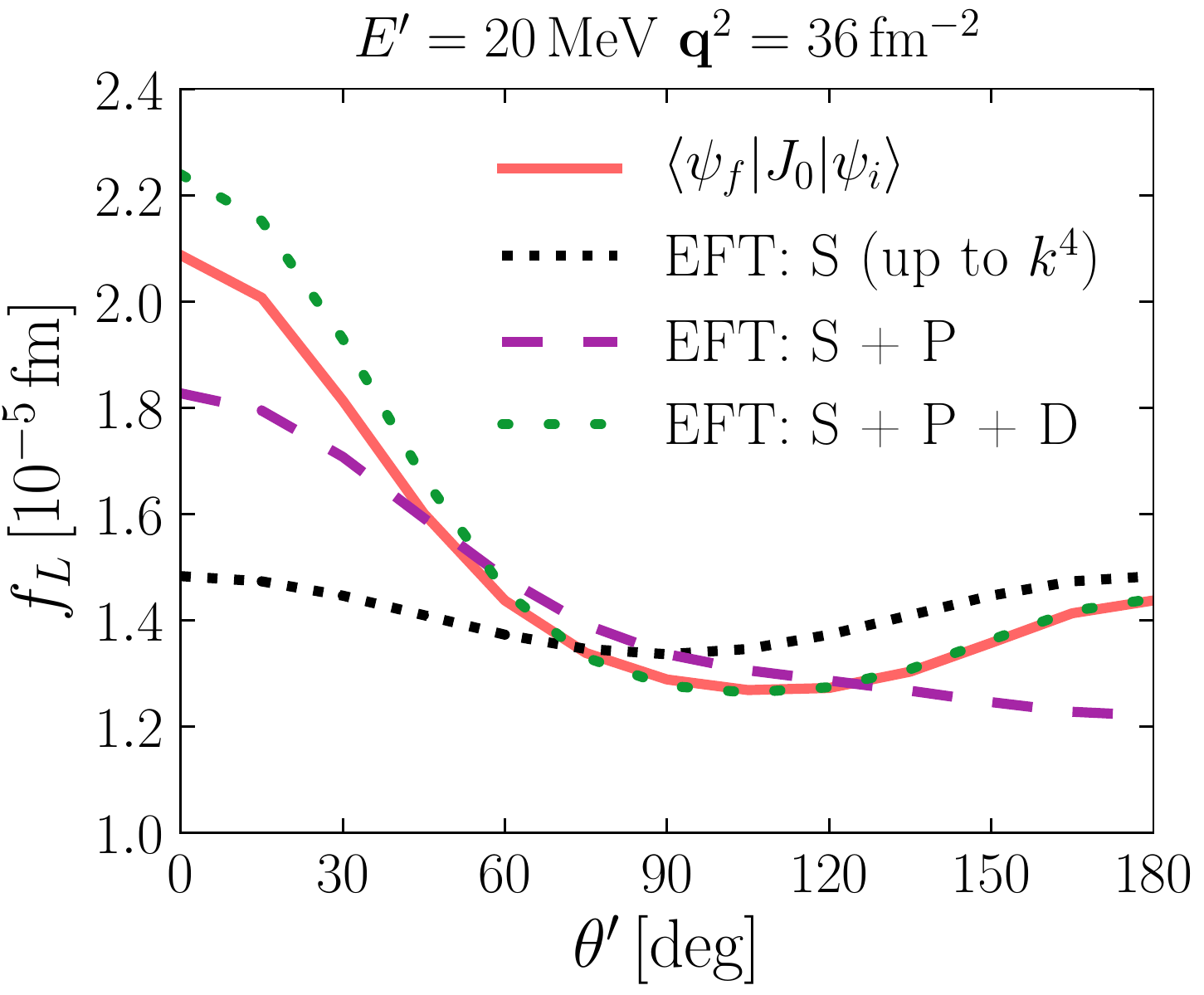}
\caption{Exact $\fL$ (solid line) for $E_{np} = 20~\rm{MeV}$ and $q^2 = 36~\rm{fm^{-2}}$ compared 
to $\fL$ obtained using the derivative expansion for the evolved current ($\lambda = 1.5~\rm{\fmi}$).  
For this kinematics 
$x_d = 1.88$, $Q^2 = 1.3~\rm{GeV^2}.$ 
}
\label{fig:evolved_simple_pic}
\end{figure}

Figure~\ref{fig:evolved_simple_pic} shows results 
for $\fL$ calculated using the derivative expansion as outlined above.
The solid line in Fig.~\ref{fig:evolved_simple_pic} is $\fL$ calculated 
in the unevolved picture.  The sparsely dotted line is $\fL$ calculated 
keeping just the $S$ channel in the final state (just the first term on 
the right-hand side of Eq.~\eqref{eq:final_state_sum_over_pw}), and for the 
derivative expansion of the evolved current we keep up to terms 
proportional to $k^4$ in Eq.~\eqref{eq:J0_lam_EFT_S}.  The dashed 
line is $\fL$ calculated from including the $S$ and $P$ channel 
terms in the final state.  In the derivative expansion for the 
$P$ channel, we keep only the leading-order (LO) linear term in 
momentum in Eq.~\eqref{eq:J0_lam_EFT_P}.  Lastly, 
the densely dotted line includes the correction 
to $\fL$ from the $D$ channel in the final state as well. 
Again, we only include the leading-order quadratic corrections in the 
derivative expansion for the evolved current in the $D$ channel.% 
\footnote{We find that the values for $\mbraket{^3\!D_1; k_1}{\Delta J}{^3\!S_1; k_2}$ 
channel are not quadratic in $k_1^2$;  however, they are about an 
order of magnitude smaller than the 
$\mbraket{^3\!D_3; k_1}{\Delta J}{^3\!S_1; k_2}$ channel.  The smallness makes the 
$^3\!D_1$ channel inconsequential and we neglect it here.}   
We find that $\fL$ calculated in the low-resolution picture through the 
derivative expansion agrees very well with the unevolved answer.  
The agreement can be made even better by going to higher-order terms 
in the derivative expansion.      

\begin{figure}[tbh!]
	\centering
	\includegraphics[width=0.9\columnwidth]
	{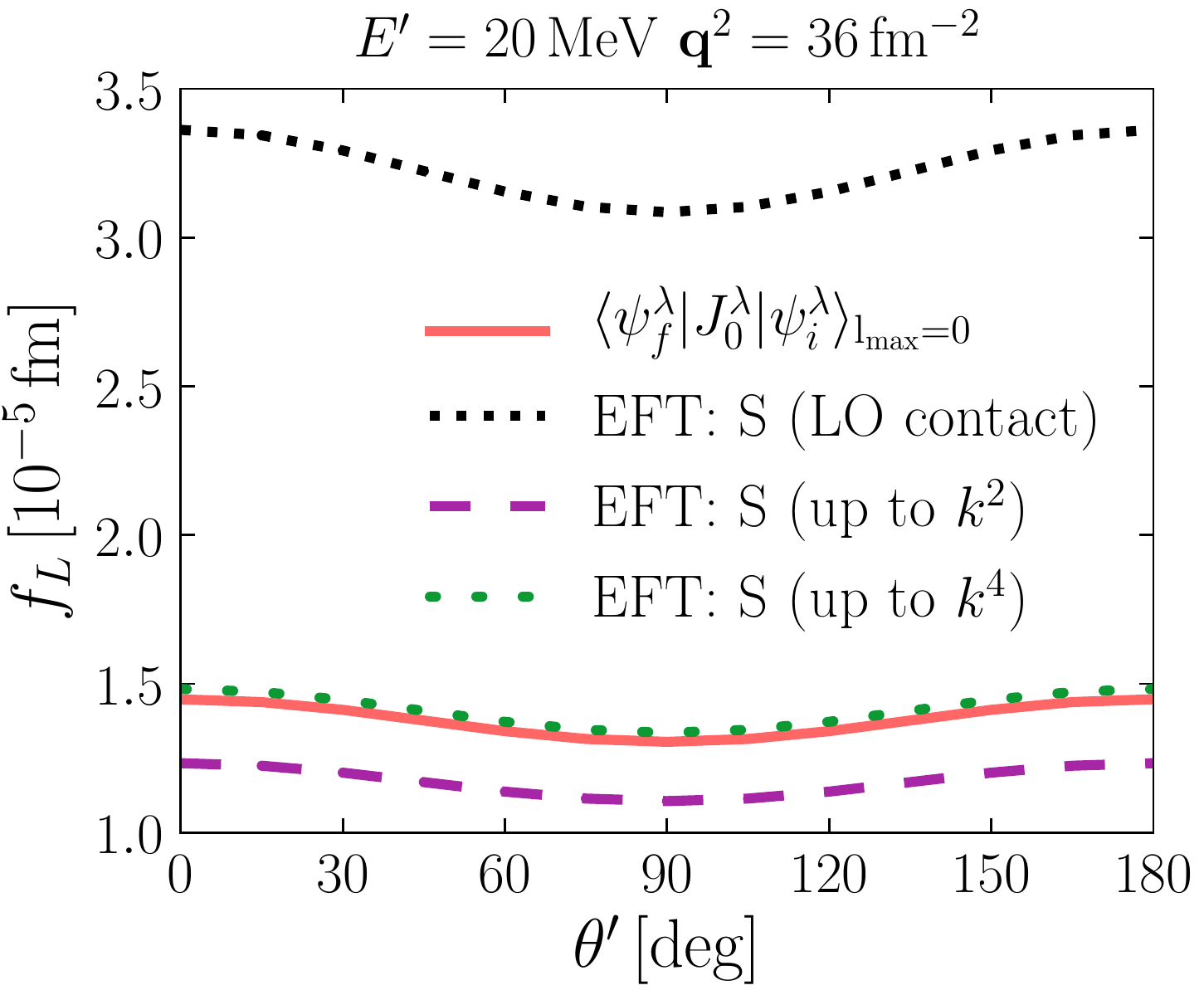}
	\caption{Convergence of the derivative expansion in the $S$ channel.  
  Here $x_d = 1.88$, $Q^2 = 1.3~\rm{GeV^2}$, and $\lambda = 1.5~\rm{\fmi}$.}
  \label{fig:convergence_der_exp_S_channel}
\end{figure}

Note that in Fig.~\ref{fig:evolved_simple_pic}, we have added the next-to-next-to-leading-order (N2LO) 
correction for the $S$-channel to the LO correction for the $P$ and $D$-channels.  
We find that the keeping only the LO correction for the $S$-channel gives a 
poor result.  We illustrate this in Fig.~\ref{fig:convergence_der_exp_S_channel}.  

The solid line in Fig.~\ref{fig:convergence_der_exp_S_channel} corresponds to 
$\fL$ calculated from the exact evolved current, but keeping only the 
$S$-channels for the initial and final states.  
That is, $\fL$ is calculated from the matrix element 
$\mbraket{{\psi_f^{\lambda}};{^3S_1}}{{J_0^{\lambda}}_{\rm exact}}
{{\psi_i^{\lambda}};{^3S_1}}$.  This is then compared to  
$\fL$ calculated from the derivative expansion for 
$\mbraket{{^3S_1}; k_1}{{J_0^{\lambda}}_{\rm exact}}
{{^3S_1}; k_2}$ going to successively higher terms in Eq.~\eqref{eq:J0_lam_EFT_S}.  
We find that the LO answer from the derivative expansion misses the 
the exact answer by about a factor of $2$, but it gets rapidly better when we 
include higher order terms in $k^2$.

\subsection{SVD expansion at low resolution}
\label{subsec:SVD_analysis}

A low-order derivative expansion of $\Delta J_0^{\lambda}(k',k)$ is most effective when the
 low-momentum components $k,k'\ll \lambda$ give the dominant contributions to the transition matrix element.  
 However, for the $\fL$ calculations, the integrals over the initial- and final-state wave functions 
 don't saturate until $k,k'\approx 1.6\lambda$.  
 In this sense, it is not surprising that the LO result in the $S$-channel is rather poor.  

A better way to arrange the expansion is suggested by the observation in Eq.~\eqref{eq:U_kq_factorization} 
that the SRG transformation approximately factorizes, $U_{\lambda}(k,q)\approx K_{\lambda}(k)Q_{\lambda}(q)$, 
for well-separated momenta $k< \lambda \ll q$~\cite{Anderson:2010aq}.  
Since it is precisely this portion of the $U_{\lambda}$-matrix, plus the smooth low-momentum 
block $U_{\lambda}(k,k')$, that enters into the construction of the evolved current operator 
for the present kinematics, one expects that a low-rank singular value decomposition (SVD) 
should efficiently capture the behavior of $\Delta J_q^{\lambda} \equiv \Delta J_0^{\lambda}(q)$.  
That is,
\beq
  \mbraket{k^\prime; ^3\!L_J}{\Delta J_q^{\lambda}}{k; ^3\!S_1} 
  \xrightarrow{\rm SVD} \sum_i c^q_{(i)} \,  j^{(i)}_{\rm left}(k^\prime) 
  j^{(i)}_{\rm right}(k)\;,
\label{eq:current_SVD}
\eeq
where $c^q_{(i)}$ are the singular values, and $j^{(i)}_{\rm left}$ and 
$j^{(i)}_{\rm right}$ 
are the left and right singular vectors respectively for the channel under 
consideration.  
As in Section~\ref{subsec:Derivative_expansion}, we keep only the 
$^3\!S_1$ channel in the deuteron evolved state and therefore $^3\!S_1$ in the 
ket in Eq.~\eqref{eq:current_SVD} in the current work.  Extending to include the 
$^3\!D_1$ channel is straightforward.  

The SVD analysis proceeds by constructing a matrix $\Delta J_q^{\lambda}(k_i, k_j)$ 
for $ 0 < k_i, k_j < \kmax$ for the given partial wave channel and performing the SVD 
to get the singular values and singular vectors.  
In the present work we choose $\kmax = 1.6 \, \lambda$, which gives a truncation error 
from the integrals over the initial- and final-state wave functions of less than $0.5 \%$.  
For the dominant $m_J=0$ channels, we find that the first singular value 
is substantially larger than all the subsequent ones. For the sub-dominant $m_J=\pm 1$ channels, 
the first two singular values are of the same order of magnitude, with a substantial falloff thereafter.  
Once we have the singular values and vectors, we can put them together to get the transition matrix 
elements and ultimately $\fL$.

\begin{figure}[tbh!]
  \centering
  \includegraphics[width=0.9\columnwidth]
  {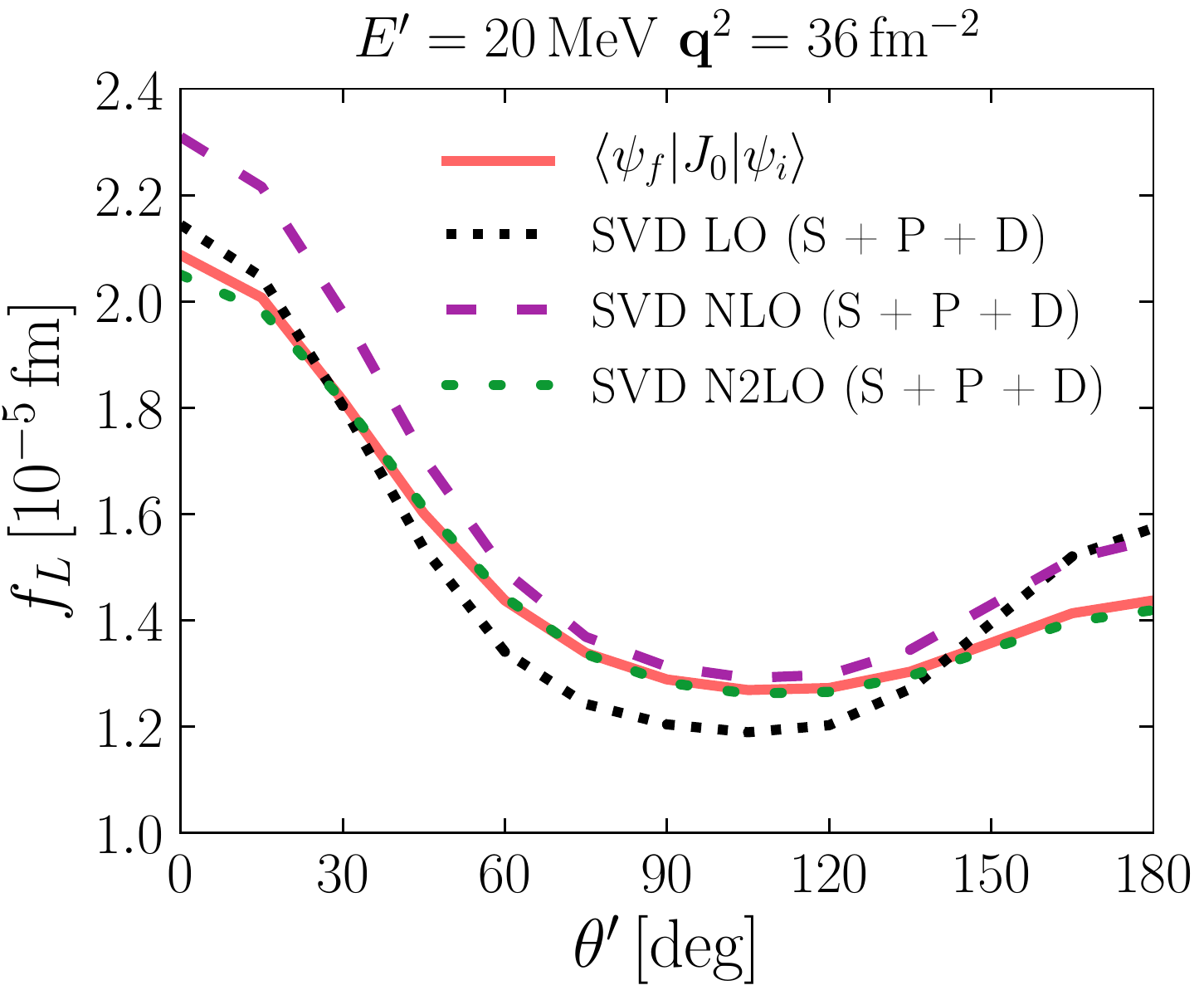}
  \caption{
  Exact $\fL$ (solid line) compared to 
  to $\fL$ obtained using the SVD expansion for the evolved current at successive orders
  for $\lambda = 1.5\,\fmi$ and $\kmax = 1.6\,\lambda$.  S+P+D indicates that we 
  have gone up to $D$-channel in the partial wave expansion of the final state. }
  \label{fig:evolved_simple_picture_SVD}
\end{figure}

\begin{figure}[tbh!]
  \centering
  \includegraphics[width=0.9\columnwidth]
  {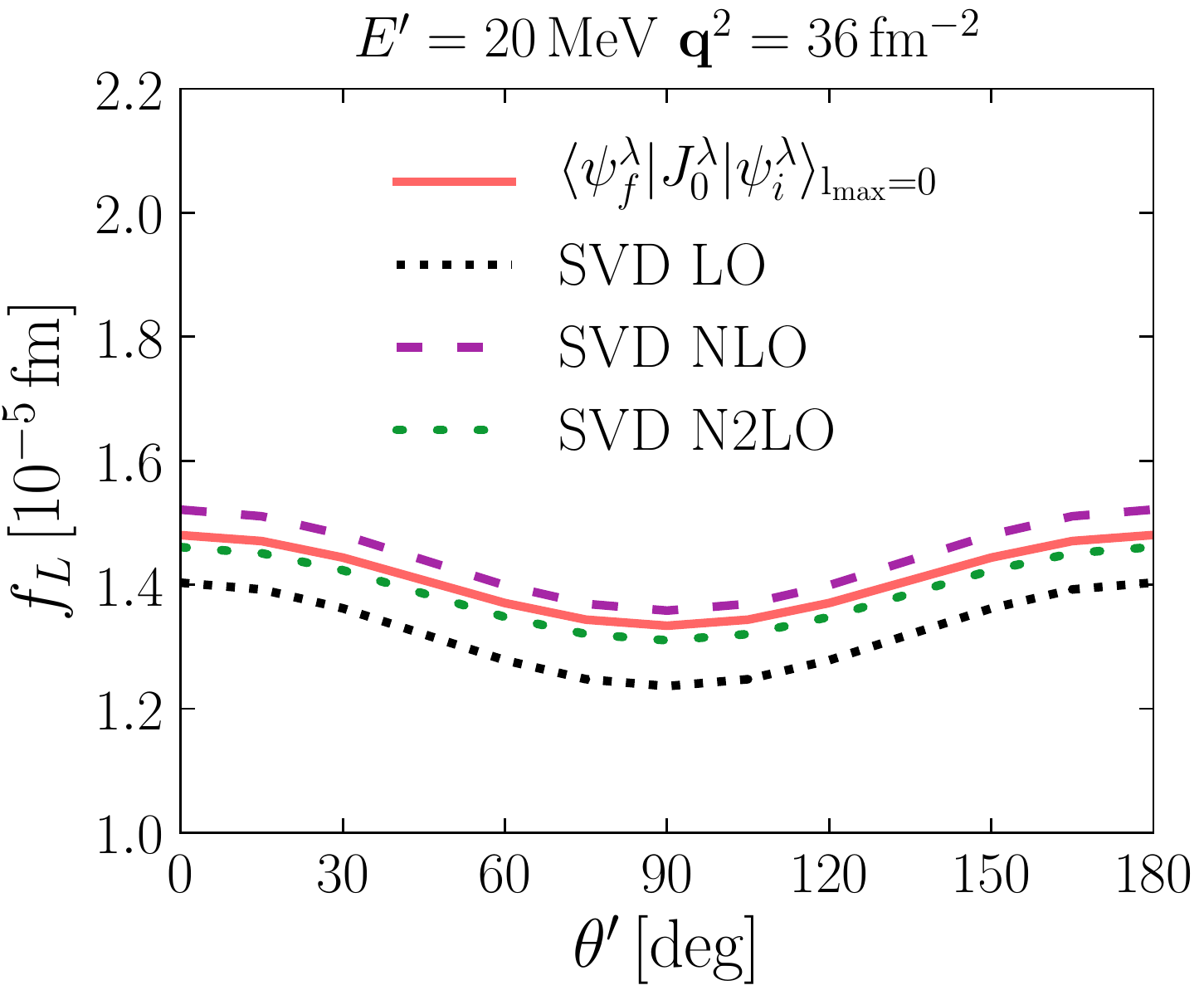}
  \caption{Convergence of the SVD expansion in the $S$ channel for $\lambda = 1.5\,\fmi$ and $\kmax = 1.6\,\lambda$.  
  Here $x_d = 1.88$, $Q^2 = 1.3~\rm{GeV^2}$.}
  \label{fig:convergence_SVD_exp_S_channel}
\end{figure}

Figure~\ref{fig:evolved_simple_picture_SVD} compares the exact $\fL$ for the same kinematics as in 
Section~\ref{subsec:Derivative_expansion} to the $\fL$ calculated using the SVD expansion for the evolved current.  
The LO result in Fig.~\ref{fig:evolved_simple_picture_SVD} corresponds to keeping 
the first term in Eq.~\eqref{eq:current_SVD} for each channel, the 
NLO result corresponds to keeping the first two terms for each channel, 
and so on.  Not surprisingly, given the smooth nature of the evolved current, we find that the SVD 
expansion quickly converges to the exact answer.  We also find that the LO SVD is far superior to 
the LO derivative expansion.  As shown in 
Fig.~\ref{fig:convergence_SVD_exp_S_channel}, in the $S$-channel, the LO SVD agrees to within 10\% of the exact result, while the LO derivative expansion of off by a factor of 2 
(cf.~Fig.~\ref{fig:convergence_der_exp_S_channel}).  
We speculate that this dramatic improvement can be understood as follows; the derivative expansion 
is an expansion in unregulated contact interactions,%
\footnote{There are no ultraviolet divergences 
because the current is sandwiched between SRG-evolved wave functions, which only have support at low momentum.}
whereas the low-momentum part of $\Delta J_q^{\lambda}$ is smooth and takes the form of a \emph{regulated}
derivative expansion.  The LO SVD is, therefore, analogous to a regulated contact interaction, 
with $\lambda$ setting the scale of the regulator.  
Moreover SVD through the shape of the singular vectors captures more of the actual physics 
of the system compared to imposing an regulator with arbitrary shape.


 \subsection{Factorization of q-dependence}
 \label{subsubsec:q_independence}

The derivative expansion for RG-evolved operators naturally factorizes the low and high momentum scales in a problem, and gives a natural explanation for why certain quantities in many-body systems (e.g., high-momentum tails of momentum distributions and structure factors) scale off the corresponding quantities in few-body systems~\cite{Bogner:2012zm}.  This separation is reflected in Eq.~\eqref{eq:J0_lam_EFT_P}, where the $q$-dependence of the evolved current operator is factorized into the LECs. We would like to check if an analogous separation holds for the SVD expansion of
$\Delta J^\lambda_q(k^\prime, k)$, in which case we expect the majority of the $q$-dependence is carried by the singular values.

\begin{figure}[tbh!]
  \centering
  \includegraphics[width=0.9\columnwidth]
  {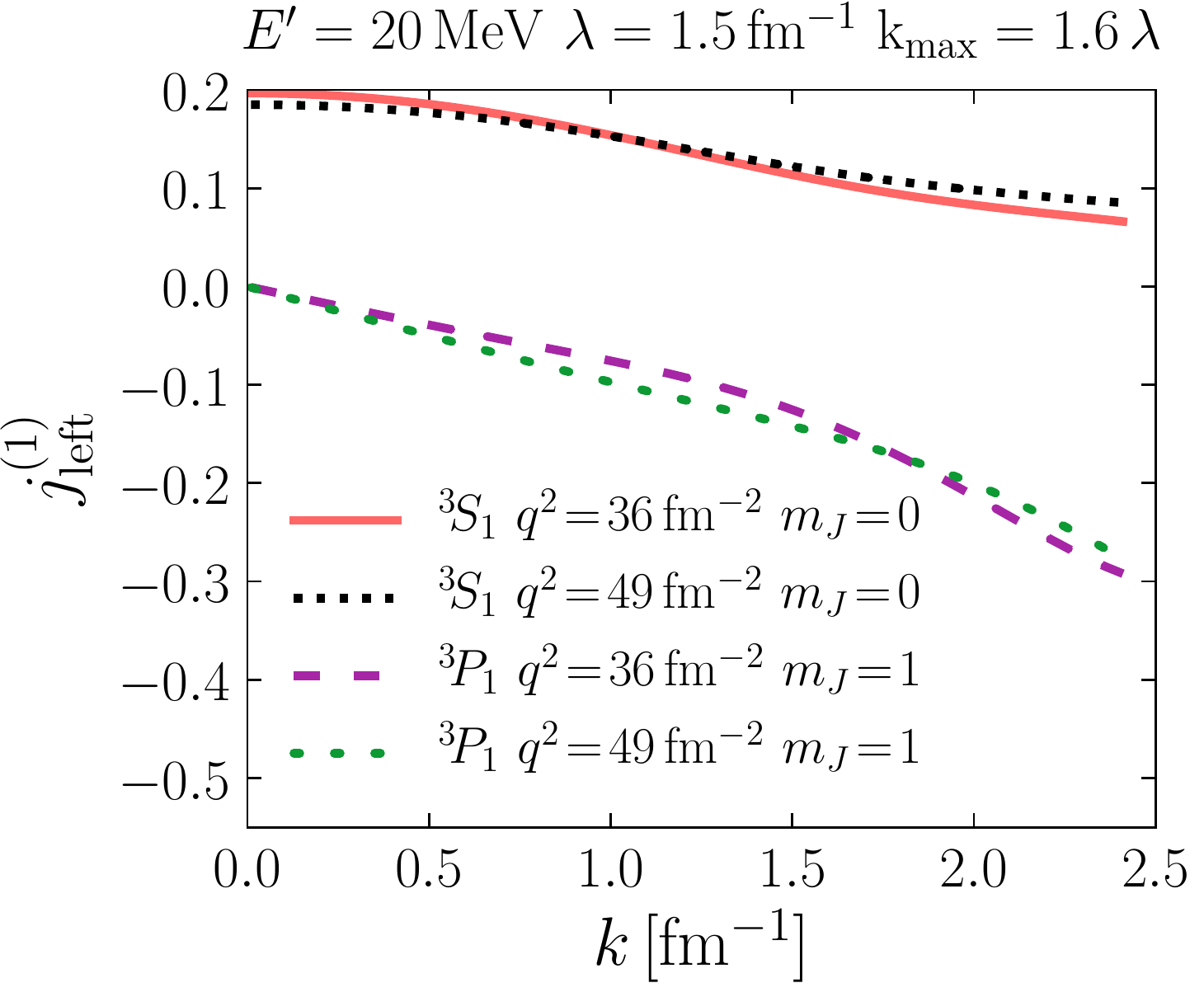}
  \caption{Comparison of singular vectors for different $q^2$ for two different channels for $\lambda = 1.5\,\fmi$ and $\kmax = 1.6\,\lambda$.}
  \label{fig:singular_vector_comparison}
\end{figure}

To check this, we look at the ratio of the LO SVD expansion $\Delta J^\lambda_q(k^\prime, k)$ at small 
$k$ for two different values of $q$ ($q^2 = 36$ fm$^{-2}$ and $49$ fm$^{-2}$), and compare it to the 
ratios of the corresponding LO singular values. 
Referring to Table~\ref{tab:ratio_Delta_J}, we see that most of the $q$-dependence is indeed 
carried by the singular values.  
Another way to demonstrate the factorized $q$-dependence is to look at the singular vectors.  
We do this in Fig.~\ref{fig:singular_vector_comparison}, where we find that the singular vectors 
for $q^2 = 36 \, \rm{fm^{-2}}$ and $q^2 = 49 \, \rm{fm^{-2}}$ 
are almost the same, indicating again that the dominant $q$-dependence is carried by the 
singular values. 

 \begin{table}
 \caption{Ratio of LO SVD matrix elements $\Delta J_q^{\lambda}(k_1,k_2)$  for two different $q$-values 
 at fixed $k_1 = 0.3\,\fmi$ and $k_2 = 0.5\,\fmi$, 
 compared to the ratio of the corresponding LO singular values. 
 The similar values in the two columns demonstrate that the $q$-dependence of $\Delta J(k_1,k_2)$ is mostly factorized in the singular values.}
 \label{tab:ratio_Delta_J}
 \begin{center}
%{\setlength{\extrarowheight}{6000pt}}
\def\arraystretch{2.2}
\begin{tabular}{ |c|c|c| }
      \hline
  \multicolumn{3}{|c|}{
  $q_1^2 = 36 {\rm\, fm^{-2}}$\, $q_2^2 = 49 {\rm\, fm^{-2}}$  \, 
  $\lambda = 1.5 {\rm \, fm^{-1}}$} \\ \hline
           &  $\dfrac{\Delta J^{\lambda~\rm{SVD~LO}}_{q_1}}{\Delta J^{\lambda~\rm{SVD~LO}}_{q_2}}$  & $\dfrac{c_{(1)}^{q_1}}{c_{(1)}^{q_2}}$ 
   \\ [1.5ex] \hline
  \makecell{$^3\!S_1 - ^3\!S_1$ \\ $(m_J = 0)$}    & 2.099  & 1.89 \\ 
  \makecell{$^3\!P_1 - ^3\!S_1$ \\ $(m_J = 1)$}    & 3.009  & 2.98   \\ \hline 
  
\end{tabular}
\end{center} 
 \end{table}

\begin{figure}[htbp]
  \centering
  \includegraphics[width=0.9\columnwidth]
  {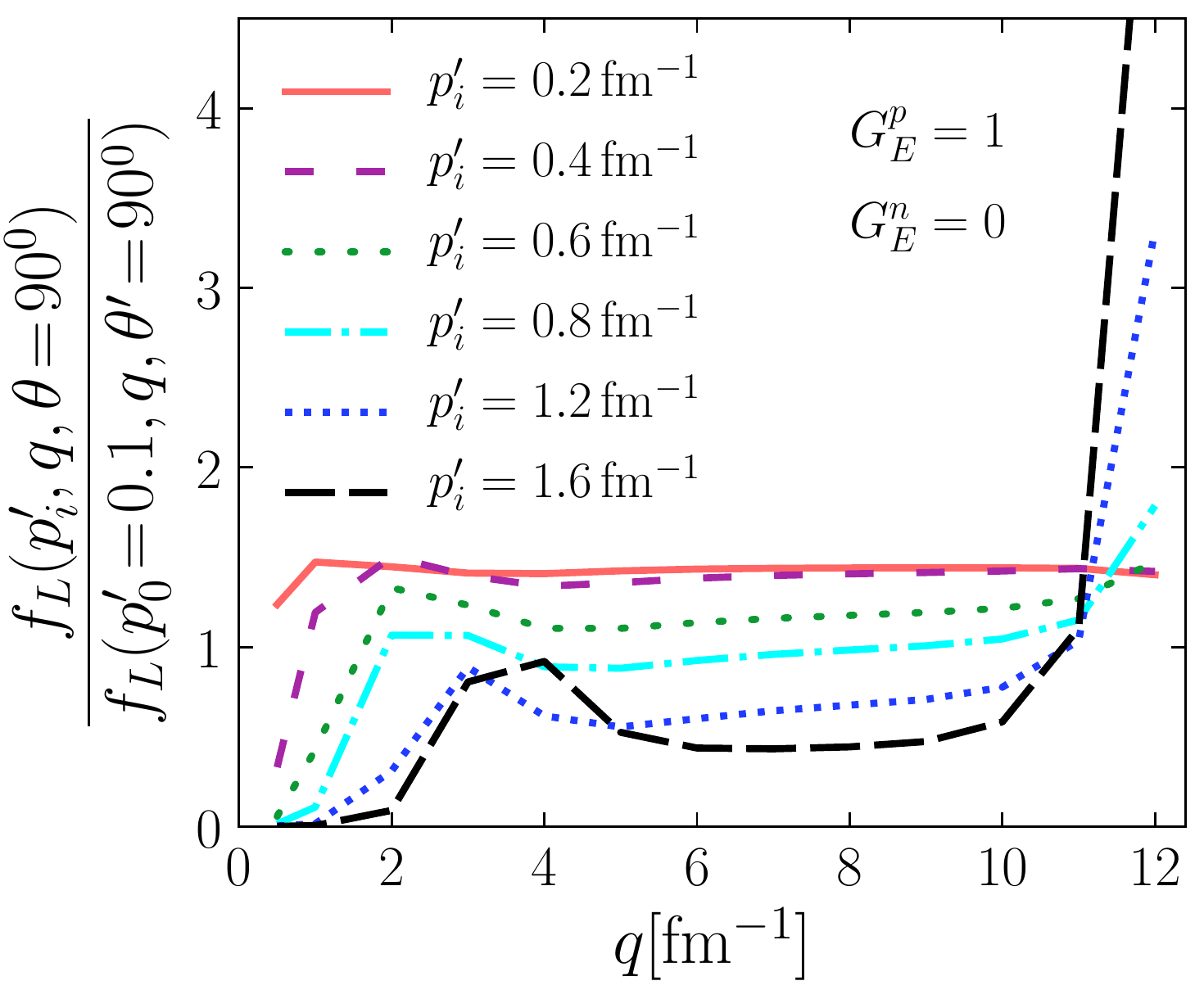}
  \caption{Demonstration that for $\pp \ll q$, the $q$ dependence 
  of $\fL$ factorizes.
  }
  \label{fig:fl_factorization}
\end{figure}

Armed with our knowledge of the factorized $q$-dependence in the low-resolution picture, the interpretation of certain observations becomes quite straightforward.  For instance, it is found that for small outgoing nucleon 
momentum ($\pp$) and large momentum transfer ($q$), the 
longitudinal structure function $\fL$ factorizes into a 
function of $\pp$ and a function of $q$.  
This is demonstrated in Fig.~\ref{fig:fl_factorization}.  
The plateau in the ratio of $\fL$ shown in 
Fig.~\ref{fig:fl_factorization} tells us that for 
$\pp \ll q$, 
\beq
  \fL(\pp, \thetap; q) \rightarrow g(\pp, \thetap) B(q)\;.
  \label{eq:fl_factorization}
\eeq
$\thetap$ here is the proton emission angle.  

\begin{figure}[htbp]
  \centering
  \begin{overpic}[width=0.8\columnwidth]{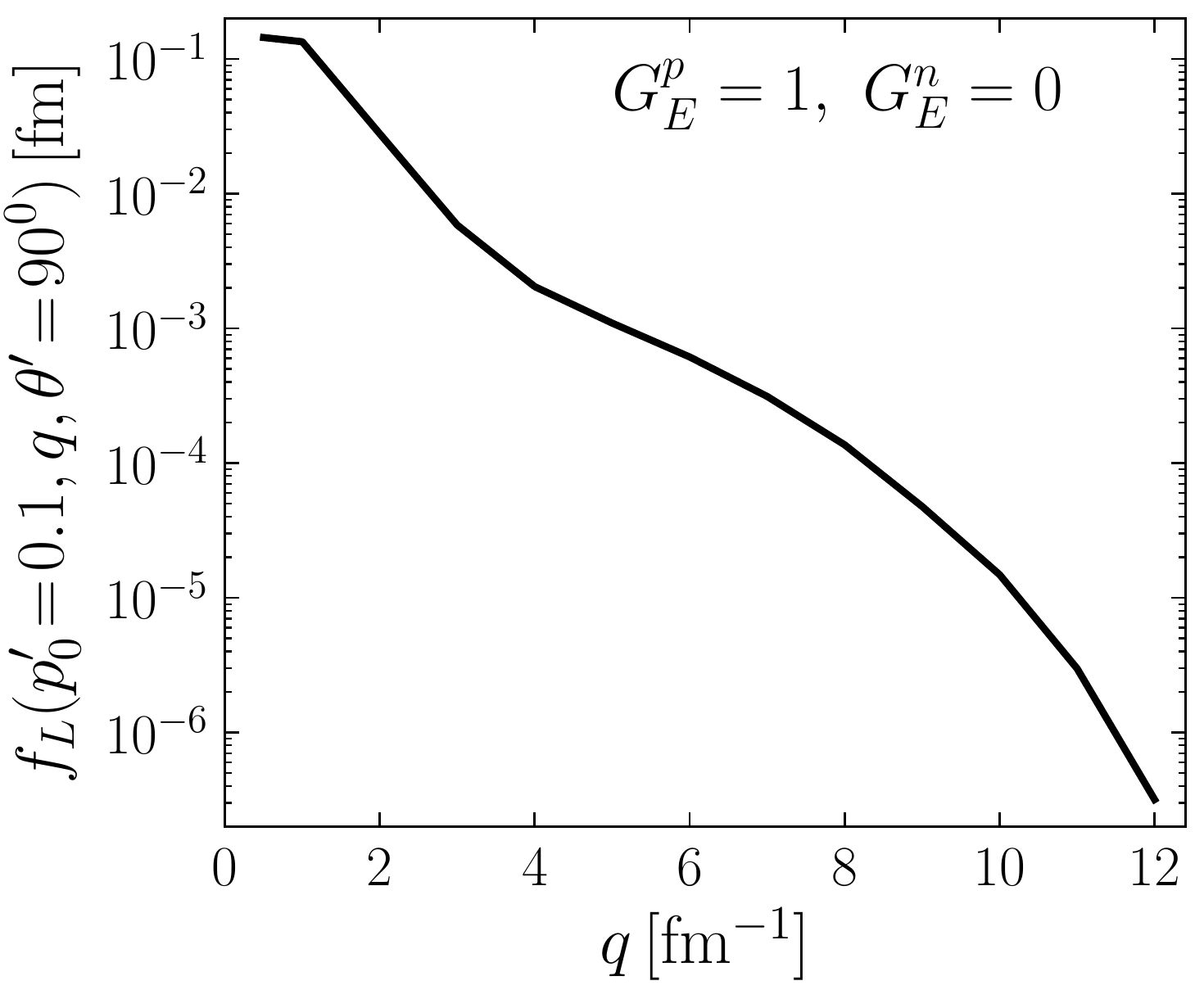}
\put(45.1,21.2){\tikz\draw[black, dashed] (0,0) rectangle (2.65,2.65);}
\end{overpic}
\caption{Demonstration that $\fL$ is a strong function of $q$.  
$Y$-axis is the denominator of the ratio plotted in 
Fig.~\ref{fig:fl_factorization}.  For the region in which the 
ratio of $\fL$ plateaus, the denominator by itself varies by over 
three orders of magnitude. 
}
  \label{fig:fl_strong_fn_q}
\end{figure}

Note that as seen in Fig.~\ref{fig:fl_strong_fn_q}, 
$\fL$ by itself is a strong function of $q$.  In the region 
where the ratio in Fig.~\ref{fig:fl_factorization} plateaus, 
the denominator of the ratio varies by over three orders of magnitude. 
Note that for simplicity, we set the electric form factors for the
proton and neutron to $1$ and $0$, respectively. 

The explanation of factorization in 
Eq.~\eqref{eq:fl_factorization} is straightforward in the 
low-resolution picture through the above SVD analysis.  
As we have seen, the LO term in the Eq.~\eqref{eq:current_SVD} 
already gives a reasonable estimate in most cases.  
Moreover, from Table~\ref{tab:ratio_Delta_J}, we find that most of the $q$-dependence
is carried by the leading singular value, so that the $p'$- and $q$-dependence is approximately factorized.\footnote{The singular values weakly depend on $\pp$ if we use the physical values for 
the form factors instead of setting them to $1$ and $0$.  This is because the 
form factors multiplying the current are a function of $Q^2$ 
(cf.~Eq.~\eqref{eq:J0_minus_analytical_delta}).} 
The factorization in the transition matrix element follows immediately and explains the plateaus in the ratios of $\fL$ observed in Fig.~\ref{fig:fl_factorization}.  

Note that the factorization in Eq.~\eqref{eq:fl_factorization} 
is the most prominent for $\thetap = 90^0$ and low $\pp$, 
because this is the 
region where the contribution from higher partial waves is minimal.  
For other angles and moderately large $\pp$ the matrix element 
$\mbraket{\psi_f^\lambda}{J_0^\lambda}{\psi_i^\lambda}$ will be a sum 
of factorized terms and therefore the $q$-dependence of 
$\fL$ won't necessarily factor out. 

We would like to note that the preliminary analysis shows that the 
SVD works for higher energies as well as long as $\pp < q/2$ (for instance 
the kinematics shown in Figs.~\ref{fig:evolved_IA_pic_49p0} and 
\ref{fig:evolved_IA_pic_36p0}).  
However, in order to compare the SVD answer to the exact answer, one 
needs to do the SVD in more partial wave channels 
(e.g., than that shown in Fig.~\ref{fig:evolved_simple_picture_SVD}) and 
in particular include the $D$-state of the deuteron.

%%%%%%%%%%%%%%%%%%%%%%%%%%%%%%%%%%%%%%%%%%%%%%%%%%%%
%%%%%%%%%%%%%%%%%%%%%%%%%%%%%%%%%%%%%%%%%%%%%%%%%%%%

\section{Summary and outlook} \label{sec:summary}

In this work, we extended our use of deuteron electrodistintegration
as a laboratory for exploring the consequences of scale dependence in
nuclear knock-out reactions.
We have embedded the analysis into a renormalization group framework, using
the SRG as a convenient tool to change the scale.
This enables us to separately study the impact of scale changes on all the
ingredients of the calculation.  
We are particularly interested
in kinematic regions of experimental interest 
where these changes are significant.

For such regions we found that working at low-resolution can have distinct advantages.  
We found that at high $Q^2$ ($Q^2 \approx 1.8~\rm{GeV^2}$) and large $x_d$ ($x_d > 1.5$), 
the local decoupling of the final state in the evolved 
picture leads to decreased contributions from final state interactions 
and thereby an increased validity of the impulse approximation.  
We also saw that the explanation of factorization in the observable 
$\fL$ becomes straightforward in the low-momentum picture.  

It is conventional wisdom that the low-resolution potentials are ill-suited 
for high-momentum transfer reactions, such as those used to probe short range
correlations in nuclei.  By focusing on a particular kinematics 
region with large momentum transfer $q^2$ and relatively small energy $E'$, 
($Q^2 = 1.3 \rm{~GeV^2}$, $x_d = 1.88$)
we demonstrated that this is not the case.  We showed that the 
relevant RG changes to the operator are tractable, which allows us to 
recover the unevolved answer in the low-resolution picture.

%\subsection{Scale dependent $S$-state contribution} 
%\label{subsec:scale_dependent_D_state_contri}

Analysis of a reaction calculation often involves understanding 
which components of the nuclear wave functions are probed.  
For example, it might be claimed that a reaction 
is sensitive to the $D$-state probability 
in the deuteron.   
In our model calculations,
such a claim would be based on the unevolved wave functions in 
Fig.~\ref{fig:deut_wfn_r_space} and \ref{fig:deut_wfn_k_space}, where we find that 
for intermediate
momenta that the $D$-state deuteron wave function has a 
higher magnitude than the $S$-state wave function.  
However, as we saw in Fig.~\ref{fig:deut_wfn_k_space}, the 
high-momentum part of the deuteron wave functions depends on the 
SRG scale.  Therefore, the claim that a certain kinematics is 
sensitive to a specific channel in the deuteron wave function is 
highly scale dependent.    

\begin{figure}[tbh!]
  \centering
  \includegraphics[width=0.9\columnwidth]
  {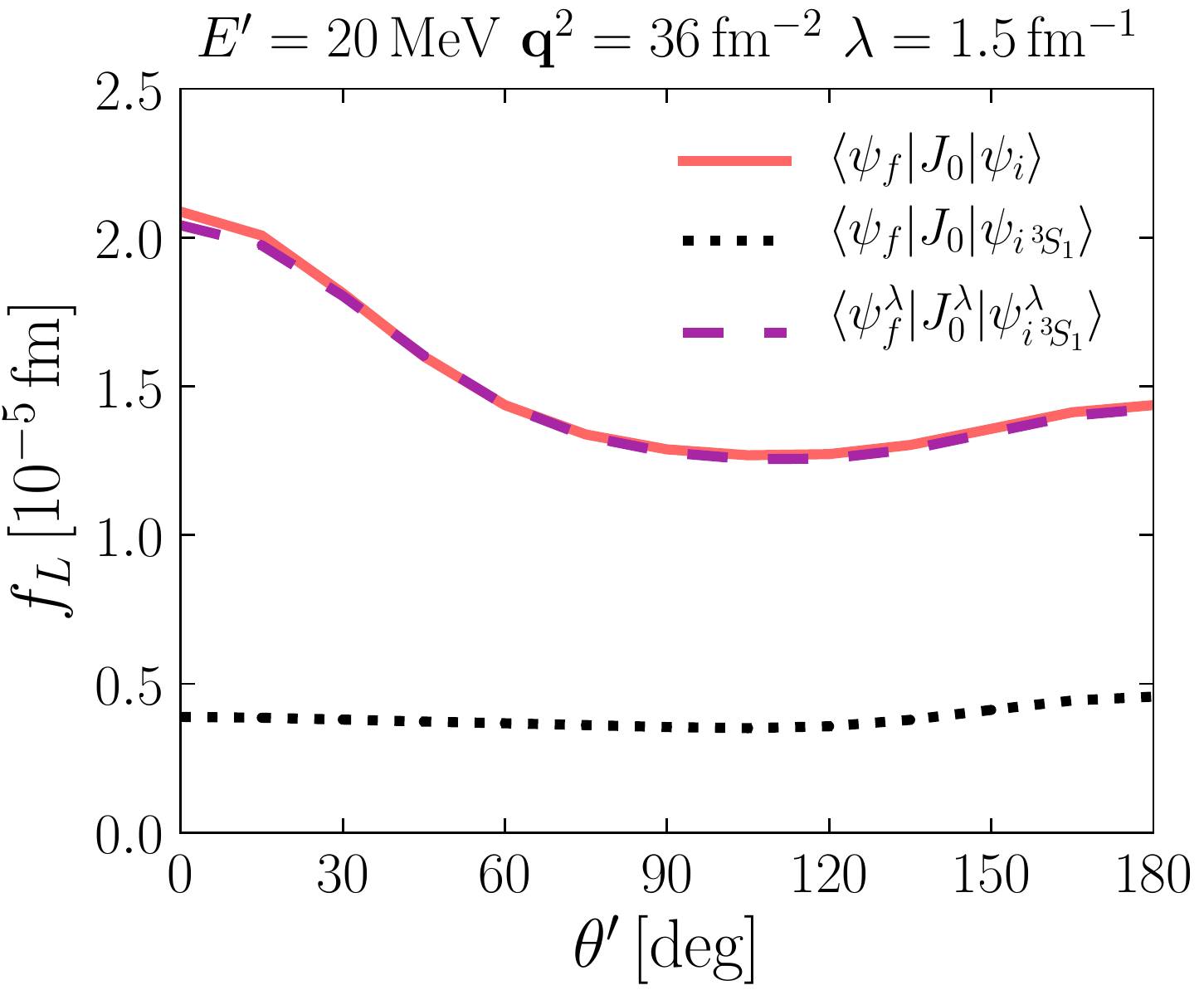}
  \caption{Contribution to $\fL$ from the deuteron $S$-state in the 
  evolved and unevolved case. Here $x_d = 1.88$, $Q^2 = 1.3~\rm{GeV^2}$.}
  \label{fig:S_state_contri_fL_theta}
\end{figure}

We see a representative demonstration of this in Fig.~\ref{fig:S_state_contri_fL_theta}
using the kinematics explored in Section~\ref{subsec:Derivative_expansion}.  
The solid line in Fig.~\ref{fig:S_state_contri_fL_theta} is  
$\fL$ in the unevolved case.  The dotted line is $\fL$ calculated 
by keeping only the $S$-state in the deuteron.  Thus keeping only 
the $S$-state is clearly a very poor approximation in the unevolved case.  
However, as indicated by the dashed line, it is a very good approximation 
for the evolved case.  

\begin{figure}[tbh!]
  \centering
  \includegraphics[width=0.9\columnwidth]
  {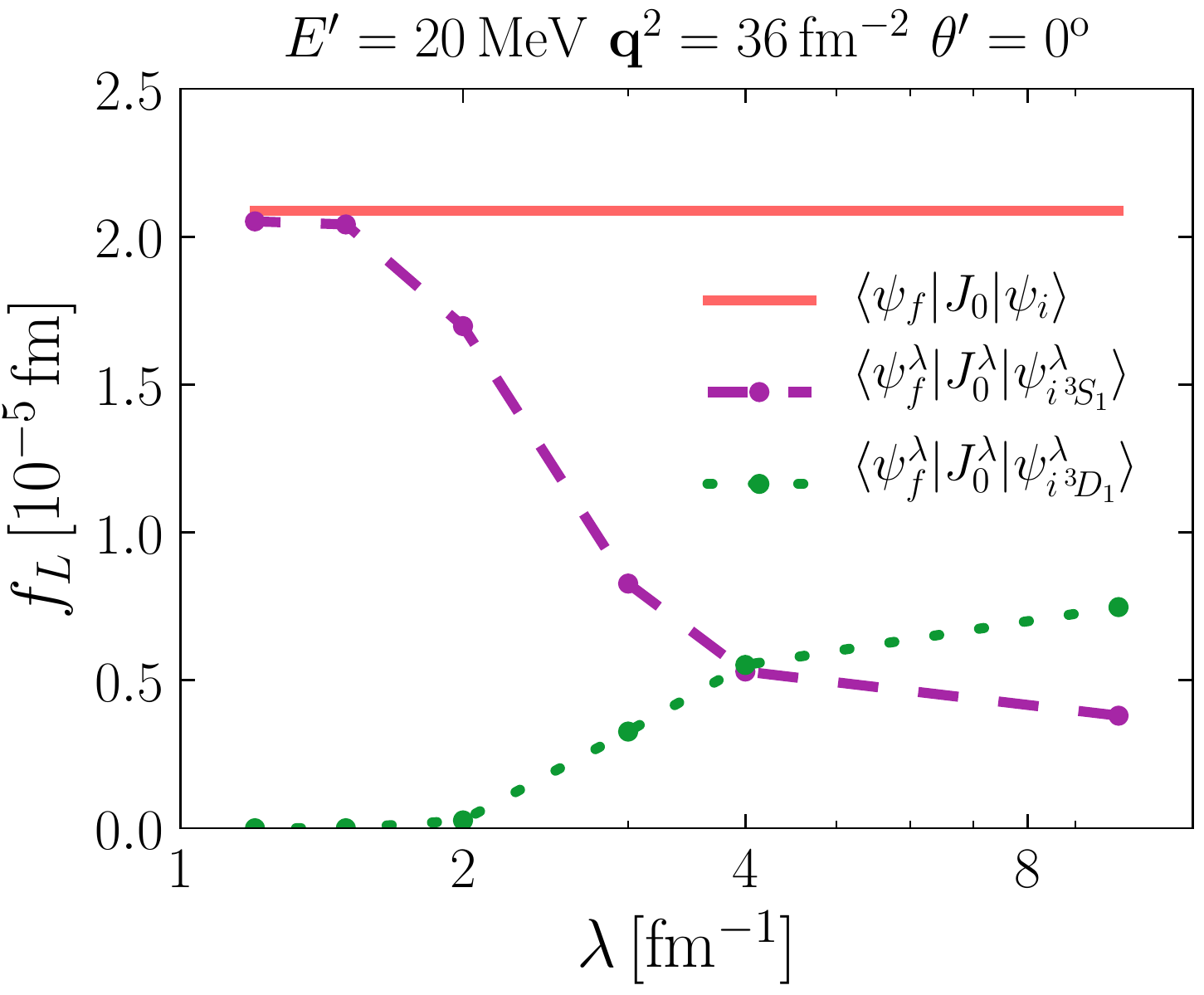}
  \caption{Contribution to $\fL$ from the deuteron $S$- and $D$-states 
  as a function of SRG $\lambda$ for the same $x_d$ and $Q^2$ as
  in Fig.~\ref{fig:S_state_contri_fL_theta} but with fixed $\theta' = 0^\circ$. }
  \label{fig:S_state_contri_fn_lambda}
\end{figure}

Figure~\ref{fig:S_state_contri_fn_lambda} shows the contribution 
to $\fL$ for $\theta'=0^\circ$ from the $S$- and $D$-channels of the deuteron as a 
function of $\lambda$.  We see that the $S$-channel 
contribution increases with SRG evolution, while the 
$D$-channel contribution is driven to zero.  We stress that the 
results in Figs.~\ref{fig:S_state_contri_fL_theta} and 
\ref{fig:S_state_contri_fn_lambda} change quantitatively 
with different kinematics.  Nonetheless, they demonstrate how an  
`intuitive' picture of probing specific parts of the 
nuclear wave function is highly scale dependent.

It has been established previously that the $D$-state probability of the deuteron, 
which is analogous to a spectroscopic
factor, is not measurable~\cite{Amado:1979zz,Friar:1979zz}.
We have seen how this is manifested in deuteron electrodisintegration
as a scale dependence under SRG evolution.
Our example demonstrates that if one tries to calculate 
the cross section at a definite 
high-resolution scale with a calculation
that is not fully consistent, one would come to the false conclusion that
this $D$-state physics is a measurable ingredient of the experiment rather than
a scale- and scheme-dependent feature.

Thus our results supply an object lesson for those seeking to extract
absolute nuclear structure information from knock-out reactions.
In a simple picture at high RG resolution based on the IA, the cross section for the specified 
kinematics and a one-body current comes dominantly from high momentum components 
of the deuteron wave function 
and the $D$-state part plays an essential role.  Thus, one imagines the reaction 
is a probe of short-range correlations and the impact of the tensor force in nuclei.
One also finds that final-state interactions are a critical ingredient, which
generally obscures what is learned.
But an analysis of the same kinematics at low RG resolution yields a very
different picture of the cross section, which instead comes dominantly
from simple wave functions and a two-body current well represented as contact
operators.
This picture implies a simple calculation of the identical cross section that does not
rely on the $D$-state part of the deuteron wave function at all.

\begin{figure}[tbh!]
  \begin{tabular}{cccc}
   \hline
     kinematics  &     before  &  after   \\ \hline
     quasielastic &
  \raisebox{-15pt}{\includegraphics[width=0.3\columnwidth]{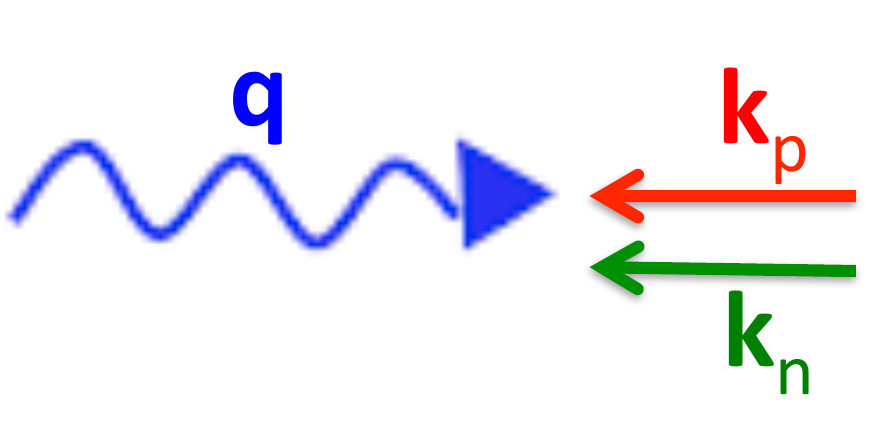}} &
  \raisebox{-15pt}{\includegraphics[width=0.23\columnwidth]{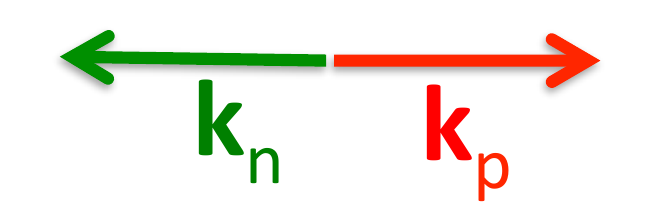}} \\
    \parbox{.2\columnwidth}{high $\vecq^2$, low $E'$ at $\lambda = \infty$} &
  \raisebox{-20pt}{\includegraphics[width=0.485\columnwidth]{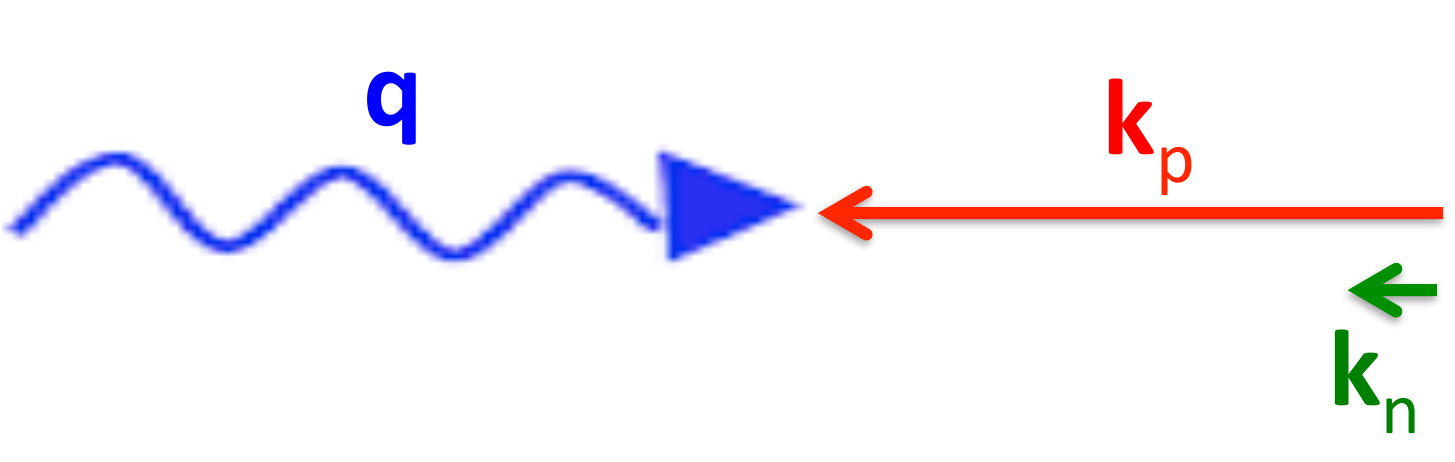}} &
  \raisebox{-10pt}{\includegraphics[width=0.13\columnwidth]{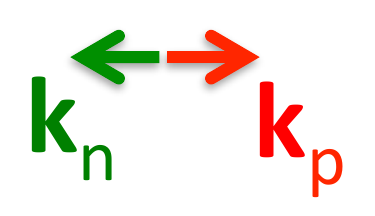}} \\
    \parbox{.2\columnwidth}{high $\vecq^2$, low $E'$ at small $\lambda$} &
  \hspace*{-22pt}%
  \raisebox{-15pt}{\includegraphics[width=0.385\columnwidth]{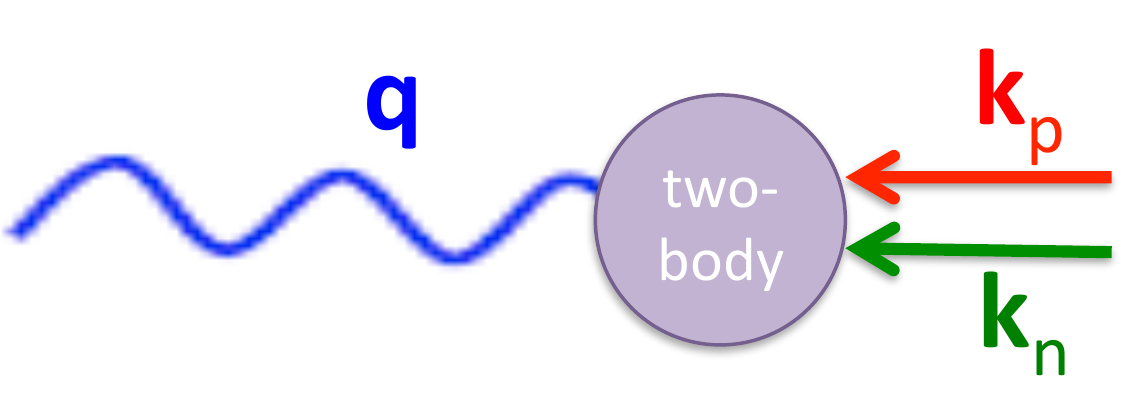}} &
  \raisebox{-10pt}{\includegraphics[width=0.13\columnwidth]{SRC_after.pdf}} \\
  \hline
  \end{tabular}
  \caption{Schematic pictures of the most important incoming and outgoing proton and neutron momenta  
  for $\theta'=0^\circ$ scattering in the impulse approximation with quasifree kinematics at any $\lambda$, and analogous pictures for
  scattering with large momentum
  transfer near threshold at high and low resolution. }
  \label{fig:cartoons}
\end{figure}

The results we have found from explicit calculations 
can be understood using simple, intuitive arguments. 
Schematic pictures of the dominant mechanisms for two classes of kinematics are
shown in Fig.~\ref{fig:cartoons}.
Recall that we are in the c.m. system of the final proton and neutron, so their momenta are always
back-to-back and equal in magnitude.  For convenience we focus on scattering angle $\theta = 0$
and consider only the photon coupling to the proton,
but the arguments here are easily generalized.  In the initial state, the photon three-momentum
and the net momenta in the deuteron must sum to zero, but we need to identify the dominant
region of proton-neutron phase space for the cross section.

The first row illustrates the generic situation for QFR kinematics and a one-body current.
The definition of the QFR dictates the final magnitudes of momenta, equal to half the initial
photon three-momentum.  The one-body current $J_0$ by assumption couples only to the proton,
so the neutron momentum is the same in initial and final state while the proton gets exactly
turned around.  That implies that the initial nucleons have a small relative momentum, which is
dominated by the relative $S$-wave and is unchanged under RG evolution, implying that there will
be little dependence on SRG $\lambda$ on the QFR.

The other example, shown for high resolution on the second line and for low resolution on the third
line, is the case we have considered in detail with 
large momentum transfer $q^2$ and relatively small final energy $E'$.  The latter dictates
the same low-relative-momentum final configuration for each resolution, 
but the dominant mechanism is forced to be very
different.
At high resolution, the one-body current mechanism requires a high-momentum proton (in that frame),
which implies a high relative momentum; in other words, an SRC configuration.  There are no
such SRCs at low resolution, but there is an induced short-range two-body current, which 
mostly just stops low relative-momentum nucleons in the deuteron.  In the former case, the $D$-state 
plays a dominant role (at least for intermediate momenta) while in the latter it is mostly
$S$-state.  
This example shows again how the kinematics alone does not always uniquely determine what
is probed in the reaction.

The present investigations are only the start of what is
needed for a thorough treatment of scale and scheme dependence for
nuclear processes.
Extensions include adding two-body currents, applications to few-body systems, 
consistent construction of operators for processes of interest from the 
RG perspective, 
and connecting to other knock-out processes.
While these are technically much more complex, we expect that many of the
basic physics observations will carry over.
An interesting follow-up will be to examine the scheme dependence that
comes from the choice of SRG generator.  An alternative to $G_s = T$ 
is to choose a form that block diagonalizes the Hamiltonian with 
respect to a specified momentum scale $\Lambda_{\textrm{bd}}$~\cite{Anderson:2008mu,Dicaire:2014fra}.
This scheme decouples physics above and below $\Lambda_{\textrm{bd}}$, 
but do not have the \emph{local} decoupling properties that were 
important for the FSI simplifications observed here.    
Work on all of these extensions is in progress.

%%%%%%%%%%%%%%%%%%%%%%%%%%%%%%%%%%%%%%%%%%%%%%%%%%%%%%%
%%%%%%%%%%%%%%%%%%%%%%%%%%%%%%%%%%%%%%%%%%%%%%%%%%%%%%%

\section*{Acknowledgements}

We would like to thank A.~Dyhdalo, K.~Hebeler, S.~König, and D.~Phillips for useful discussions.
This work was supported in part by the National Science Foundation under 
Grant Nos.~PHY-1306250, PHY-1404159, and PHY-1614460, 
the NUCLEI SciDAC Collaboration under DOE Grants DE-SC0008533 and DE-SC0008511, 
and the Double-Beta Decay and Fundamental Symmetries Topical Collaboration under DOE Grant DE-SC0015376.

\bigskip

\appendix

\section{Kinematic variables transformation}
\label{app:kinematic_variables_transformation}

The analysis of electron scattering on nuclei is often presented using the Lorentz-invariant kinematic variables 
Bjorken $x$ and the virtual photon four-momentum squared, $Q^2 \equiv -q^2$ \cite{Hen:2016kwk}.  
We note that $x$ and $Q^2$ provide a complete kinematic specification 
for inclusive scattering from an unpolarized target but not 
(exclusive) deuteron electrodisintegration,  
although there is a one-to-one mapping between the center-of-mass variables%
\footnote{In other sections for notational brevity we set 
$\mbf{q}^2_{\rm c.m.} \equiv q^2$.  Here we keep the subscript c.m. to avoid any 
ambiguities.}
$(\Ep, \mbf{q}_{\rm c.m.}^2)$ and $(x, Q^2)$, which we provide below.  
The relative angle $\thetap$ of the outgoing proton relative to the photon 
provides us additional information in the deuteron electrodisintegration case.  
For example, in the impulse approximation $\thetap$ directly specifies 
which part of the deuteron wave function is probed.  The $\thetap$ 
dependence is integrated over in the inclusive case. 

The generalized Bjorken $x_A$ for an $A$-body nuclear target is given by 
\beq
  x_A = \frac{Q^2}{2 q \cdot P_A} A \;,
\eeq
where $q^\mu$ is the virtual photon four-momentum and $P_A^\mu$ is the 
four-momentum of the target nucleus.  
The range of $x_A$ is
$0 \leq x_A \leq A$, with $x_A = 1$ characterizing elastic scattering. 
For a deuteron target in the laboratory frame, 
\beq
  x_d = \frac{Q^2}{\omega_{\rm lab} M_d} \;,
  \label{eq:xd_def}
\eeq
where $M_d$ is the mass of the deuteron.

\begin{figure}[tbp!]
\centering
\includegraphics[width=0.85\columnwidth]
{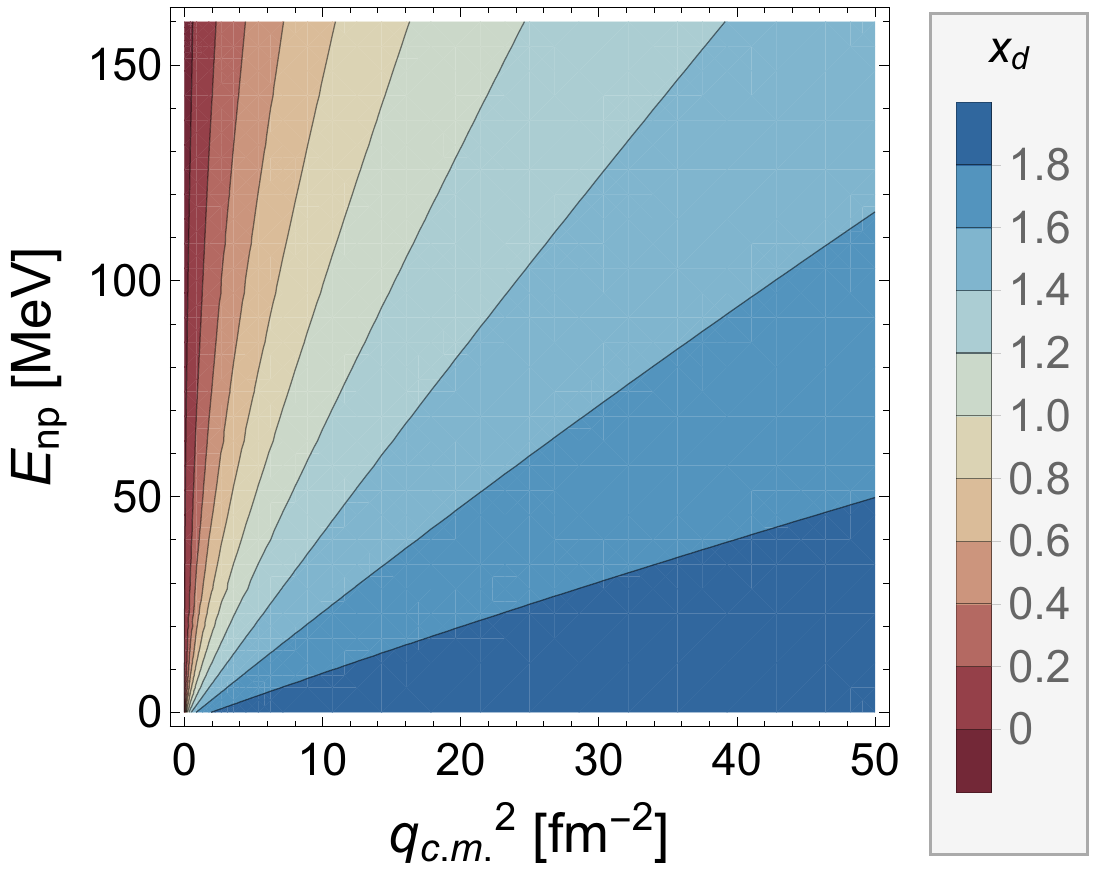}
\caption{$Q^2$ values as a function of $\Ep$ and $\mbf{q}^2_{\rm c.m.}$.}
\label{fig:Qsq_plot}
\end{figure}

\begin{figure}[tbp!]
\centering
\includegraphics[width=0.85\columnwidth]
{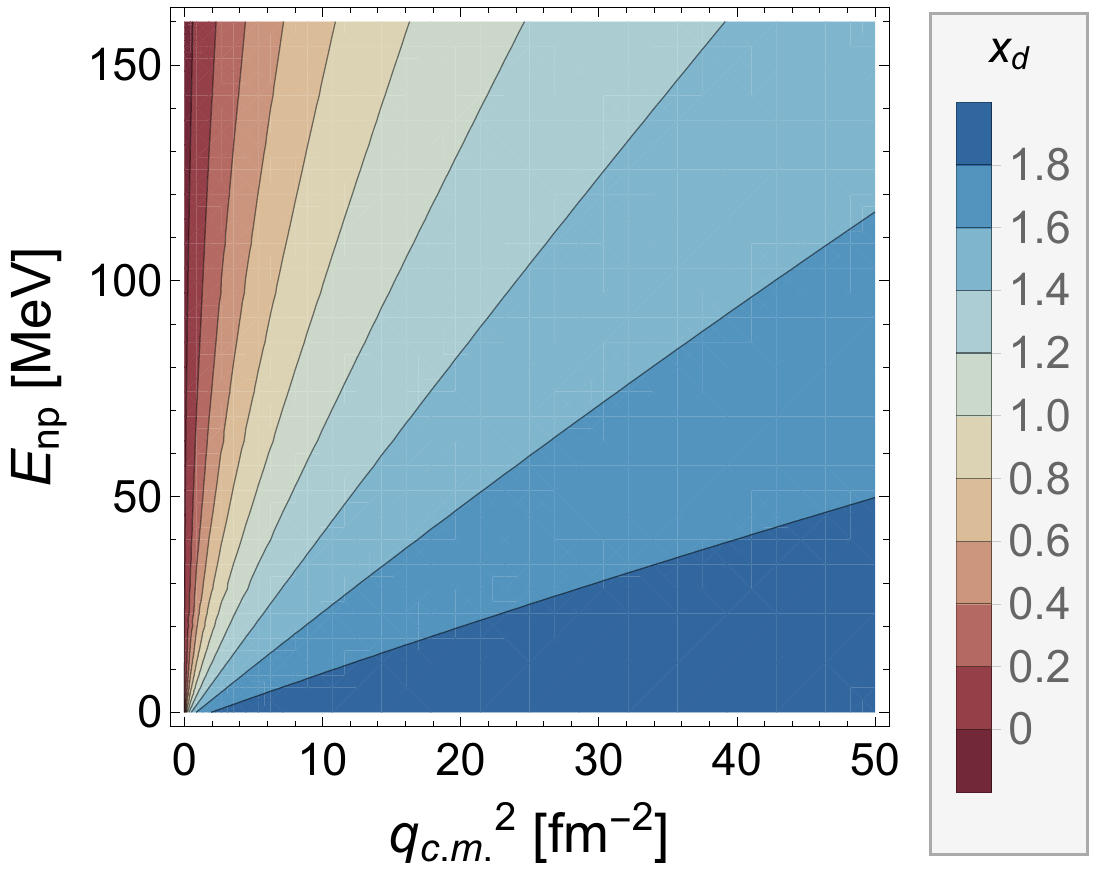}
\caption{$x_d$ values as a function of $\Ep$ and $\mbf{q}^2_{\rm c.m.}$.}
\label{fig:xd_plot}
\end{figure}

Given $E^\prime$ and $\mbf{q}^2_{\rm c.m.}$, we can evaluate $Q^2$ in the
center-of-mass frame,
\beq
  Q^2 = \mbf{q}^2_{\rm c.m.} - \omega_{\rm c.m.}^2 \;, 
  \label{eq:Qsq_def}
\eeq
and use
\beq
  \omega_{\rm c.m.} = E^\prime + 2\, M_{np} - \sqrt{M_d^2 + \mbf{q}^2_{\rm c.m.}}\;,
\eeq
where $M_{np}$ is the average neutron-proton mass. 
To evaluate $x_d$ from Eq.~\eqref{eq:xd_def},
we express $\omega_{\rm lab}$ in terms of 
center-of-mass quantities~\cite{Fabian:1979kx},
\beq
  \omega_{\rm lab} = \frac{E_d^{\rm c.m.}}{M_d} \left(\omega_{\rm c.m.} + 
  \frac{\mbf{q}^2_{\rm c.m.}}{E_d^{\rm c.m.}}\right) \;,
  \label{eq:omega_lab_def}
\eeq 
where $E_d^{\rm c.m.}$ is the deuteron energy in the center-of-mass frame of 
the outgoing particles given by 
\beq
  E_d^{\rm c.m.} = \sqrt{\mbf{q}^2_{\rm c.m.} + M_d^2}\;.
\eeq
Using Eqs.~\eqref{eq:xd_def}, \eqref{eq:Qsq_def}, and \eqref{eq:omega_lab_def}, we 
can map the c.m. variables to the Lorentz-invariant variables.  The corresponding values 
for $Q^2$ and $x_d$ as functions of $\Ep$ and $\mbf{q}^2_{\rm c.m.}$ are 
shown in Figs.~\ref{fig:Qsq_plot} and \ref{fig:xd_plot}. 
The kinematically allowed regions are $Q^2 \geq 0$ and $0 \leq x_d \leq 2$.


\begin{thebibliography}{45}%
\makeatletter
\providecommand \@ifxundefined [1]{%
 \@ifx{#1\undefined}
}%
\providecommand \@ifnum [1]{%
 \ifnum #1\expandafter \@firstoftwo
 \else \expandafter \@secondoftwo
 \fi
}%
\providecommand \@ifx [1]{%
 \ifx #1\expandafter \@firstoftwo
 \else \expandafter \@secondoftwo
 \fi
}%
\providecommand \natexlab [1]{#1}%
\providecommand \enquote  [1]{``#1''}%
\providecommand \bibnamefont  [1]{#1}%
\providecommand \bibfnamefont [1]{#1}%
\providecommand \citenamefont [1]{#1}%
\providecommand \href@noop [0]{\@secondoftwo}%
\providecommand \href [0]{\begingroup \@sanitize@url \@href}%
\providecommand \@href[1]{\@@startlink{#1}\@@href}%
\providecommand \@@href[1]{\endgroup#1\@@endlink}%
\providecommand \@sanitize@url [0]{\catcode `\\12\catcode `\$12\catcode
  `\&12\catcode `\#12\catcode `\^12\catcode `\_12\catcode `\%12\relax}%
\providecommand \@@startlink[1]{}%
\providecommand \@@endlink[0]{}%
\providecommand \url  [0]{\begingroup\@sanitize@url \@url }%
\providecommand \@url [1]{\endgroup\@href {#1}{\urlprefix }}%
\providecommand \urlprefix  [0]{URL }%
\providecommand \Eprint [0]{\href }%
\providecommand \doibase [0]{http://dx.doi.org/}%
\providecommand \selectlanguage [0]{\@gobble}%
\providecommand \bibinfo  [0]{\@secondoftwo}%
\providecommand \bibfield  [0]{\@secondoftwo}%
\providecommand \translation [1]{[#1]}%
\providecommand \BibitemOpen [0]{}%
\providecommand \bibitemStop [0]{}%
\providecommand \bibitemNoStop [0]{.\EOS\space}%
\providecommand \EOS [0]{\spacefactor3000\relax}%
\providecommand \BibitemShut  [1]{\csname bibitem#1\endcsname}%
\let\auto@bib@innerbib\@empty
%</preamble>
\bibitem [{\citenamefont {Furnstahl}\ and\ \citenamefont
  {Schwenk}(2010)}]{Furnstahl:2010wd}%
  \BibitemOpen
  \bibfield  {author} {\bibinfo {author} {\bibfnamefont {R.~J.}\ \bibnamefont
  {Furnstahl}}\ and\ \bibinfo {author} {\bibfnamefont {A.}~\bibnamefont
  {Schwenk}},\ }\href {\doibase 10.1088/0954-3899/37/6/064005} {\bibfield
  {journal} {\bibinfo  {journal} {J. Phys. G}\ }\textbf {\bibinfo {volume}
  {37}},\ \bibinfo {pages} {064005} (\bibinfo {year} {2010})},\ \Eprint
  {http://arxiv.org/abs/1001.0328} {arXiv:1001.0328 [nucl-th]} \BibitemShut
  {NoStop}%
%%CITATION = 1001.0328;%%
\bibitem [{\citenamefont {Furnstahl}(2013)}]{Furnstahl:2013dsa}%
  \BibitemOpen
  \bibfield  {author} {\bibinfo {author} {\bibfnamefont {R.~J.}\ \bibnamefont
  {Furnstahl}},\ }in\ \href
  {http://inspirehep.net/record/1255127/files/arXiv:1309.5771.pdf} {\emph
  {\bibinfo {booktitle} {{Proceedings, International Conference on Nuclear
  Theory in the Supercomputing Era (NTSE-2013)}}}}\ (\bibinfo {year} {2013})\
  p.\ \bibinfo {pages} {371},\ \Eprint {http://arxiv.org/abs/1309.5771}
  {arXiv:1309.5771 [nucl-th]} \BibitemShut {NoStop}%
%%CITATION = ARXIV:1309.5771;%%
\bibitem [{\citenamefont {Sterman}\ \emph {et~al.}(1995)\citenamefont
  {Sterman}, \citenamefont {Smith}, \citenamefont {Collins}, \citenamefont
  {Whitmore}, \citenamefont {Brock}, \citenamefont {Huston}, \citenamefont
  {Pumplin}, \citenamefont {Tung}, \citenamefont {Weerts}, \citenamefont
  {Yuan}, \citenamefont {Kuhlmann}, \citenamefont {Mishra}, \citenamefont
  {Morf\'{\i}n}, \citenamefont {Olness}, \citenamefont {Owens}, \citenamefont
  {Qiu},\ and\ \citenamefont {Soper}}]{RevModPhys.67.157}%
  \BibitemOpen
  \bibfield  {author} {\bibinfo {author} {\bibfnamefont {G.}~\bibnamefont
  {Sterman}}, \bibinfo {author} {\bibfnamefont {J.}~\bibnamefont {Smith}},
  \bibinfo {author} {\bibfnamefont {J.~C.}\ \bibnamefont {Collins}}, \bibinfo
  {author} {\bibfnamefont {J.}~\bibnamefont {Whitmore}}, \bibinfo {author}
  {\bibfnamefont {R.}~\bibnamefont {Brock}}, \bibinfo {author} {\bibfnamefont
  {J.}~\bibnamefont {Huston}}, \bibinfo {author} {\bibfnamefont
  {J.}~\bibnamefont {Pumplin}}, \bibinfo {author} {\bibfnamefont {W.-K.}\
  \bibnamefont {Tung}}, \bibinfo {author} {\bibfnamefont {H.}~\bibnamefont
  {Weerts}}, \bibinfo {author} {\bibfnamefont {C.-P.}\ \bibnamefont {Yuan}},
  \bibinfo {author} {\bibfnamefont {S.}~\bibnamefont {Kuhlmann}}, \bibinfo
  {author} {\bibfnamefont {S.}~\bibnamefont {Mishra}}, \bibinfo {author}
  {\bibfnamefont {J.~G.}\ \bibnamefont {Morf\'{\i}n}}, \bibinfo {author}
  {\bibfnamefont {F.}~\bibnamefont {Olness}}, \bibinfo {author} {\bibfnamefont
  {J.}~\bibnamefont {Owens}}, \bibinfo {author} {\bibfnamefont
  {J.}~\bibnamefont {Qiu}}, \ and\ \bibinfo {author} {\bibfnamefont {D.~E.}\
  \bibnamefont {Soper}},\ }\href {\doibase 10.1103/RevModPhys.67.157}
  {\bibfield  {journal} {\bibinfo  {journal} {Rev. Mod. Phys.}\ }\textbf
  {\bibinfo {volume} {67}},\ \bibinfo {pages} {157} (\bibinfo {year}
  {1995})}\BibitemShut {NoStop}%
\bibitem [{\citenamefont {Collins}(2013)}]{Collins:2011zzd}%
  \BibitemOpen
  \bibfield  {author} {\bibinfo {author} {\bibfnamefont {J.}~\bibnamefont
  {Collins}},\ }\href {http://www.cambridge.org/de/knowledge/isbn/item5756723}
  {\emph {\bibinfo {title} {{Foundations of perturbative QCD}}}}\ (\bibinfo
  {publisher} {Cambridge University Press},\ \bibinfo {address} {Cambridge},\
  \bibinfo {year} {2013})\BibitemShut {NoStop}%
%%CITATION = INSPIRE-922696;%%
\bibitem [{\citenamefont {Bedaque}\ and\ \citenamefont {van
  Kolck}(2002)}]{Bedaque:2002mn}%
  \BibitemOpen
  \bibfield  {author} {\bibinfo {author} {\bibfnamefont {P.~F.}\ \bibnamefont
  {Bedaque}}\ and\ \bibinfo {author} {\bibfnamefont {U.}~\bibnamefont {van
  Kolck}},\ }\href@noop {} {\bibfield  {journal} {\bibinfo  {journal} {Ann.
  Rev. Nucl. Part. Sci.}\ }\textbf {\bibinfo {volume} {52}},\ \bibinfo {pages}
  {339} (\bibinfo {year} {2002})},\ \Eprint
  {http://arxiv.org/abs/nucl-th/0203055} {nucl-th/0203055} \BibitemShut
  {NoStop}%
%%CITATION = NUCL-TH/0203055;%%
\bibitem [{\citenamefont {Epelbaum}\ \emph {et~al.}(2009)\citenamefont
  {Epelbaum}, \citenamefont {Hammer},\ and\ \citenamefont
  {Mei{\ss}ner}}]{Epelbaum:2008ga}%
  \BibitemOpen
  \bibfield  {author} {\bibinfo {author} {\bibfnamefont {E.}~\bibnamefont
  {Epelbaum}}, \bibinfo {author} {\bibfnamefont {H.-W.}\ \bibnamefont
  {Hammer}}, \ and\ \bibinfo {author} {\bibfnamefont {U.-G.}\ \bibnamefont
  {Mei{\ss}ner}},\ }\href {\doibase 10.1103/RevModPhys.81.1773} {\bibfield
  {journal} {\bibinfo  {journal} {Rev. Mod. Phys.}\ }\textbf {\bibinfo {volume}
  {81}},\ \bibinfo {pages} {1773} (\bibinfo {year} {2009})},\ \Eprint
  {http://arxiv.org/abs/0811.1338} {arXiv:0811.1338 [nucl-th]} \BibitemShut
  {NoStop}%
%%CITATION = ARXIV:0811.1338;%%
\bibitem [{\citenamefont {Machleidt}\ and\ \citenamefont
  {Entem}(2011)}]{Machleidt:2011zz}%
  \BibitemOpen
  \bibfield  {author} {\bibinfo {author} {\bibfnamefont {R.}~\bibnamefont
  {Machleidt}}\ and\ \bibinfo {author} {\bibfnamefont {D.~R.}\ \bibnamefont
  {Entem}},\ }\href {\doibase 10.1016/j.physrep.2011.02.001} {\bibfield
  {journal} {\bibinfo  {journal} {Phys. Rept.}\ }\textbf {\bibinfo {volume}
  {503}},\ \bibinfo {pages} {1} (\bibinfo {year} {2011})},\ \Eprint
  {http://arxiv.org/abs/1105.2919} {arXiv:1105.2919 [nucl-th]} \BibitemShut
  {NoStop}%
%%CITATION = ARXIV:1105.2919;%%
\bibitem [{\citenamefont {Epelbaum}\ and\ \citenamefont
  {Meissner}(2012)}]{Epelbaum:2012vx}%
  \BibitemOpen
  \bibfield  {author} {\bibinfo {author} {\bibfnamefont {E.}~\bibnamefont
  {Epelbaum}}\ and\ \bibinfo {author} {\bibfnamefont {U.-G.}\ \bibnamefont
  {Meissner}},\ }\href {\doibase 10.1146/annurev-nucl-102010-130056} {\bibfield
   {journal} {\bibinfo  {journal} {Ann. Rev. Nucl. Part. Sci.}\ }\textbf
  {\bibinfo {volume} {62}},\ \bibinfo {pages} {159} (\bibinfo {year} {2012})},\
  \Eprint {http://arxiv.org/abs/1201.2136} {arXiv:1201.2136 [nucl-th]}
  \BibitemShut {NoStop}%
%%CITATION = ARXIV:1201.2136;%%
\bibitem [{\citenamefont {Bogner}\ \emph {et~al.}(2010)\citenamefont {Bogner},
  \citenamefont {Furnstahl},\ and\ \citenamefont {Schwenk}}]{Bogner:2009bt}%
  \BibitemOpen
  \bibfield  {author} {\bibinfo {author} {\bibfnamefont {S.~K.}\ \bibnamefont
  {Bogner}}, \bibinfo {author} {\bibfnamefont {R.~J.}\ \bibnamefont
  {Furnstahl}}, \ and\ \bibinfo {author} {\bibfnamefont {A.}~\bibnamefont
  {Schwenk}},\ }\href {\doibase 10.1016/j.ppnp.2010.03.001} {\bibfield
  {journal} {\bibinfo  {journal} {Prog. Part. Nucl. Phys.}\ }\textbf {\bibinfo
  {volume} {65}},\ \bibinfo {pages} {94} (\bibinfo {year} {2010})},\ \Eprint
  {http://arxiv.org/abs/0912.3688} {arXiv:0912.3688 [nucl-th]} \BibitemShut
  {NoStop}%
%%CITATION = ARXIV:0912.3688;%%
\bibitem [{\citenamefont {Alvioli}\ \emph {et~al.}(2013)\citenamefont
  {Alvioli}, \citenamefont {Ciofi Degli~Atti}, \citenamefont {Kaptari},
  \citenamefont {Mezzetti},\ and\ \citenamefont {Morita}}]{Alvioli:2013qyz}%
  \BibitemOpen
  \bibfield  {author} {\bibinfo {author} {\bibfnamefont {M.}~\bibnamefont
  {Alvioli}}, \bibinfo {author} {\bibfnamefont {C.}~\bibnamefont {Ciofi
  Degli~Atti}}, \bibinfo {author} {\bibfnamefont {L.~P.}\ \bibnamefont
  {Kaptari}}, \bibinfo {author} {\bibfnamefont {C.~B.}\ \bibnamefont
  {Mezzetti}}, \ and\ \bibinfo {author} {\bibfnamefont {H.}~\bibnamefont
  {Morita}},\ }\href {\doibase 10.1142/S021830131330021X} {\bibfield  {journal}
  {\bibinfo  {journal} {Int. J. Mod. Phys.}\ }\textbf {\bibinfo {volume}
  {E22}},\ \bibinfo {pages} {1330021} (\bibinfo {year} {2013})},\ \Eprint
  {http://arxiv.org/abs/1306.6235} {arXiv:1306.6235 [nucl-th]} \BibitemShut
  {NoStop}%
%%CITATION = ARXIV:1306.6235;%%
\bibitem [{\citenamefont {Ryckebusch}\ \emph {et~al.}(2015)\citenamefont
  {Ryckebusch}, \citenamefont {Cosyn},\ and\ \citenamefont
  {Vanhalst}}]{Vanhalst:2014cqa}%
  \BibitemOpen
  \bibfield  {author} {\bibinfo {author} {\bibfnamefont {J.}~\bibnamefont
  {Ryckebusch}}, \bibinfo {author} {\bibfnamefont {W.}~\bibnamefont {Cosyn}}, \
  and\ \bibinfo {author} {\bibfnamefont {M.}~\bibnamefont {Vanhalst}},\ }\href
  {\doibase 10.1088/0954-3899/42/5/055104} {\bibfield  {journal} {\bibinfo
  {journal} {J. Phys.}\ }\textbf {\bibinfo {volume} {G42}},\ \bibinfo {pages}
  {055104} (\bibinfo {year} {2015})},\ \Eprint {http://arxiv.org/abs/1405.3814}
  {arXiv:1405.3814 [nucl-th]} \BibitemShut {NoStop}%
%%CITATION = ARXIV:1405.3814;%%
\bibitem [{\citenamefont {Atti}(2015)}]{Atti:2015eda}%
  \BibitemOpen
  \bibfield  {author} {\bibinfo {author} {\bibfnamefont {C.~C.~d.}\
  \bibnamefont {Atti}},\ }\href {\doibase 10.1016/j.physrep.2015.06.002}
  {\bibfield  {journal} {\bibinfo  {journal} {Phys. Rept.}\ }\textbf {\bibinfo
  {volume} {590}},\ \bibinfo {pages} {1} (\bibinfo {year} {2015})}\BibitemShut
  {NoStop}%
%%CITATION = PRPLC,590,1;%%
\bibitem [{\citenamefont {Hen}\ \emph {et~al.}(2017)\citenamefont {Hen},
  \citenamefont {Miller}, \citenamefont {Piasetzky},\ and\ \citenamefont
  {Weinstein}}]{Hen:2016kwk}%
  \BibitemOpen
  \bibfield  {author} {\bibinfo {author} {\bibfnamefont {O.}~\bibnamefont
  {Hen}}, \bibinfo {author} {\bibfnamefont {G.~A.}\ \bibnamefont {Miller}},
  \bibinfo {author} {\bibfnamefont {E.}~\bibnamefont {Piasetzky}}, \ and\
  \bibinfo {author} {\bibfnamefont {L.~B.}\ \bibnamefont {Weinstein}},\ }\href
  {\doibase 10.1103/RevModPhys.89.045002} {\bibfield  {journal} {\bibinfo
  {journal} {Rev. Mod. Phys.}\ }\textbf {\bibinfo {volume} {89}},\ \bibinfo
  {pages} {045002} (\bibinfo {year} {2017})},\ \Eprint
  {http://arxiv.org/abs/1611.09748} {arXiv:1611.09748 [nucl-ex]} \BibitemShut
  {NoStop}%
%%CITATION = ARXIV:1611.09748;%%
\bibitem [{\citenamefont {Furnstahl}(2012)}]{Furnstahl:2012fn}%
  \BibitemOpen
  \bibfield  {author} {\bibinfo {author} {\bibfnamefont {R.}~\bibnamefont
  {Furnstahl}},\ }\href {\doibase 10.1016/j.nuclphysbps.2012.06.005} {\bibfield
   {journal} {\bibinfo  {journal} {Nucl. Phys. Proc. Suppl.}\ }\textbf
  {\bibinfo {volume} {228}},\ \bibinfo {pages} {139} (\bibinfo {year}
  {2012})},\ \Eprint {http://arxiv.org/abs/1203.1779} {arXiv:1203.1779
  [nucl-th]} \BibitemShut {NoStop}%
%%CITATION = ARXIV:1203.1779;%%
\bibitem [{\citenamefont {Duguet}\ \emph {et~al.}(2015)\citenamefont {Duguet},
  \citenamefont {Hergert}, \citenamefont {Holt},\ and\ \citenamefont
  {Somà}}]{Duguet:2014tua}%
  \BibitemOpen
  \bibfield  {author} {\bibinfo {author} {\bibfnamefont {T.}~\bibnamefont
  {Duguet}}, \bibinfo {author} {\bibfnamefont {H.}~\bibnamefont {Hergert}},
  \bibinfo {author} {\bibfnamefont {J.~D.}\ \bibnamefont {Holt}}, \ and\
  \bibinfo {author} {\bibfnamefont {V.}~\bibnamefont {Somà}},\ }\href
  {\doibase 10.1103/PhysRevC.92.034313} {\bibfield  {journal} {\bibinfo
  {journal} {Phys. Rev.}\ }\textbf {\bibinfo {volume} {C92}},\ \bibinfo {pages}
  {034313} (\bibinfo {year} {2015})},\ \Eprint {http://arxiv.org/abs/1411.1237}
  {arXiv:1411.1237 [nucl-th]} \BibitemShut {NoStop}%
%%CITATION = ARXIV:1411.1237;%%
\bibitem [{\citenamefont {Braaten}\ and\ \citenamefont
  {Platter}(2008)}]{Braaten:2008uh}%
  \BibitemOpen
  \bibfield  {author} {\bibinfo {author} {\bibfnamefont {E.}~\bibnamefont
  {Braaten}}\ and\ \bibinfo {author} {\bibfnamefont {L.}~\bibnamefont
  {Platter}},\ }\href {\doibase 10.1103/PhysRevLett.100.205301} {\bibfield
  {journal} {\bibinfo  {journal} {Phys. Rev. Lett.}\ }\textbf {\bibinfo
  {volume} {100}},\ \bibinfo {pages} {205301} (\bibinfo {year} {2008})},\
  \Eprint {http://arxiv.org/abs/0803.1125} {arXiv:0803.1125 [cond-mat.other]}
  \BibitemShut {NoStop}%
%%CITATION = ARXIV:0803.1125;%%
\bibitem [{\citenamefont {Hofmann}\ \emph {et~al.}(2013)\citenamefont
  {Hofmann}, \citenamefont {Barth},\ and\ \citenamefont
  {Zwerger}}]{Hofmann:2013oia}%
  \BibitemOpen
  \bibfield  {author} {\bibinfo {author} {\bibfnamefont {J.}~\bibnamefont
  {Hofmann}}, \bibinfo {author} {\bibfnamefont {M.}~\bibnamefont {Barth}}, \
  and\ \bibinfo {author} {\bibfnamefont {W.}~\bibnamefont {Zwerger}},\ }\href
  {\doibase 10.1103/PhysRevB.87.235125} {\bibfield  {journal} {\bibinfo
  {journal} {Phys. Rev.}\ }\textbf {\bibinfo {volume} {B87}},\ \bibinfo {pages}
  {235125} (\bibinfo {year} {2013})},\ \Eprint {http://arxiv.org/abs/1304.2891}
  {arXiv:1304.2891 [cond-mat.str-el]} \BibitemShut {NoStop}%
%%CITATION = ARXIV:1304.2891;%%
\bibitem [{\citenamefont {Bogner}\ \emph {et~al.}(2007)\citenamefont {Bogner},
  \citenamefont {Furnstahl},\ and\ \citenamefont {Perry}}]{Bogner:2006pc}%
  \BibitemOpen
  \bibfield  {author} {\bibinfo {author} {\bibfnamefont {S.~K.}\ \bibnamefont
  {Bogner}}, \bibinfo {author} {\bibfnamefont {R.~J.}\ \bibnamefont
  {Furnstahl}}, \ and\ \bibinfo {author} {\bibfnamefont {R.~J.}\ \bibnamefont
  {Perry}},\ }\href@noop {} {\bibfield  {journal} {\bibinfo  {journal} {Phys.
  Rev. C}\ }\textbf {\bibinfo {volume} {75}},\ \bibinfo {pages} {061001}
  (\bibinfo {year} {2007})},\ \Eprint {http://arxiv.org/abs/nucl-th/0611045}
  {nucl-th/0611045} \BibitemShut {NoStop}%
%%CITATION = NUCL-TH/0611045;%%
\bibitem [{\citenamefont {Furnstahl}\ and\ \citenamefont
  {Hebeler}(2013)}]{Furnstahl:2013oba}%
  \BibitemOpen
  \bibfield  {author} {\bibinfo {author} {\bibfnamefont {R.~J.}\ \bibnamefont
  {Furnstahl}}\ and\ \bibinfo {author} {\bibfnamefont {K.}~\bibnamefont
  {Hebeler}},\ }\href {\doibase 10.1088/0034-4885/76/12/126301} {\bibfield
  {journal} {\bibinfo  {journal} {Rept. Prog. Phys.}\ }\textbf {\bibinfo
  {volume} {76}},\ \bibinfo {pages} {126301} (\bibinfo {year} {2013})},\
  \Eprint {http://arxiv.org/abs/1305.3800} {arXiv:1305.3800 [nucl-th]}
  \BibitemShut {NoStop}%
%%CITATION = ARXIV:1305.3800;%%
\bibitem [{\citenamefont {Binder}\ \emph {et~al.}(2014)\citenamefont {Binder},
  \citenamefont {Langhammer}, \citenamefont {Calci},\ and\ \citenamefont
  {Roth}}]{Binder:2013xaa}%
  \BibitemOpen
  \bibfield  {author} {\bibinfo {author} {\bibfnamefont {S.}~\bibnamefont
  {Binder}}, \bibinfo {author} {\bibfnamefont {J.}~\bibnamefont {Langhammer}},
  \bibinfo {author} {\bibfnamefont {A.}~\bibnamefont {Calci}}, \ and\ \bibinfo
  {author} {\bibfnamefont {R.}~\bibnamefont {Roth}},\ }\href {\doibase
  10.1016/j.physletb.2014.07.010} {\bibfield  {journal} {\bibinfo  {journal}
  {Phys. Lett.}\ }\textbf {\bibinfo {volume} {B736}},\ \bibinfo {pages} {119}
  (\bibinfo {year} {2014})},\ \Eprint {http://arxiv.org/abs/1312.5685}
  {arXiv:1312.5685 [nucl-th]} \BibitemShut {NoStop}%
%%CITATION = ARXIV:1312.5685;%%
\bibitem [{\citenamefont {Roth}\ \emph
  {et~al.}(2014{\natexlab{a}})\citenamefont {Roth}, \citenamefont {Calci},
  \citenamefont {Langhammer},\ and\ \citenamefont {Binder}}]{Roth:2013fqa}%
  \BibitemOpen
  \bibfield  {author} {\bibinfo {author} {\bibfnamefont {R.}~\bibnamefont
  {Roth}}, \bibinfo {author} {\bibfnamefont {A.}~\bibnamefont {Calci}},
  \bibinfo {author} {\bibfnamefont {J.}~\bibnamefont {Langhammer}}, \ and\
  \bibinfo {author} {\bibfnamefont {S.}~\bibnamefont {Binder}},\ }\href
  {\doibase 10.1103/PhysRevC.90.024325} {\bibfield  {journal} {\bibinfo
  {journal} {Phys. Rev.}\ }\textbf {\bibinfo {volume} {C90}},\ \bibinfo {pages}
  {024325} (\bibinfo {year} {2014}{\natexlab{a}})},\ \Eprint
  {http://arxiv.org/abs/1311.3563} {arXiv:1311.3563 [nucl-th]} \BibitemShut
  {NoStop}%
%%CITATION = ARXIV:1311.3563;%%
\bibitem [{\citenamefont {Roth}\ \emph
  {et~al.}(2014{\natexlab{b}})\citenamefont {Roth}, \citenamefont {Calci},
  \citenamefont {Langhammer},\ and\ \citenamefont {Binder}}]{Roth:2014vla}%
  \BibitemOpen
  \bibfield  {author} {\bibinfo {author} {\bibfnamefont {R.}~\bibnamefont
  {Roth}}, \bibinfo {author} {\bibfnamefont {A.}~\bibnamefont {Calci}},
  \bibinfo {author} {\bibfnamefont {J.}~\bibnamefont {Langhammer}}, \ and\
  \bibinfo {author} {\bibfnamefont {S.}~\bibnamefont {Binder}},\ }\href
  {\doibase 10.1007/s00601-014-0860-0} {\bibfield  {journal} {\bibinfo
  {journal} {Few Body Syst.}\ }\textbf {\bibinfo {volume} {55}},\ \bibinfo
  {pages} {659} (\bibinfo {year} {2014}{\natexlab{b}})}\BibitemShut {NoStop}%
%%CITATION = FBSYE,55,659;%%
\bibitem [{\citenamefont {Schuster}\ \emph {et~al.}(2014)\citenamefont
  {Schuster}, \citenamefont {Quaglioni}, \citenamefont {Johnson}, \citenamefont
  {Jurgenson},\ and\ \citenamefont {Navratil}}]{Schuster:2014lga}%
  \BibitemOpen
  \bibfield  {author} {\bibinfo {author} {\bibfnamefont {M.~D.}\ \bibnamefont
  {Schuster}}, \bibinfo {author} {\bibfnamefont {S.}~\bibnamefont {Quaglioni}},
  \bibinfo {author} {\bibfnamefont {C.~W.}\ \bibnamefont {Johnson}}, \bibinfo
  {author} {\bibfnamefont {E.~D.}\ \bibnamefont {Jurgenson}}, \ and\ \bibinfo
  {author} {\bibfnamefont {P.}~\bibnamefont {Navratil}},\ }\href {\doibase
  10.1103/PhysRevC.90.011301} {\bibfield  {journal} {\bibinfo  {journal} {Phys.
  Rev.}\ }\textbf {\bibinfo {volume} {C90}},\ \bibinfo {pages} {011301}
  (\bibinfo {year} {2014})},\ \Eprint {http://arxiv.org/abs/1402.7106}
  {arXiv:1402.7106 [nucl-th]} \BibitemShut {NoStop}%
%%CITATION = ARXIV:1402.7106;%%
\bibitem [{\citenamefont {Neff}\ \emph {et~al.}(2015)\citenamefont {Neff},
  \citenamefont {Feldmeier},\ and\ \citenamefont {Horiuchi}}]{Neff:2015xda}%
  \BibitemOpen
  \bibfield  {author} {\bibinfo {author} {\bibfnamefont {T.}~\bibnamefont
  {Neff}}, \bibinfo {author} {\bibfnamefont {H.}~\bibnamefont {Feldmeier}}, \
  and\ \bibinfo {author} {\bibfnamefont {W.}~\bibnamefont {Horiuchi}},\ }\href
  {\doibase 10.1103/PhysRevC.92.024003} {\bibfield  {journal} {\bibinfo
  {journal} {Phys. Rev.}\ }\textbf {\bibinfo {volume} {C92}},\ \bibinfo {pages}
  {024003} (\bibinfo {year} {2015})},\ \Eprint
  {http://arxiv.org/abs/1506.02237} {arXiv:1506.02237 [nucl-th]} \BibitemShut
  {NoStop}%
%%CITATION = ARXIV:1506.02237;%%
\bibitem [{\citenamefont {Neff}\ and\ \citenamefont
  {Feldmeier}(2016)}]{Neff:2016ajx}%
  \BibitemOpen
  \bibfield  {author} {\bibinfo {author} {\bibfnamefont {T.}~\bibnamefont
  {Neff}}\ and\ \bibinfo {author} {\bibfnamefont {H.}~\bibnamefont
  {Feldmeier}},\ }\href@noop {} {\  (\bibinfo {year} {2016})},\ \Eprint
  {http://arxiv.org/abs/1610.04066} {arXiv:1610.04066 [nucl-th]} \BibitemShut
  {NoStop}%
%%CITATION = ARXIV:1610.04066;%%
\bibitem [{\citenamefont {Parzuchowski}\ \emph {et~al.}(2017)\citenamefont
  {Parzuchowski}, \citenamefont {Stroberg}, \citenamefont {Navrátil},
  \citenamefont {Hergert},\ and\ \citenamefont
  {Bogner}}]{Parzuchowski:2017wcq}%
  \BibitemOpen
  \bibfield  {author} {\bibinfo {author} {\bibfnamefont {N.~M.}\ \bibnamefont
  {Parzuchowski}}, \bibinfo {author} {\bibfnamefont {S.~R.}\ \bibnamefont
  {Stroberg}}, \bibinfo {author} {\bibfnamefont {P.}~\bibnamefont {Navrátil}},
  \bibinfo {author} {\bibfnamefont {H.}~\bibnamefont {Hergert}}, \ and\
  \bibinfo {author} {\bibfnamefont {S.~K.}\ \bibnamefont {Bogner}},\
  }\href@noop {} {\  (\bibinfo {year} {2017})},\ \Eprint
  {http://arxiv.org/abs/1705.05511} {arXiv:1705.05511 [nucl-th]} \BibitemShut
  {NoStop}%
%%CITATION = ARXIV:1705.05511;%%
\bibitem [{\citenamefont {Hergert}\ \emph {et~al.}(2017)\citenamefont
  {Hergert}, \citenamefont {Bogner}, \citenamefont {Lietz}, \citenamefont
  {Morris}, \citenamefont {Novario}, \citenamefont {Parzuchowski},\ and\
  \citenamefont {Yuan}}]{Hergert2017}%
  \BibitemOpen
  \bibfield  {author} {\bibinfo {author} {\bibfnamefont {H.}~\bibnamefont
  {Hergert}}, \bibinfo {author} {\bibfnamefont {S.~K.}\ \bibnamefont {Bogner}},
  \bibinfo {author} {\bibfnamefont {J.~G.}\ \bibnamefont {Lietz}}, \bibinfo
  {author} {\bibfnamefont {T.~D.}\ \bibnamefont {Morris}}, \bibinfo {author}
  {\bibfnamefont {S.~J.}\ \bibnamefont {Novario}}, \bibinfo {author}
  {\bibfnamefont {N.~M.}\ \bibnamefont {Parzuchowski}}, \ and\ \bibinfo
  {author} {\bibfnamefont {F.}~\bibnamefont {Yuan}},\ }\enquote {\bibinfo
  {title} {In-medium similarity renormalization group approach to the nuclear
  many-body problem},}\ in\ \href {\doibase 10.1007/978-3-319-53336-0_10}
  {\emph {\bibinfo {booktitle} {An Advanced Course in Computational Nuclear
  Physics: Bridging the Scales from Quarks to Neutron Stars}}},\ \bibinfo
  {editor} {edited by\ \bibinfo {editor} {\bibfnamefont {M.}~\bibnamefont
  {Hjorth-Jensen}}, \bibinfo {editor} {\bibfnamefont {M.~P.}\ \bibnamefont
  {Lombardo}}, \ and\ \bibinfo {editor} {\bibfnamefont {U.}~\bibnamefont {van
  Kolck}}}\ (\bibinfo  {publisher} {Springer International Publishing},\
  \bibinfo {address} {Cham},\ \bibinfo {year} {2017})\ pp.\ \bibinfo {pages}
  {477--570}\BibitemShut {NoStop}%
\bibitem [{\citenamefont {Bacca}\ and\ \citenamefont
  {Pastore}(2014)}]{Bacca:2014tla}%
  \BibitemOpen
  \bibfield  {author} {\bibinfo {author} {\bibfnamefont {S.}~\bibnamefont
  {Bacca}}\ and\ \bibinfo {author} {\bibfnamefont {S.}~\bibnamefont
  {Pastore}},\ }\href {\doibase 10.1088/0954-3899/41/12/123002} {\bibfield
  {journal} {\bibinfo  {journal} {J. Phys.}\ }\textbf {\bibinfo {volume}
  {G41}},\ \bibinfo {pages} {123002} (\bibinfo {year} {2014})},\ \Eprint
  {http://arxiv.org/abs/1407.3490} {arXiv:1407.3490 [nucl-th]} \BibitemShut
  {NoStop}%
%%CITATION = ARXIV:1407.3490;%%
\bibitem [{\citenamefont {Navrátil}\ \emph {et~al.}(2016)\citenamefont
  {Navrátil}, \citenamefont {Quaglioni}, \citenamefont {Hupin}, \citenamefont
  {Romero-Redondo},\ and\ \citenamefont {Calci}}]{Navratil:2016ycn}%
  \BibitemOpen
  \bibfield  {author} {\bibinfo {author} {\bibfnamefont {P.}~\bibnamefont
  {Navrátil}}, \bibinfo {author} {\bibfnamefont {S.}~\bibnamefont
  {Quaglioni}}, \bibinfo {author} {\bibfnamefont {G.}~\bibnamefont {Hupin}},
  \bibinfo {author} {\bibfnamefont {C.}~\bibnamefont {Romero-Redondo}}, \ and\
  \bibinfo {author} {\bibfnamefont {A.}~\bibnamefont {Calci}},\ }\href
  {\doibase 10.1088/0031-8949/91/5/053002} {\bibfield  {journal} {\bibinfo
  {journal} {Phys. Scripta}\ }\textbf {\bibinfo {volume} {91}},\ \bibinfo
  {pages} {053002} (\bibinfo {year} {2016})},\ \Eprint
  {http://arxiv.org/abs/1601.03765} {arXiv:1601.03765 [nucl-th]} \BibitemShut
  {NoStop}%
%%CITATION = ARXIV:1601.03765;%%
\bibitem [{\citenamefont {More}\ \emph {et~al.}(2015)\citenamefont {More},
  \citenamefont {König}, \citenamefont {Furnstahl},\ and\ \citenamefont
  {Hebeler}}]{More:2015tpa}%
  \BibitemOpen
  \bibfield  {author} {\bibinfo {author} {\bibfnamefont {S.~N.}\ \bibnamefont
  {More}}, \bibinfo {author} {\bibfnamefont {S.}~\bibnamefont {König}},
  \bibinfo {author} {\bibfnamefont {R.~J.}\ \bibnamefont {Furnstahl}}, \ and\
  \bibinfo {author} {\bibfnamefont {K.}~\bibnamefont {Hebeler}},\ }\href
  {\doibase 10.1103/PhysRevC.92.064002} {\bibfield  {journal} {\bibinfo
  {journal} {Phys. Rev.}\ }\textbf {\bibinfo {volume} {C92}},\ \bibinfo {pages}
  {064002} (\bibinfo {year} {2015})},\ \Eprint
  {http://arxiv.org/abs/1510.04955} {arXiv:1510.04955 [nucl-th]} \BibitemShut
  {NoStop}%
%%CITATION = ARXIV:1510.04955;%%
\bibitem [{\citenamefont {Yang}\ and\ \citenamefont
  {Phillips}(2013)}]{Yang:2013rza}%
  \BibitemOpen
  \bibfield  {author} {\bibinfo {author} {\bibfnamefont {C.-J.}\ \bibnamefont
  {Yang}}\ and\ \bibinfo {author} {\bibfnamefont {D.~R.}\ \bibnamefont
  {Phillips}},\ }\href {\doibase 10.1140/epja/i2013-13122-8} {\bibfield
  {journal} {\bibinfo  {journal} {Eur. Phys. J. A}\ }\textbf {\bibinfo {volume}
  {49}},\ \bibinfo {pages} {122} (\bibinfo {year} {2013})},\ \Eprint
  {http://arxiv.org/abs/1305.7279} {arXiv:1305.7279 [nucl-th]} \BibitemShut
  {NoStop}%
%%CITATION = ARXIV:1305.7279;%%
\bibitem [{\citenamefont {Dainton}\ \emph {et~al.}(2014)\citenamefont
  {Dainton}, \citenamefont {Furnstahl},\ and\ \citenamefont
  {Perry}}]{Dainton:2013axa}%
  \BibitemOpen
  \bibfield  {author} {\bibinfo {author} {\bibfnamefont {B.}~\bibnamefont
  {Dainton}}, \bibinfo {author} {\bibfnamefont {R.~J.}\ \bibnamefont
  {Furnstahl}}, \ and\ \bibinfo {author} {\bibfnamefont {R.~J.}\ \bibnamefont
  {Perry}},\ }\href {\doibase 10.1103/PhysRevC.89.014001} {\bibfield  {journal}
  {\bibinfo  {journal} {Phys. Rev.}\ }\textbf {\bibinfo {volume} {C89}},\
  \bibinfo {pages} {014001} (\bibinfo {year} {2014})},\ \Eprint
  {http://arxiv.org/abs/1310.6690} {arXiv:1310.6690 [nucl-th]} \BibitemShut
  {NoStop}%
%%CITATION = ARXIV:1310.6690;%%
\bibitem [{\citenamefont {Wiringa}\ \emph {et~al.}(1995)\citenamefont
  {Wiringa}, \citenamefont {Stoks},\ and\ \citenamefont
  {Schiavilla}}]{Wiringa:1994wb}%
  \BibitemOpen
  \bibfield  {author} {\bibinfo {author} {\bibfnamefont {R.~B.}\ \bibnamefont
  {Wiringa}}, \bibinfo {author} {\bibfnamefont {V.~G.~J.}\ \bibnamefont
  {Stoks}}, \ and\ \bibinfo {author} {\bibfnamefont {R.}~\bibnamefont
  {Schiavilla}},\ }\href {\doibase 10.1103/PhysRevC.51.38} {\bibfield
  {journal} {\bibinfo  {journal} {Phys. Rev. C}\ }\textbf {\bibinfo {volume}
  {51}},\ \bibinfo {pages} {38} (\bibinfo {year} {1995})},\ \Eprint
  {http://arxiv.org/abs/nucl-th/9408016} {arXiv:nucl-th/9408016} \BibitemShut
  {NoStop}%
%%CITATION = NUCL-TH/9408016;%%
\bibitem [{\citenamefont {Mukhamedzhanov}\ and\ \citenamefont
  {Kadyrov}(2010)}]{Mukhamedzhanov:2010hc}%
  \BibitemOpen
  \bibfield  {author} {\bibinfo {author} {\bibfnamefont {A.~M.}\ \bibnamefont
  {Mukhamedzhanov}}\ and\ \bibinfo {author} {\bibfnamefont {A.~S.}\
  \bibnamefont {Kadyrov}},\ }\href {\doibase 10.1103/PhysRevC.82.051601}
  {\bibfield  {journal} {\bibinfo  {journal} {Phys. Rev.}\ }\textbf {\bibinfo
  {volume} {C82}},\ \bibinfo {pages} {051601} (\bibinfo {year} {2010})},\
  \Eprint {http://arxiv.org/abs/1005.3788} {arXiv:1005.3788 [nucl-th]}
  \BibitemShut {NoStop}%
%%CITATION = ARXIV:1005.3788;%%
\bibitem [{\citenamefont {Gezerlis}\ \emph {et~al.}(2014)\citenamefont
  {Gezerlis}, \citenamefont {Tews}, \citenamefont {Epelbaum}, \citenamefont
  {Freunek}, \citenamefont {Gandolfi}, \citenamefont {Hebeler}, \citenamefont
  {Nogga},\ and\ \citenamefont {Schwenk}}]{Gezerlis:2014zia}%
  \BibitemOpen
  \bibfield  {author} {\bibinfo {author} {\bibfnamefont {A.}~\bibnamefont
  {Gezerlis}}, \bibinfo {author} {\bibfnamefont {I.}~\bibnamefont {Tews}},
  \bibinfo {author} {\bibfnamefont {E.}~\bibnamefont {Epelbaum}}, \bibinfo
  {author} {\bibfnamefont {M.}~\bibnamefont {Freunek}}, \bibinfo {author}
  {\bibfnamefont {S.}~\bibnamefont {Gandolfi}}, \bibinfo {author}
  {\bibfnamefont {K.}~\bibnamefont {Hebeler}}, \bibinfo {author} {\bibfnamefont
  {A.}~\bibnamefont {Nogga}}, \ and\ \bibinfo {author} {\bibfnamefont
  {A.}~\bibnamefont {Schwenk}},\ }\href {\doibase 10.1103/PhysRevC.90.054323}
  {\bibfield  {journal} {\bibinfo  {journal} {Phys. Rev.}\ }\textbf {\bibinfo
  {volume} {C90}},\ \bibinfo {pages} {054323} (\bibinfo {year} {2014})},\
  \Eprint {http://arxiv.org/abs/1406.0454} {arXiv:1406.0454 [nucl-th]}
  \BibitemShut {NoStop}%
%%CITATION = ARXIV:1406.0454;%%
\bibitem [{\citenamefont {Epelbaum}\ \emph {et~al.}(2015)\citenamefont
  {Epelbaum}, \citenamefont {Krebs},\ and\ \citenamefont
  {Mei{\ss}ner}}]{Epelbaum:2014sza}%
  \BibitemOpen
  \bibfield  {author} {\bibinfo {author} {\bibfnamefont {E.}~\bibnamefont
  {Epelbaum}}, \bibinfo {author} {\bibfnamefont {H.}~\bibnamefont {Krebs}}, \
  and\ \bibinfo {author} {\bibfnamefont {U.~G.}\ \bibnamefont {Mei{\ss}ner}},\
  }\href {\doibase 10.1103/PhysRevLett.115.122301} {\bibfield  {journal}
  {\bibinfo  {journal} {Phys. Rev. Lett.}\ }\textbf {\bibinfo {volume} {115}},\
  \bibinfo {pages} {122301} (\bibinfo {year} {2015})},\ \Eprint
  {http://arxiv.org/abs/1412.4623} {arXiv:1412.4623 [nucl-th]} \BibitemShut
  {NoStop}%
%%CITATION = ARXIV:1412.4623;%%
\bibitem [{\citenamefont {Piarulli}\ \emph {et~al.}(2015)\citenamefont
  {Piarulli}, \citenamefont {Girlanda}, \citenamefont {Schiavilla},
  \citenamefont {Navarro~Pérez}, \citenamefont {Amaro},\ and\ \citenamefont
  {Ruiz~Arriola}}]{Piarulli:2014bda}%
  \BibitemOpen
  \bibfield  {author} {\bibinfo {author} {\bibfnamefont {M.}~\bibnamefont
  {Piarulli}}, \bibinfo {author} {\bibfnamefont {L.}~\bibnamefont {Girlanda}},
  \bibinfo {author} {\bibfnamefont {R.}~\bibnamefont {Schiavilla}}, \bibinfo
  {author} {\bibfnamefont {R.}~\bibnamefont {Navarro~Pérez}}, \bibinfo
  {author} {\bibfnamefont {J.~E.}\ \bibnamefont {Amaro}}, \ and\ \bibinfo
  {author} {\bibfnamefont {E.}~\bibnamefont {Ruiz~Arriola}},\ }\href {\doibase
  10.1103/PhysRevC.91.024003} {\bibfield  {journal} {\bibinfo  {journal} {Phys.
  Rev.}\ }\textbf {\bibinfo {volume} {C91}},\ \bibinfo {pages} {024003}
  (\bibinfo {year} {2015})},\ \Eprint {http://arxiv.org/abs/1412.6446}
  {arXiv:1412.6446 [nucl-th]} \BibitemShut {NoStop}%
%%CITATION = ARXIV:1412.6446;%%
\bibitem [{\citenamefont {Entem}\ \emph {et~al.}(2017)\citenamefont {Entem},
  \citenamefont {Machleidt},\ and\ \citenamefont {Nosyk}}]{Entem:2017gor}%
  \BibitemOpen
  \bibfield  {author} {\bibinfo {author} {\bibfnamefont {D.~R.}\ \bibnamefont
  {Entem}}, \bibinfo {author} {\bibfnamefont {R.}~\bibnamefont {Machleidt}}, \
  and\ \bibinfo {author} {\bibfnamefont {Y.}~\bibnamefont {Nosyk}},\
  }\href@noop {} {\  (\bibinfo {year} {2017})},\ \Eprint
  {http://arxiv.org/abs/1703.05454} {arXiv:1703.05454 [nucl-th]} \BibitemShut
  {NoStop}%
%%CITATION = ARXIV:1703.05454;%%
\bibitem [{\citenamefont {Anderson}\ \emph {et~al.}(2010)\citenamefont
  {Anderson}, \citenamefont {Bogner}, \citenamefont {Furnstahl},\ and\
  \citenamefont {Perry}}]{Anderson:2010aq}%
  \BibitemOpen
  \bibfield  {author} {\bibinfo {author} {\bibfnamefont {E.}~\bibnamefont
  {Anderson}}, \bibinfo {author} {\bibfnamefont {S.}~\bibnamefont {Bogner}},
  \bibinfo {author} {\bibfnamefont {R.}~\bibnamefont {Furnstahl}}, \ and\
  \bibinfo {author} {\bibfnamefont {R.}~\bibnamefont {Perry}},\ }\href
  {\doibase 10.1103/PhysRevC.82.054001} {\bibfield  {journal} {\bibinfo
  {journal} {Phys. Rev. C}\ }\textbf {\bibinfo {volume} {82}},\ \bibinfo
  {pages} {054001} (\bibinfo {year} {2010})},\ \Eprint
  {http://arxiv.org/abs/1008.1569} {arXiv:1008.1569 [nucl-th]} \BibitemShut
  {NoStop}%
%%CITATION = ARXIV:1008.1569;%%
\bibitem [{\citenamefont {Fabian}\ and\ \citenamefont
  {Arenhovel}(1979)}]{Fabian:1979kx}%
  \BibitemOpen
  \bibfield  {author} {\bibinfo {author} {\bibfnamefont {W.}~\bibnamefont
  {Fabian}}\ and\ \bibinfo {author} {\bibfnamefont {H.}~\bibnamefont
  {Arenhovel}},\ }\href {\doibase 10.1016/0375-9474(79)90599-2} {\bibfield
  {journal} {\bibinfo  {journal} {Nucl. Phys.}\ }\textbf {\bibinfo {volume}
  {A314}},\ \bibinfo {pages} {253} (\bibinfo {year} {1979})}\BibitemShut
  {NoStop}%
%%CITATION = NUPHA,A314,253;%%
\bibitem [{\citenamefont {Bogner}\ and\ \citenamefont
  {Roscher}(2012)}]{Bogner:2012zm}%
  \BibitemOpen
  \bibfield  {author} {\bibinfo {author} {\bibfnamefont {S.}~\bibnamefont
  {Bogner}}\ and\ \bibinfo {author} {\bibfnamefont {D.}~\bibnamefont
  {Roscher}},\ }\href {\doibase 10.1103/PhysRevC.86.064304} {\bibfield
  {journal} {\bibinfo  {journal} {Phys. Rev. C}\ }\textbf {\bibinfo {volume}
  {86}},\ \bibinfo {pages} {064304} (\bibinfo {year} {2012})},\ \Eprint
  {http://arxiv.org/abs/1208.1734} {arXiv:1208.1734 [nucl-th]} \BibitemShut
  {NoStop}%
%%CITATION = ARXIV:1208.1734;%%
\bibitem [{\citenamefont {Amado}(1979)}]{Amado:1979zz}%
  \BibitemOpen
  \bibfield  {author} {\bibinfo {author} {\bibfnamefont {R.~D.}\ \bibnamefont
  {Amado}},\ }\href {\doibase 10.1103/PhysRevC.19.1473} {\bibfield  {journal}
  {\bibinfo  {journal} {Phys. Rev. C}\ }\textbf {\bibinfo {volume} {19}},\
  \bibinfo {pages} {1473} (\bibinfo {year} {1979})}\BibitemShut {NoStop}%
%%CITATION = PHRVA,C19,1473;%%
\bibitem [{\citenamefont {Friar}(1979)}]{Friar:1979zz}%
  \BibitemOpen
  \bibfield  {author} {\bibinfo {author} {\bibfnamefont {J.~L.}\ \bibnamefont
  {Friar}},\ }\href {\doibase 10.1103/PhysRevC.20.325} {\bibfield  {journal}
  {\bibinfo  {journal} {Phys. Rev. C}\ }\textbf {\bibinfo {volume} {20}},\
  \bibinfo {pages} {325} (\bibinfo {year} {1979})}\BibitemShut {NoStop}%
%%CITATION = PHRVA,C20,325;%%
\bibitem [{\citenamefont {Anderson}\ \emph {et~al.}(2008)\citenamefont
  {Anderson} \emph {et~al.}}]{Anderson:2008mu}%
  \BibitemOpen
  \bibfield  {author} {\bibinfo {author} {\bibfnamefont {E.}~\bibnamefont
  {Anderson}} \emph {et~al.},\ }\href {\doibase 10.1103/PhysRevC.77.037001}
  {\bibfield  {journal} {\bibinfo  {journal} {Phys. Rev. C}\ }\textbf {\bibinfo
  {volume} {77}},\ \bibinfo {pages} {037001} (\bibinfo {year} {2008})},\
  \Eprint {http://arxiv.org/abs/0801.1098} {arXiv:0801.1098 [nucl-th]}
  \BibitemShut {NoStop}%
%%CITATION = 0801.1098;%%
\bibitem [{\citenamefont {Dicaire}\ \emph {et~al.}(2014)\citenamefont
  {Dicaire}, \citenamefont {Omand},\ and\ \citenamefont
  {Navratil}}]{Dicaire:2014fra}%
  \BibitemOpen
  \bibfield  {author} {\bibinfo {author} {\bibfnamefont {N.~M.}\ \bibnamefont
  {Dicaire}}, \bibinfo {author} {\bibfnamefont {C.}~\bibnamefont {Omand}}, \
  and\ \bibinfo {author} {\bibfnamefont {P.}~\bibnamefont {Navratil}},\ }\href
  {\doibase 10.1103/PhysRevC.90.034302} {\bibfield  {journal} {\bibinfo
  {journal} {Phys. Rev.}\ }\textbf {\bibinfo {volume} {C90}},\ \bibinfo {pages}
  {034302} (\bibinfo {year} {2014})},\ \Eprint {http://arxiv.org/abs/1406.1815}
  {arXiv:1406.1815 [nucl-th]} \BibitemShut {NoStop}%
%%CITATION = ARXIV:1406.1815;%%
\end{thebibliography}%
\end{document}